%% file: thesis.tex
\documentclass[12pt,a4paper,titlepage]{book}
\usepackage{graphicx}
\usepackage[latin1]{inputenc}
\usepackage{amssymb}
\usepackage[reqno]{amsmath}
\usepackage{a4wide}
\usepackage{fancyhdr}
\usepackage{marvosym}
\usepackage{sectsty}
\usepackage{cancel}
\usepackage{palatino}
\usepackage{mathpazo}
\usepackage{mathptmx}
\usepackage{epsfig}
\usepackage{amssymb}
\usepackage{color}
\definecolor{grey}{rgb}{.85,.85,.85}
\newcommand{\graybox}[1]{\colorbox{grey}%
{\parbox[t]{.95\textwidth}{#1}}}

\newcommand{\Tr}{\textrm{Tr}}

\newcommand{\csquarebox}[1]{\fbox{\hbox to #1{\vbox to #1{}\hss}}}
\newcommand{\squarebox}{\csquarebox{0.3cm}}
\newtheorem{definition}{Definition}
\newtheorem{lemma}{Lemma}
\newtheorem{theorem}{Theorem}
\newtheorem{postulate}{Postulate}
\newtheorem{corollary}{Corollary}
\begin{document}

\sectionfont{\fontfamily{phv}\selectfont}
\subsectionfont{\fontfamily{phv}\selectfont}
\subsubsectionfont{\fontfamily{phv}\selectfont}

\pagenumbering{roman}
\begin{titlepage}

\begin{center}
\vspace*{30mm}
{\Huge Unambiguous State Discrimination\\
of two density matrices\\
in Quantum Information Theory\\[3cm]}

{\large Den Naturwissenschaftlichen Fakult\"aten\\
der Friedrich-Alexander-Universit\"at Erlangen-N\"urnberg \\
zur\\
Erlangung des Doktorgrades\\[6cm]}

{\large vorgelegt von\\ Philippe Raynal \\ aus Lyon, Frankreich\\[6cm]

Quantum Information Theory Group\\ Theoretische Physik I\\[1.5cm]

Lehrstuhl f\"ur Optik\\ Institut f\"ur Optik, Information und Photonik\\ Max Planck Forschungsgruppe\\[1.5cm]

Erlangen 2006
}

\end{center}

\newpage

\noindent Als Dissertation genehmigt von den naturwissenschaftlichen Fakult\"aten der Universit\"at Erlangen-N\"urnberg\\[15cm]
\begin{tabbing}
Vorsitzender der Promotionskommission:xxxxxxx \=  \kill  
Tag der m\"undlichen Pr\"ufung: \>  16.08.2006\\
\\
Vorsitzender der Promotionskommission: \> Prof.~Dr.~D.~P.~H\"ader \\
Erstberichterstatter: \> Prof.~Dr.~N.~L\"utkenhaus \\
Zweitberichterstatter: \> Prof.~Dr.~D. Bru\ss \\
\end{tabbing}
\end{titlepage}

\pagestyle{fancy}
\newcommand{\helv}{%
\fontfamily{phv}\fontseries{b}\fontsize{12}{14}\selectfont}

\renewcommand{\chaptermark}[1]%
{\markboth{\MakeUppercase{\thechapter.\ #1}}{}}
\renewcommand{\sectionmark}[1]%
{\markright{\MakeUppercase{\thesection.\ #1}}}

\fancyhf{}
\fancyhead[LE,RO]{\rm\thepage}
\fancyhead[LO]{\nouppercase{\rm\rightmark}}
\fancyhead[RE]{\nouppercase{\rm\leftmark}}
\fancyhead[LE,RO]{\helv \thepage}
\fancyhead[LO]{\nouppercase{\helv \rightmark}}
\fancyhead[RE]{\nouppercase{\helv \leftmark}}
\fancypagestyle{plain}{\fancyhead{}\renewcommand{\headrulewidth}{0pt}}

\renewcommand{\chaptermark}[1]{\markboth{#1}{}} 
\cleardoublepage
\addcontentsline{toc}{chapter}{Abstract}
\input{Abstract.tex}
\cleardoublepage
\addcontentsline{toc}{chapter}{Zusammenfassung}
\input{Zusammenfassung.tex}
\cleardoublepage
\tableofcontents
\listoffigures
\mainmatter

\renewcommand{\chaptermark}[1]{\markboth{\thechapter.\ #1}{}}  

\input{chapter1.tex}
\input{chapter2.tex}

\input{chapter4.tex}
\input{chapter5.tex}
\input{chapter6_correction.tex}
\input{BB84.tex}
\input{conclusion.tex}
\input{Appendix.tex}
\cleardoublepage
\addcontentsline{toc}{chapter}{Bibliography}
\bibliography{thesis}
\bibliographystyle{unsrt}
\cleardoublepage
\addcontentsline{toc}{chapter}{Curriculum Vitae}
\input{Lebenslauf.tex}
\end{document}

%% file: Abstract.tex
\chapter*{Abstract}

Quantum state discrimination is a fundamental task in quantum information theory. The signals are usually nonorthogonal quantum states, which implies 
that they can not be perfectly distinguished. One possible discrimination strategy is the so-called Unambiguous State Discrimination (USD) where the states are successfully identified with non-unit probability, but without error. The optimal USD measurement has been extensively studied in the case of pure states, especially for any pair of pure states. Recently, the problem of unambiguously discriminating mixed quantum states has attracted much attention. In the case of a pair of generic mixed states, no complete solution is known. In this thesis, we first present reduction theorems for optimal unambiguous discrimination of two generic density matrices. We show that this problem can be reduced to that of two density matrices that have the same rank $r$ in a 2$r$-dimensional Hilbert space. These reduction theorems also allow us to reduce USD problems to simpler ones for which the solution might be known. As an application, we consider the unambiguous comparison of $n$ linearly independent pure states with a simple symmetry. Moreover, lower bounds on the optimal failure probability have been derived. For two mixed states they are given in terms of the fidelity. Here we give tighter bounds as well as necessary and sufficient conditions for two mixed states to reach these bounds. We also construct the corresponding optimal measurement. With this result, we provide analytical solutions for unambiguously discriminating a class of generic mixed states. This goes beyond known results which are all reducible to some pure state case. We however show that examples exist where the bounds cannot be reached. Next, we derive properties on the rank and the spectrum of an optimal USD measurement. This finally leads to a second class of exact solutions. Indeed we present the optimal failure probability as well as the optimal measurement for unambiguously discriminating any pair of geometrically uniform mixed states in four dimensions. This class of problems includes for example the discrimination of both the basis and the bit value mixed states in the BB84 QKD protocol with coherent states. 

%% file: Zusammenfassung.tex
\chapter*{Zusammenfassung}

Quantenzustandsunterscheidung ist eine fundamentale Aufgabe der Quanteninformationstheorie. Die Signale sind normalerweise nicht-orthogonale Quantenzust\"ande, d.h. sie k\"onnen nicht perfekt unterschieden werden. Eine der m\"oglichen Unterscheidungsstrategien ist die so genannte Eindeutige Zustandsunterschiedung (Unambiguous State Discrimination - USD), bei der die Zust\"ande mit einer Wahrscheinlichkeit kleiner als eins erfolgreich erkannt werden, allerdings fehlerfrei. Optimale USD-Messungen f\"ur reine Zust\"ande sind ausf\"uhrlich untersucht worden, insbesondere f\"ur jedes Paar von reinen Zust\"anden. Vor kurzem hat die Aufgabenstellung der eindeutigen Zustandsunterscheidung gemischter Zust\"ande viel Aufmerksamkeit auf sich gezogen. Im Falle eines Paares von allgemeinen gemischten Zust\"anden ist keine vollst\"andige L\"osung bekannt. In dieser Doktorarbeit legen wir zuerst Reduktionstheoreme f\"ur optimale eindeutige Unterscheidung von zwei allgemeinen Dichtematrizen vor. Wir zeigen, dass diese Aufgabenstellung reduziert werden kann auf diejenige von zwei Matrizen, die denselben Rang $r$ in einem 2$r$-dimensionalen Hibert-Raum haben. Diese Reduktionstheoreme erm\"oglichen uns ebenfalls, USD-Aufgaben auf einfachere zur\"uckzuf\"uhren, f\"ur die die L\"osung m\"oglicherweise bekannt ist. Der eindeutige Vergleich von $n$ linear abh\"angigen reinen Zust\"anden mit einfacher Symmetrie wird als Anwendung behandelt. Dar\"uber hinaus wurden untere Grenzen f\"ur die optimale Fehlerwahrscheinlichkeit entwickelt. F\"ur zwei gemischte Zust\"ande werden diese in Form der Fidelity angegeben. Hier geben wir engere Grenzen an, ebenso wie notwendige und ausreichende Bedingungen f\"ur zwei gemischte Zust\"ande, diese Grenzen zu erreichen. Wir konstruieren ebenfalls die entsprechende optimale Messung. Zusammen mit diesem Ergebnis pr\"asentieren wir analytische L\"osungen f\"ur die eindeutige Unterscheidung einer Kategorie allgemeiner gemischter Zust\"ande. Dies geht \"uber bekannte Ergebnisse hinaus, die alle auf reine Zust\"ande zur\"uckführbar sind. Wir zeigen allerdings, dass es Beispiele gibt, bei denen die Grenzen nicht erreicht werden k\"onnen. Als n\"achstes leiten wir Eigenschaften des Rangs und des Spektrums einer optimalen USD-Messung her. Dies führt schließlich zu einer zweiten Kategorie exakter L\"osungen. Wir zeigen die optimale Fehlerwahrscheinlichkeit auf, ebenso wie die optimale Messung, um jedes Paar geometrisch gleichf\"ormiger gemischter Zust\"ande in vier Dimensionen zu unterscheiden. Diese Kategorie von Aufgabenstellungen schließt zum Beispiel die Unterscheidung von sowohl der basis- als auch der bit value-gemischten Zust\"ande des BB84-QKD-Protokolls mit koh\"arenten Zust\"anden ein.

%% file: chapter1.tex
\chapter{Prologue} \label{prologue}

Physics attempts to describe the world with the language of mathematics. Given a system an observer summarizes his knowledge in an abstract mathematical object, the so-called 'state'. At a given point in time this observer may decide to acquire information about the system. Such an acquisition of information is called a measurement. In that sense, Quantum Mechanics is concerned with knowledge, and the two pillars of Quantum Mechanics are {\it states} and {\it measurements}.

Information Theory started in the late 1940's boosted by the second world war and its needs for communication and computational power. Information Theory addresses the fundamental questions of the transmission, processing and coding of information.

It is therefore quite natural that Quantum Mechanics and Information Theory finally merge to describe the production, the transmission and the detection of information as well as its processing and coding. Quantum Information Theory was born.

\section{Quantum Information Theory}
Since no information-theoretic formulation\footnote{See the work of R. Clifton, J. Bub and H. Halvorson or the work of A. Grinbaum for two appealing attempts.} is yet available, Quantum Information Theory (QIT) is formulated on the basis of four postulates that mathematically describe a physical system, its evolution and measurements that can be performed on it.
Let us now review these four postulates \cite{nielsen-chuang}.

\begin{postulate}
Hilbert space\\
\graybox{ Associated to any isolated quantum system is a Hilbert space known as the {\it state space} of the system. The system is completely described by a unit vector $|\Psi\rangle$ called the {\it state vector} in the {\it state space}.}
\end{postulate}

\begin{postulate}
Unitary evolution\\
\graybox{The evolution of a {\it closed} (i.e.\ an isolated system having no interaction with the environment) quantum system is described by a unitary transformation.
That is, if $|\Psi\rangle$ is the state at time $t$, and $|\Psi'\rangle$ is the state at time $t'$, then $|\Psi'\rangle=U|\Psi\rangle$ for some unitary operator $U$ which depends only on $t$ and $t'$.}
\end{postulate}

\begin{postulate}
Measurement\\
\graybox{A measurement is described by a collection $\{M_m\}$ of measurement operators. These operators are acting on the state space of the system being measured. The index $m$ refers to the measurement outcomes that may occur in the experiment. If the state of the quantum system is $|\Psi\rangle$ immediately before the measurement then the probability that result $m$ occurs is given by $p(m) =\langle\Psi| M_m^\dagger M_m|\Psi\rangle,$ and the state of the system after the measurement is $\frac{M_m|\Psi\rangle}{\sqrt{\langle\Psi| M_m^\dagger M_m|\Psi\rangle}}$.

Moreover the measurement operators satisfy the {\it completeness} equation, $\sum_m M_m^\dagger M_m ={\mathbb 1}$.}
\end{postulate}
Note that in Quantum Information Theory the measurement operators $\{M_m\}$ are often called Kraus operators \cite{kraus83a}.

\begin{postulate}
Composite system\\
\graybox{The state space of a composite quantum system is the tensor product of the state spaces of the component quantum systems. That is, if we have systems numbered $1$ through $n$, and system number $i$ is prepared in the state $|\Psi_i\rangle$, then the joint state of the total system is $|\Psi_1\rangle \otimes  |\Psi_2\rangle \otimes \dots \otimes |\Psi_n\rangle$.}
\end{postulate}

Note that, unlike in Quantum Mechanics, observables do not have a crucial role in Quantum Information Theory. Moreover, in general, we can consider the state of a system to be not only a vector state but a classical mixture of vector states. The notion of density matrices then is useful as we will see in the next subsection. Measurements are the core of Quantum Information theory because it is through a measurement that we learn information about a system. Therefore, we also introduce the mathematical language used to described a measurement.

\subsection{Ensemble of quantum states and density matrix}

Let us suppose a quantum system is in the state $|\Psi_i\rangle$ chosen in a set of states $\{|\Psi_i\rangle\}$. We can imagine that the appearance probabilities $\eta_i$ of each state of the set are in general different. We then summarize our knowledge on the system with the ensemble $\{|\Psi_i\rangle, \eta_i\}$. It is called an {\it ensemble of the system}. If the ensemble is composed of only one state (and of course its {\it a priori} probability equals $1$), the state is called {\it pure}. If not, one speaks of {\it mixed} states that is to say a classical mixture of pure states. To efficiently describe a {\it mixed} state, we use an operator instead of a vector state, the so-called density matrix.

\begin{definition} Density matrix\\
Let us consider a system with ensemble $\{|\Psi_i\rangle, \eta_i\}$. The state of the system can then be described in a compact form by the density matrix
\begin{eqnarray}
\rho=\sum_i \eta_i |\Psi_i\rangle\langle\Psi_i|.
\end{eqnarray}
\end{definition}

Such a density matrix possesses the three important properties
\begin{eqnarray}
\Tr(\rho)&=&1  \,\,\, {\rm(Normalization)},\\
\rho &\ge& 0  \,\,\, {\rm(Positivity)},\\
\Tr(\rho^2)=1 &\Rightarrow& \rho=|\Psi\rangle \langle \Psi| \,\,\, {\rm(Purity)},
\end{eqnarray}
where $\ge 0$ means positive semi-definite. Actually, the state ensemble of a system is not unique.

\begin{theorem}
Unitary freedom in the state ensemble\\
The sets $\{|\Psi_i\rangle, \eta_i\}$ and $\{|\Phi_i\rangle, \nu_i\}$ generate the same density matrix if and only if there exists a unitary transformation $U$ such that
\begin{eqnarray}
\sqrt{\nu_i}|\Phi_i\rangle=\sum_j U_{ij}\sqrt{\eta_j}|\Psi_i\rangle.
\end{eqnarray}
\end{theorem}

Equivalently,
\begin{corollary} Unitary freedom in the state ensemble of a density matrix\\
The two density matrices $\sum_i \eta_i |\Psi_i\rangle\langle\Psi_i|$ and $\sum_i \nu_i |\Phi_i\rangle\langle\Phi_i|$ describe the same state if and only if there exists a unitary transformation $U$ such that
\begin{eqnarray}
\sqrt{\nu_i}|\Phi_i\rangle=\sum_j U_{ij}\sqrt{\eta_j}|\Psi_i\rangle.
\end{eqnarray}
\end{corollary}

\subsection{Generalized measurements - POVM}

The third postulate of QIT, and its measurement operators $E_m$, can be used to define the positive semi-definite operators $E_m=M_m^\dagger M_m$. The set $\{E_m\}_m$ is called a Positive Operator-Valued Measure (POVM) \cite{helstrom76a,kraus83a,peres93a} and each operator $E_m$, a POVM element. On one hand, the fact that the probabilities $p(m)=\langle\Psi|E_m|\Psi\rangle$ are real and positive is expressed by the positivity of the POVM elements $\{E_m\}_m$. On the other hand, the fact that probabilities add up to one is expressed by the completeness relation $\sum_m E_m={\mathbb 1}$. Indeed, the sum of the probability $p(m)$ is $\sum_m p(m) = \sum_m \langle\Psi|E_m|\Psi\rangle=\langle\Psi| \sum_m E_m|\Psi\rangle=\langle\Psi|\Psi\rangle=1$. An important property of a POVM element is that its spectrum is upper bounded by $1$. Otherwise, it is clear that the expectation value $\langle\Psi|E_m|\Psi\rangle$ would exceed unity which contradicts the requirement that a probability is less than $1$. We finally give a general definition of a POVM.

\begin{definition} POVM\\
A Positive Operator-Valued Measurement (POVM) is a set of positive semi-definite operators $\{E_m\}_m$ such that
\begin{eqnarray}
E_k &\ge& 0 \,\,\, {\rm(Positivity)}\\
\sum_k E_k&=& {\mathbb 1}\,\,\, {\rm(Completeness \,\,\, relation)}
\end{eqnarray}
The probability to obtain the outcome $k$ for a given state $\rho_i$ is then given by
\begin{eqnarray}
p(k|i)=\Tr(E_k\rho_i).
\end{eqnarray}
\end{definition}

In the previous formula, $\Tr(.)$ stands for the trace. A POVM is also called a generalized measurement since it is the most general description of a measurement. Indeed, projective measurements, usually encountered in Quantum Mechanics are, in the above formalism, merely a special case where $E_m E_n=\delta_{mn}$, $E_m^2=E_m$. Such a projective measurement is called a Projection Valued Measure (PVM). Nevertheless, a generalized measurement can also be described by a projective measurement on an {\it enlarged} Hilbert space. A generalized measurement is then seen as a special case of projective measurements. The two pictures finally are equivalent as long as the Hilbert space is not fixed. This is made precise in the following theorem due to Naimark \cite{naimark40a,nagy}.

\begin{theorem} Naimark's extension\\
Given $\{E_k\}$ a POVM on a Hilbert space $\cal H$, it exists an embedding of $\cal H$ into a larger Hilbert space $\cal K$ such that the measure can be described by  projections onto orthogonal subspaces in $\cal K$. That is, there exist a Hilbert space $\cal K$, an embedding $\cal E$ such that ${\cal E }{\cal H} ={\cal K}$ and a PVM $\{R_k\}$ in $\cal K$, such that with P, the projection defined by $P{\cal K} ={\cal H}$, $E_k = PR_kP, \, \forall k$.
\end{theorem}

\subsection{Definitions and notations}

Here we briefly fix some notations. Throughout this thesis, we will make an extensive use of the support ${\cal S}_{P}:={\rm support}(P)$ of a Hermitian operator $P$. The support of a Hermitian operator is defined as the subspace spanned by its eigenvectors. We can moreover define the kernel ${\cal K}_P:={\rm kernel}(P)$ of a Hermitian operator $P$ as the subspace orthogonal to its support. We also denote $r_P:={\rm rank}(P)={\rm dim}({\cal S}_{P})$, the rank of $P$.

Next we define in a Hilbert space $\cal H$ the sum and the intersection of two Hilbert subspaces ${\cal H}_1$ and ${\cal H}_2$. The sum ${\cal H}_1 + {\cal H}_2$ of the subspaces ${\cal H}_1$ and ${\cal H}_2$ is defined to be the set consisting of all sums of the form $a_1+a_2$, where $a_1 \in {\cal H}_1$ and $a_2 \in {\cal H}_2$. ${\cal H}_1 + {\cal H}_2$ is a Hilbert subspace of $\cal H$. The intersection ${\cal H}_1 \cap {\cal H}_2$ is defined to be the set consisting of all the elements $a$, where $a \in {\cal H}_1$ and $a \in {\cal H}_2$. ${\cal H}_1 \cap {\cal H}_2$ is a Hilbert subspace of $\cal H$. The complementary orthogonal subspace ( or orthogonal complement) of a subspace $\cal S$ in $\cal H$, written ${\cal S}^{\perp}$, is the set of all the elements of $\cal H$ orthogonal to $\cal S$ with respect to the usual euclidean inner product. We then have $\cal H=\cal S \oplus {\cal S}^{\perp}$, the direct sum of the two orthogonal subspaces. Note that we use indifferently the notation ${\cal K}_{P}$ or ${\cal S}_{P}^\perp$ for a Hermitian operator $P$.

We need to define a positive semi-definite operator. A Hermitian operator $A$ acting on ${\cal H}$ is positive semi-definite if and only if $\langle \Psi |A| \Psi \rangle \ge 0$, for all $| \Psi \rangle$ in ${\cal H}$. In other words, a Hermitian operator is positive if and only if all its eigenvalues are positive or zero. We use the notation $A \ge 0$ to say that an operator $A$ is positive semi-definite. For such a positive semi-definite operator $A$. We can define its unique square root $\sqrt{A}$ and decompose it into the form $A=MM^{\dagger}$ with $M=\sqrt{A}U$, for any unitary matrix $U$. Since the states $\rho_i$ and the POVM elements $E_k$ are positive semi-definite operators, we can introduce their square root and use the previous decomposition.

\section{Unambiguous Quantum State Discrimination}

A quantum state describes what we know about a quantum system. Given a single copy of a quantum system which can be prepared in several known quantum states, our aim is to determine in which state the system is. This can be well understood in a communication context where only a single copy of the system is given and only a single shot-measurement is performed. This is in contrast with usual experiments in physics where many copies of a system are measured to get the probability distribution of the system. In quantum state discrimination (see \cite{chefles00a} for a review of quantum state discrimination), no statistics is built since only a single-shot measurement is performed on a single copy of the system. Actually there are fundamental limitations to the precision with which the state of the system can be determined with a single measurement. Whenever the possible quantum states are nonorthogonal, perfect discrimination of the states becomes impossible. This can be understood from the intuition that two non-orthogonal states have some probability to behave the same way. More precisely, if a quantum system is prepared in one of the two state $|\Psi\rangle$ and $|\Phi\rangle$, which are neither identical nor orthogonal, there is no measurement that perfectly determines in which state the system is. Mathematically, a measurement, that perfectly determines in which state the system is, is composed of two outcomes (i.e.\ two POVM elements) $E_{\Psi}$ and $E_{\Phi}$ that identify $|\Psi\rangle$ and $|\Phi\rangle$ respectively with no errors. This means, in terms of probabilities, that
\begin{eqnarray}
\langle \Psi | E_{\Psi} | \Psi \rangle&=&1,\\
\langle \Phi | E_{\Phi} |\Phi \rangle&=&1,\\
\langle \Psi | E_{\Phi} | \Psi \rangle&=&0,\\
\langle \Phi | E_{\Psi} |\Phi \rangle&=&0.
\end{eqnarray}
If we express $|\Phi\rangle$ in the basis $\{|\Psi\rangle,|\Psi^\perp\rangle\}$, Eqn.(1.11) becomes
\begin{eqnarray}
( \langle \Psi | \Phi \rangle^* \langle \Psi | + \langle \Psi^\perp | \Phi \rangle^* \langle \Psi^\perp |) E_{\Phi} ( \langle \Psi | \Phi \rangle |\Psi \rangle + \langle \Psi^\perp | \Phi \rangle |\Psi^\perp \rangle)=1
\end{eqnarray}
where $*$ stands for complex conjugation. With the help of Eqn.(1.12) which is equivalent to $E_{\Phi} | \Psi \rangle=0$ since $E_{\Phi} \ge 0$ (see proof in Appendix A), we obtain
\begin{eqnarray}\label{max}
|\langle \Phi | \Psi^\perp \rangle|^2 \langle \Psi^\perp | E_{\Phi} | \Psi^\perp \rangle=1.
\end{eqnarray}
Since the spectrum of $E_{\Phi}$ is upper bounded by $1$, $\langle \Psi^\perp | E_{\Phi} | \Psi^\perp \rangle \le 1$ and Eqn.(\ref{max}) is fulfilled only if $|\langle \Phi | \Psi^\perp \rangle|^2=1$ which contradicts the assumption that $|\Psi\rangle$ and $|\Phi\rangle$ are non-orthogonal.\\

The immediate consequence of this limited precision is to resort to various state discrimination strategies depending on what one really wants to learn about the state. Given a strategy, we finally have to optimize the measurement with respect to some criteria.\\
\begin{figure}[h!]
  \centering
  \fbox{\includegraphics[width=14cm]{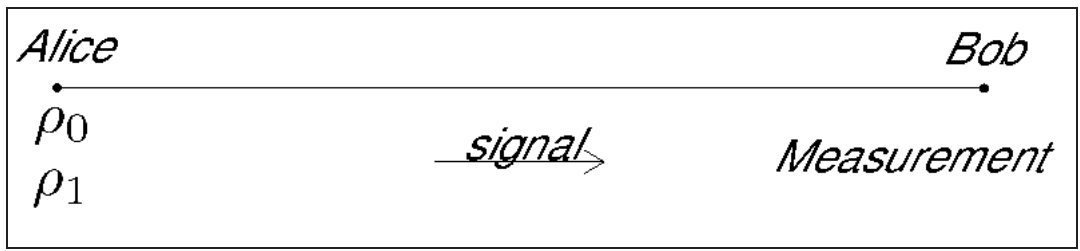}}
  \caption{Two parties Alice and Bob want to communicate}
  \label{ab}
\end{figure}

The basic scenario involves two parties Alive and Bob who want to communicate (see Fig.~\ref{ab}). Alice prepares a quantum system in a state, member of a set of states known by Bob. In general Alice does not prepare each state with the same probability. We speak of an {\it a priori} probability. She sends a quantum system to Bob who performs a measurement in order to obtain the information he wants. In other words, a state ensemble of a quantum system is given and we want to determine the state of that system. In his famous book published in 1976 \cite{helstrom76a}, Helstrom established the mathematical bases of such detection tasks.  He introduced the notion of {\it Bayes' cost function} which can describe any discrimination strategy. The idea is the following.
For each possible outcome $k$ conditioned on a signal state $j$, a price to pay $C_{kj}$ is associated. If $C_{kj}$ is positive, Bob has to pay Alice. If $C_{kj}$ is negative, Bob earns money. To set up a strategy corresponds to give the {\it Bayes' cost matrix} $C_{kj}$. Related to this matrix, the {\it Bayes' cost function}, given by
\begin{eqnarray}
C=\sum_{kj} \eta_j C_{kj} p(k|j),
\end{eqnarray}
represents the total price that Bob has to pay to Alice. Information about a state is represented by an outcome $k$ conditioned on a signal state $j$. It then appears clear that, depending on which information really matters to Bob and Alice, the strategy or, equivalently, the {\it Bayes' cost matrix} $C_{kj}$ will change. The aim for Bob is of course to minimize the prize he has to pay to Alice. To minimize the {\it Bayes' cost function} $C$ while the {\it a priori} probability $\eta_j$ and the states $\rho_j$ are fixed, Bob is only free to change his measurement. In this thesis, we play the role of Bob who wants to find the optimal measurement to lose a minimal amount of money.

\begin{figure}[h!]
  \centering
  \fbox{\includegraphics[width=14cm]{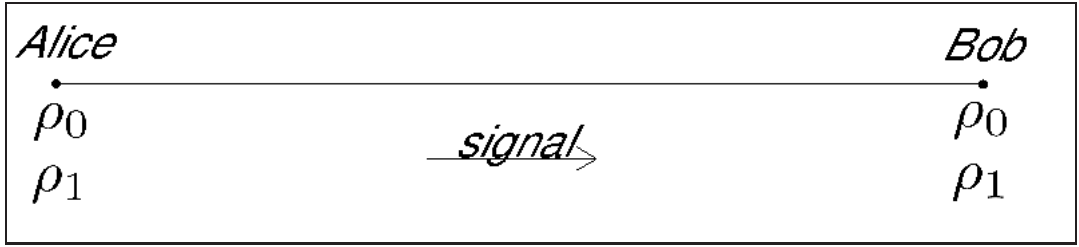}}
  \caption{Two possible outcomes in the scenario of Minimum Error Discrimination}
  \label{med}
\end{figure}

The {\it Bayes' cost matrix} $C_{kj}$ depends on the strategy adopted by Alice and Bob. For instance, Bob might want to know which state was sent with the minimum error probability. This strategy is called {\em Minimum Error Discrimination} (MED) \cite{helstrom76a} - see Fig.~\ref{med}. In MED, the {\it Bayes' cost matrix} $C_{kj}$ is given by 
\begin{eqnarray}
C_{kj}=\left\{\begin{array}{cc}
0 & k=j,\\
1 & k \neq j.
\end{array}\right.
\end{eqnarray}

\begin{figure}[h!]
  \centering
  \fbox{\includegraphics[width=10cm]{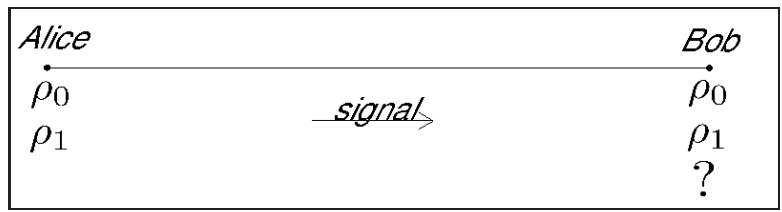}}
  \caption{Three possible outcomes in the scenario of Unambiguous State Discrimination}
  \label{usd}
\end{figure}

Alternatively, one might consider an error-free discrimination of the signal states. In this strategy, the measurement can either correctly identify the state or send out a flag stating that it failed to identify the state. A correct identification of the state is called a conclusive result while a failure to identify the state is known as an inconclusive result usually denoted by '?' or 'don't know'. The objective then is to minimize the probability of inconclusive result, the so-called failure probability. This strategy is called {\em Unambiguous State Discrimination} (USD) - see Fig.~\ref{usd}. The coefficients of the non square ($j=0,1$ and $k=0,1,?$) are {\it Bayes' cost matrix} $C_{kj}$ are
\begin{eqnarray}
C_{kj}=\left\{\begin{array}{cc}
0 & k=j,\\
1 & k=?, \forall j, \\
\infty & otherwise.
\end{array}\right.
\end{eqnarray}
Note that the coefficients $C_{k\neq j}$ where $k,j=0,1$ are set to infinity in order to impose the error-free conditions $p(k|j \neq k)=0$, $k,j=0,1$ to obtain a non diverging {\it Bayes' cost function}.\\

We can list another task related to state discrimination where we are given a finite number of identical copies of an unknown state in a $d$-dimensional Hilbert space. Our goal is to estimate the actual state with the maximum accuracy, which is often quantified by the fidelity between the actual state and the estimated state (see chapter 2 for a definition of the fidelity). Since the state to estimate can be any state in the $d$-dimensional Hilbert space, one has to average the accuracy over all the possible states of the $d$-dimensional Hilbert space. This scenario is known as {\em Quantum State Estimation} \cite{massar95,bruss99b} (see Ref.~\cite{holevo73a,jozsa94b,fuchs95a} for other scenarios).\\

Let us add another comment. The fact that non-orthogonal quantum states are not perfectly distinguishable also has benefits. It leads in particular to secure Quantum Key Distribution (QKD) in a cryptographic context \cite{ekert94a}. The security in classical computer science is ensure by the complexity of some task like factorization of big prime numbers. In QKD, the security is due to the quantum laws of Nature and does not anymore rely on the assumption of eavesdropper's limited computational power.\\


In general, the optimal measurements for a given strategy depends on the quantum states and the {\it a priori} probability of their appearance. For a given strategy and a given state ensemble, the task is to find the measurement which minimizes the {\it Bayes' cost function}. Such a measurement (it might not be unique) is called an {\it optimal measurement}.\\

In this thesis, we are interested in the unambiguous discrimination of two known mixed quantum states. Therefore the task is to find an optimal measurement that minimizes the failure probability. The problem of unambiguously discriminating pure states with equal {\it a priori} probabilities was formulated in 1987 by Dieks \cite{dieks88a} and Ivanovic \cite{ivanovic87a} and elegantly solved by Peres \cite{peres88a}. Seven years later, Jaeger and Shimony presented the general solution for two pure states with different {\it a priori} probabilities \cite{jaeger95a}. Shortly after this result, Chefles and Barnett showed that only linearly independent pure states can be unambiguously discriminated \cite{chefles98b}. Finally Chefles provided the optimal failure probability and its corresponding optimal measurement in the case of $n$ symmetric states \cite{chefles98a}. The enumeration of analytical results for USD of pure states scenarios already ends here even if an algorithm for the case of three pure states was proposed by Peres and Terno in 1998 \cite{peres98a}. In fact, since Sun's work in 2002 \cite{sun02b,eldar03b}, it is  known that USD (of both pure and mixed states) is a convex optimization problem \cite{vandenberghe96a,vandenberghe04,bental01}. Mathematically, this means that the quantity to optimize as well as the constraints on the unknowns are convex functions. Practically, this means that the optimal solution can be extremely efficiently computed. This is therefore a very useful tool. Nevertheless our aim is to understand the structure of USD, to relate it to neat and relevant quantities and to find analytical solutions.\\

The case of mixed states recently attracted more attention. But until this present work, no optimal measurements for mixed states has been found unless the USD problem can be reduced to some known pure state case. This reduction comes from simple geometrical considerations and can be summarized in three theorems. Important examples of such reducible problems are {\em Unambiguous State Discrimination of two mixed states with one-dimensional kernel} {\cite{rudolph03a}}, {\em Unambiguous State Comparison} \cite{barnett03a,kleinmann05,herzog05a} (see Ref.~\cite{barnett03a,jex04a,jex04b} for the unambiguous comparison of unknown states), {\em State Filtering} \cite{sun02a,bergou03a,herzog05b} and {\em Unambiguous Discrimination of two subspaces} \cite{bergou06}. This four cases are all reducible to some pure state case and can therefore be solved. To specify that a USD problem is not reducible by means of our three reduction theorems, we use the expression 'USD of generic density matrices'. Lower and upper bounds on the failure probability to unambiguously discriminate two density matrices are also known. In 2004, Eldar derived necessary and sufficient conditions for the optimality of a USD POVM \cite{eldar04a}. Unfortunately these conditions appear rather difficult to solve. In contrast to the MED problem, which is already solved for any pair of mixed states \cite{helstrom76a,herzog04}, the optimal USD of mixed states is an open problem.\\

\section{Results}

We outline here the six main results derived in this thesis.\\

1) {\bf Three reduction theorems to reduce the dimension of a USD problem}\\

2) {\bf Unambiguous comparison of $n$ pure states with a simple symmetry}\\

3) {\bf First class of exact solutions}\\

4) {\bf Second class of exact solutions}\\

5) {\bf A fourth, incomplete, reduction theorem}\\

6) {\bf USD and BB84-type QKD protocol}


\subsubsection{Three reduction theorems to reduce the dimension of a USD problem [Chapter 3]}
As seen in the previous section, only few analytical optimal solutions in Unambiguous State Discrimination are known. For pure states scenarios, only two classes of exact solutions have been provided so far. They are the solutions for USD of two pure states \cite{jaeger95a} and USD of $n$ linearly independent symmetric pure states \cite{chefles98a}. In the case of mixed states, there are actually four known solutions: {\em unambiguous discrimination of two mixed states with one-dimensional kernel} {\cite{rudolph03a}}, {\em unambiguous comparison of two pure states} {\cite{barnett03a,kleinmann05,herzog05a}}, {\em state filtering} \cite{sun02a,bergou03a,herzog05b} and {\em unambiguous discrimination of two subspaces} \cite{bergou06}. It seems surprising that research on USD of pure states has been less successful than work on USD of mixed states! A solution to this apparent paradox is given by our first result. Indeed these four optimal solutions in USD of mixed states only require the optimal solution for USD of two pure states. More generally, we  prove that the problem of discriminating any two density matrices can be reduced to the problem of discriminating two density matrices of the same rank $r$ in a $2r$-dimensional Hilbert space. This introduces the notion of {\it standard} USD problem. Such a standard USD problem is proposed as a starting point for any further theoretical investigation on USD. That way, we can avoid to deal with trivial or already known classes of solutions. The reductions are of three types and can be summarized in three theorems. In few words, the reduction theorems work as follows. In a first reduction theorem, we split off any common subspace between the supports of the two density matrices $\rho_0$ and $\rho_1$. In a second reduction theorem, we eliminate, if present, the part of the support of $\rho_1$ which is orthogonal to the support of $\rho_0$ and {\it vice versa}. In a third reduction theorem, if two density matrices are block diagonal, we decompose the global USD problem into decoupled unambiguous discrimination tasks on each block.

\subsubsection{Unambiguous comparison of $n$ pure states with a simple symmetry [Chapter 3]}
We are given $n$ pure quantum states $\{|\Psi_i\rangle\}$ which occur with {\it a priori} probabilities $\{p_i\}$. We would like to know without error whether these states are all identical or not. Actually the task of unambiguously comparing any two pure states can be elegantly solved by use of the second and third reduction theorems, as Kleinmann {\it et al.} showed in \cite{kleinmann05}. Stimulated by their idea, we investigate the case of $n$ pure states having some simple symmetry. In fact we prove that the comparison of $n$ linearly independent pure states with equal {\it a priori} probabilities and equal and real overlaps can be reduced to $n$ unambiguous discriminations of two pure states and then be solved. The question to know whether any unambiguous comparison of {\it pure} states is always reducible to some pure state cases remains opened. Let us add here that, as Kleinmann {\it et al.} indicated in \cite{kleinmann05}, the unambiguous comparison of {\it mixed} states is generally not reducible to some pure states case.\\

In this thesis, we provide two classes of exact solutions for unambiguously discriminating two {\it generic} density matrices. These two classes are the only two classes known until now.

\subsubsection{First class of exact solutions [Chapter 4]}
We consider the problem of unambiguously discriminating two density matrices $\rho_0$ and $\rho_1$ with {\it a priori} probabilities $\eta_0$ and $\eta_1$. We define the fidelity of the two states as $F=\Tr(\sqrt{\sqrt{\rho_0}\rho_1\sqrt{\rho_0}})$. We provide three lower bounds on the failure probability in three regimes of the ratio between the {\it a priori} probabilities defined as $\sqrt{\frac{\eta_1}{\eta_0}} \le \frac{\Tr(P_1 \rho_0)}{F}$, $\frac{\Tr(P_1 \rho_0)}{F}\le \sqrt{\frac{\eta_1}{\eta_0}} \le \frac{F}{\Tr(P_0 \rho_1)}$ and $\frac{F}{\Tr(P_0 \rho_1)} \le \sqrt{\frac{\eta_1}{\eta_0}}$. For each regime, we give necessary and sufficient conditions for the failure probability of unambiguously discriminating two mixed states to reach the bound. With that result, we give the optimal USD POVM of a wide class of pairs of mixed states. This class corresponds to pairs of mixed states for which the lower bound on the failure probability is saturated. This is the first analytical solution for unambiguous discrimination of generic mixed states. This goes beyond known results which are all reducible to some pure state case. Note that any pair of mixed state does not saturate the bounds. The necessary and sufficient conditions take the simple form of the positivity of the two operators $\rho_0-\alpha \sqrt{\sqrt{\rho_0}\rho_1\sqrt{\rho_0}}$ and $\rho_1-\frac{1}{\alpha} \sqrt{\sqrt{\rho_1}\rho_0\sqrt{\rho_1}}$ where $\alpha$ equals $\frac{\Tr(P_1 \rho_0)}{F}$, $\sqrt{\frac{\eta_1}{\eta_0}}$ and $\frac{F}{\Tr(P_0 \rho_1)}$ in the first, second and third regime, respectively.

\subsubsection{Second class of exact solutions [Chapter 5]}
We derive a second class of exact solutions. This class corresponds to any pair of {\it geometrically uniform} mixed states without overlapping supports in a four dimensional Hilbert space. In short, two {\it geometrically uniform} mixed states are two unitary similar density matrices $\rho_0$ and $\rho_1=U\rho_0 U$ where the unitary matrix $U$ is an involution i.e.\ $U^2={\mathbb 1}$. We find that only three options for the expression of the failure probability exist. First, if the operators $\rho_0-\sqrt{\frac{\eta_1}{\eta_0}} F_0$ and $\rho_1-\sqrt{\frac{\eta_0}{\eta_1}} F_1$ are positive semi-definite, then the pair of density matrices falls in the first class of exact solutions. If this is not the case, either the operator $P_0^\perp U P_0^\perp$ has one positive and one negative eigenvalue 
or it has two eigenvalues of the same sign. In the former case, we can give the optimal failure probability in terms of the eigenvalues and eigenvectors of $P_0^\perp U P_0^\perp$. In the later case, no unambiguous discrimination is possible and the failure probability simply equals unity. For these three cases, we provide the optimal failure probability as well as the optimal measurement.

\subsubsection{A fourth, incomplete, reduction theorem [Chapter 5]}
The two USD POVM elements $E_0$ and $E_1$ have a rank less or equal to the rank of ${\cal S}_{\rho_1}^\perp$ and ${\cal S}_{\rho_0}^\perp$, respectively. This defines the notion of maximum rank of $E_0$ and $E_1$. We establish a theorem stating that if the two operators $\rho_0-\sqrt{\frac{\eta_1}{\eta_0}} F_0$ and $\rho_1-\sqrt{\frac{\eta_0}{\eta_1}}F_1$ are not positive semi-definite then the two USD POVM elements $E_0$ and $E_1$ can not have both maximum rank. A corollary can be derived assuming a standard USD problem. In that case, if the two operators $\rho_0-\sqrt{\frac{\eta_1}{\eta_0}} F_0$ and $\rho_1-\sqrt{\frac{\eta_0}{\eta_1}}F_1$ are not positive semi-definite then there exist one eigenvector of $E_?$ with eigenvalue $1$ and one eigenvector of either $E_0$ or $E_1$ with eigenvalue $1$, too. From the completeness relation fulfilled by the measurement operators, it follows that we can split off the two-dimensional subspace spanned by these two eigenvectors from the original USD problem. This could lead to a fourth reduction theorem. 'Could' because it remains to fully characterize these two eigenvectors cited above. So far, we can only prove their existence. If one could characterize them, a way to solve analytically any USD problem would be available. Indeed, we start from a general USD problem of two mixed states. We use the three first reduction theorems to bring it to standard form. We then check the positivity of the two operators $\rho_0-\sqrt{\frac{\eta_1}{\eta_0}} F_0$ and $\rho_1-\sqrt{\frac{\eta_0}{\eta_1}}F_1$. If the positivity is confirmed, then the pair of density matrices falls in the first class of exact solutions. If the two operators are not positive, we can use the fourth reduction theorem to get rid of two dimensions corresponding to the two eigenvectors mentioned above. At that point, we check the positivity of the two operators $\rho_0'-\sqrt{\frac{\eta_1'}{\eta_0'}} F_0'$ and $\rho_1'-\sqrt{\frac{\eta_0'}{\eta_1'}} F_1'$ of the reduced problem. We see here a constructive way to solve any USD problem. If the two operators $\rho_0'-\sqrt{\frac{\eta_1'}{\eta_0'}} F_0'$ and $\rho_1'-\sqrt{\frac{\eta_0'}{\eta_1'}} F_1'$ of the reduced problems never turn out to be positive, we end up with only two pure states and we can therefore always find the optimal measurement. The full characterization of the two eigenvectors involved in this incomplete reduction theorem is of great importance.

\subsubsection{USD and BB84-type QKD protocol [Chapter 6]}
The Bennett and Brassard 1984 cryptographic protocol \cite{bennett84a} provides a method to distribute a private key between two parties and allow an unconditionally secure communication. We consider in this thesis the implementation of a BB84-type QKD protocol that uses weak coherent pulses with a phase reference \cite{dusek00a}. In that context, two important questions related to unambiguous state discrimination can be addressed. First, 'With what probability can an eavesdropper unambiguously distinguish the {\it basis} of the signal?' and second 'With what probability can an eavesdropper unambiguously determine which {\it bit value} is sent without being interested in the knowledge of the basis?' These two questions can be translated in some unambiguous discrimination task of two {\it geometrically uniform} mixed states in a four dimensional Hilbert space. We answer these two questions providing useful insights for further investigations on practical implementations of Quantum Key Distribution protocols.\\

The structure of this thesis is the following. In chapter 2, we mathematically define the problem of USD. We then review the known results on unambiguous discrimination: unambiguous discrimination two pure states, unambiguous discrimination of $n$ symmetric states and a few general properties. In chapter 3, we present our three reduction theorems. They allow us to solve special tasks in quantum information theory such as, e.g.\ state filtering, unambiguous discrimination of two pure states, unambiguous discrimination of $n$ pure states with a simple symmetry and unambiguous discrimination of two subspaces. All these tasks are related to the unambiguous discrimination of two mixed states which can be reduced to the unambiguous discrimination of some pure states only. We also define a {\it standard} form as a starting point for further investigations in USD. In chapter 4, we derive lower bounds on the failure probability $Q$ as well as necessary and sufficient conditions for the failure probability to reach those bounds. This provides a first class of exact solutions for unambiguous discrimination of two {\it generic} mixed states. This class corresponds to pairs of mixed states for which the lower bound (one for each of the three regimes depending on the ratio between the {\it a priori} probabilities) on the failure probability $Q$ is saturated. For this class we give the corresponding optimal USD measurement. In chapter 5, we derive a fourth, incomplete, reduction theorem which, together with the first three reduction theorems aims to solve in a constructive way any USD problem of two density matrices. Moreover we derive a second class of exact solutions. This class corresponds to any pair of two geometrically uniform states in four dimensions. In chapter 6, we give two examples of such an unambiguous discrimination of two {\it geometrically uniform} states in four dimensions. These examples are related to the implementation of the Bennett and Brassard 1984 cryptographic protocol. In the last chapter, we summarize our results and propose directions for further research on USD of two density matrices.

%% file: chapter2.tex
\chapter{Optimal Unambiguous State Discrimination} \label{optUSD}
The optimal USD measurement is known for two {\it pure-state} cases. On one hand, the optimal failure probability as well as the corresponding optimal measurement were provided by Jaeger and Shimony for any pair of two pure states with arbitrary {\it a priori} probabilities \cite{jaeger95a}. On the other hand, Chefles found the optimal failure probability and the corresponding optimal measurement for unambiguously discriminating $n$ linearly independent symmetric pure states \cite{chefles98a}. We present the basic properties of a USD measurement before reviewing the solution to these two {\it pure-state} scenarios.

\section{The USD measurement}
We consider a set of $n \in {\mathbb N}$ known quantum states $\{ \rho_i \}$, $i=1,..,n$, with their  {\it a priori} probabilities $\{ \eta_i \}$. We are looking for a measurement that either identifies a state uniquely (conclusive result) or fails to identify it (inconclusive result). The goal is to minimize the probability of inconclusive result. The measurements involved are typically generalized measurements \cite{kraus83a} described by a POVM which consists in a set of positive semi-definite operators $\{E_k\}$ that satisfies the completeness relation $\sum_k E_k = {\mathbb 1}$ on the Hilbert space spanned by the states. The probability to obtain the outcome $k$ for a given signal $\rho_i$ is then given by $p(k|i)={\rm Tr}(\rho_i E_k)$. We will often refer to the states of the quantum system as {\it signal} states or even {\it signals}. This comes from the context of communication where the possible states of a quantum system correspond to the different signals sent to communicate.

Let us now mathematically define what an Unambiguous State Discrimination Measurement is, its corresponding failure probability, and the notion of optimality.

\begin{definition}
A measurement described by a POVM $\{E_k\}$ is called an Unambiguous State Discrimination Measurement (USDM) on a set of states $\{\rho_i\}$ if and only if the following conditions are satisfied:
\begin{itemize}
\item The POVM contains the elements $\{E_?,E_1, \dots E_{n}\}$ where $n$ is the number of different signals in the set of states. The element $E_?$ is connected to an inconclusive result, while the other elements $E_i$, $i=1,..,n$ , correspond to an identification of the state $\rho_i$. 
\item No states are wrongly identified, that is ${\rm Tr}(\rho_i E_k) = 0 \quad\quad \forall i \neq k \quad i,k=1,...,n$.
\end{itemize}
\end{definition}
Each USD Measurement gives rise to a failure probability, that is, the rate of inconclusive results. This can be calculated as
\begin{eqnarray}
Q[ \{E_k\}] := \sum_i \eta_i {\rm Tr}(\rho_i E_?).
\end{eqnarray}

\begin{definition}
A measurement described by a POVM $\{E^{opt}_k\}$ is called an Optimal Unambiguous State Discrimination Measurement (OptUSDM) on a set of states $\{\rho_i\}$ with the corresponding {\it a priori} probabilities $\{\eta_i\}$ if and only if the following conditions are satisfied
\begin{itemize}
\item The POVM $\{E^{opt}_k\}$ is a USD measurement on $\{\rho_i\}$
\item The probability of inconclusive results is minimal, that is $Q[\{E^{opt}_k\}] = \min Q[\{E_k\}]$ where the minimum is taken over all USDM.
\end{itemize}
\end{definition}

Unambiguous state discrimination is an error-free discrimination. This implies a strong constraint on the measurement. The fact that the outcome $E_k$ can only be triggered by the state $\rho_k$ implies that the support of $E_k$ is orthogonal to the supports of all the mixed states other than $\rho_k$. This is a strong constraint for any USD measurement, not only the optimal one. To see that fact rigorously we need the following lemma.

\begin{lemma}\label{trab}
For any positive semi-definite operators $A$ and $B$, ${\rm Tr}(AB)=0$ if and only if the support of the two positive semi-definite operators are orthogonal
\begin{eqnarray}
{\rm Tr}(AB)=0 \Leftrightarrow S_A \perp S_B.
\end{eqnarray}
\end{lemma}

Since a USD POVM satisfies ${\rm Tr}(E_k\rho_i)=\Tr(E_k \rho_k) \delta_{ki}$ to be an error-free measurement, a corollary of Lemma \ref{trab} can be derived.

\begin{corollary}
A USD measurement described by the POVM $\{E_k\}$ on $n$ density matrices $\{\rho_i\}$ is such that
\begin{eqnarray}
S_{E_k} \perp {S_{\rho_{i \ne k}}},\,\,\, \forall i,k=1,\dots,n. 
\end{eqnarray}
\end{corollary}

USD measurements are very sensitive in the sense that a small variation of a mixed state overthrows completely the error-free character of the already existing measurement. This is true for any USD measurement, not only the optimal ones. Let us now prove Lemma 1.

\paragraph*{\bf Proof of Lemma 1}
If $A$ and $B$ are positive semi-definite operators, they are
diagonalizable with eigenvalues $\alpha_i >0 \quad (i=1,...,rank(A))$ and $\beta_j >0 \quad (j=1,...,rank(B))$. Thus
\begin{eqnarray}
{\rm Tr}(AB) &=& {\rm Tr}(\sum_i \alpha_i |\Psi_i\rangle \langle \Psi_i|  \sum_j \beta_j|\Phi_i\rangle \langle \Phi_i|) \nonumber \\
&=& \sum_{ij} \alpha_i \beta_j |\langle \Psi_i|\Phi_j \rangle|^2 \;
\end{eqnarray}
vanishes if and only if $\{|\Phi_i \rangle\}$ and $\{|\Psi_j \rangle\}$ span orthogonal subspaces.\hfill $\blacksquare$


\section{Solution for two pure states}
\begin{figure}[h!]
  \centering
  \fbox{\includegraphics[width=14cm]{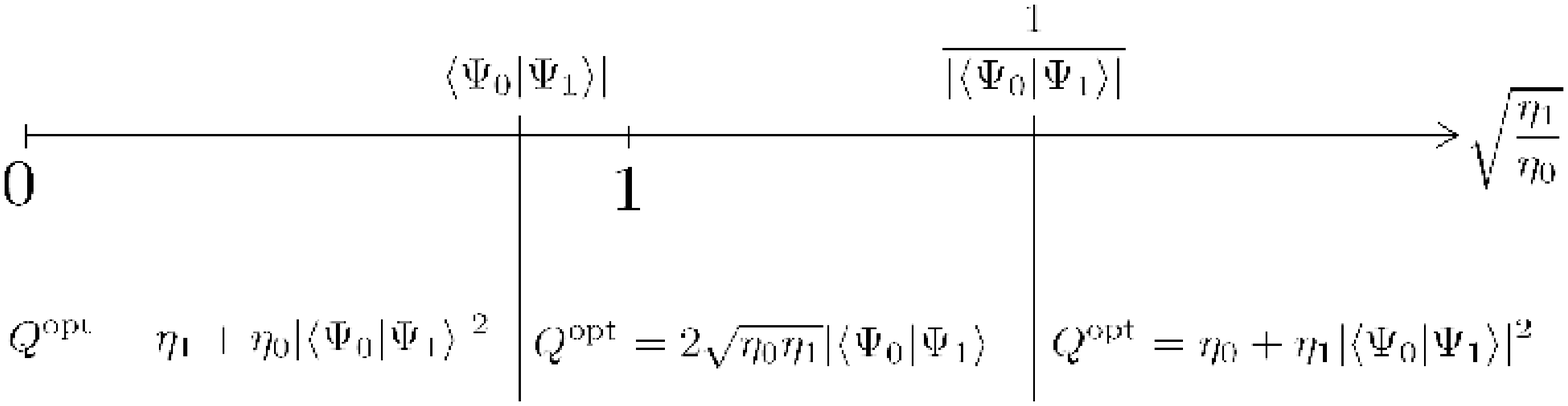}}
  \caption{Optimal failure probability for USD of two pure states}
  \label{usd2psbis}
\end{figure}

In the simple case of two pure states $|\Psi_0\rangle$ and $|\Psi_1\rangle$ with arbitrary {\it a priori} probabilities $\eta_0$ and $\eta_1$, the optimal failure probabilities (see Fig.~\ref{usd2psbis}) to unambiguously discriminate them is given by

\begin{eqnarray}
Q^{\mathrm{opt}} = \eta_1 +\eta_0|\langle\Psi_0|\Psi_1\rangle|^2 \, \,\,\, \mathrm{for} \,\,\,  \sqrt{\frac{\eta_1}{\eta_0}} \le |\langle\Psi_0|\Psi_1\rangle|,\\ 
Q^{\mathrm{opt}} = 2\sqrt{\eta_0\eta_1}|\langle\Psi_0|\Psi_1\rangle| \,\,\, \mathrm{for} \,\,\, |\langle\Psi_0|\Psi_1\rangle| \le \sqrt{\frac{\eta_1}{\eta_0}} \le \frac{1}{|\langle\Psi_0|\Psi_1\rangle|},\\ 
Q^{\mathrm{opt}} = \eta_0  +\eta_1 |\langle\Psi_0|\Psi_1\rangle|^2 \,\,\, \mathrm{for} \,\,\, \frac{1}{|\langle\Psi_0|\Psi_1\rangle|} \le \sqrt{\frac{\eta_1}{\eta_0}}.
\end{eqnarray}
This result was derived by Jaeger and Shimony in 1995. When the two {\it a priori} probabilities are equal, it reduces to the well known equation \begin{eqnarray}
Q^{\mathrm{opt}} = |\langle\Psi_0|\Psi_1\rangle|.
\end{eqnarray}
This solution is known as the Ivanovic-Diesk-Peres (IDP) limit since 1988.

\begin{figure}[h!]
  \centering
  \includegraphics[width=10cm]{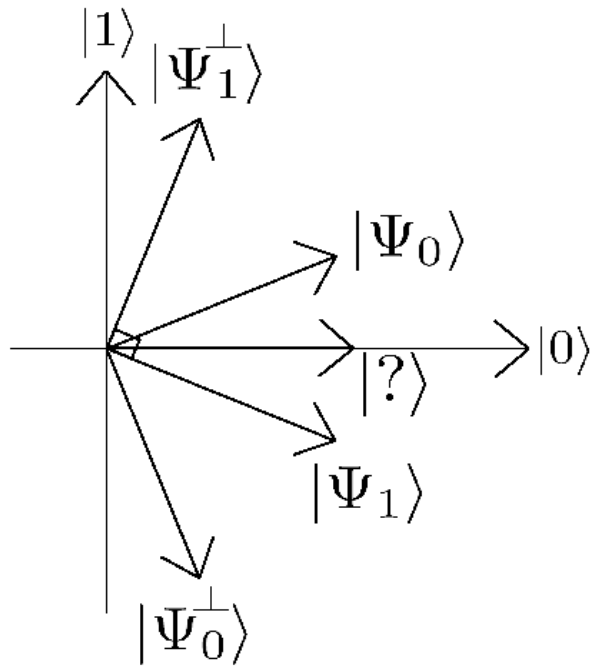}
  \caption{Basis vectors $|\Psi_1^\perp \rangle $, $|\Psi_0^\perp \rangle $ and $|? \rangle$ of the three POVM elements $E_0$, $E_1$ and $E_?$ for the optimal USD measurement of two pure states when $\langle\Psi_0|\Psi_1\rangle \ge 0$ and $|\langle\Psi_0|\Psi_1\rangle| \le \sqrt{\frac{\eta_1}{\eta_0}} \le \frac{1}{|\langle\Psi_0|\Psi_1\rangle|}$}
  \label{usd2psmeas}
\end{figure}

The optimal measurement (see Fig.~\ref{usd2psmeas}) that realizes these optimal failure probabilities is given by
\begin{eqnarray}
\begin{array}{l}
E_0=|\Psi_1^\perp \rangle \langle \Psi_1^\perp|\\
E_1=0\\
E_?=|\Psi_1 \rangle \langle \Psi_1|
\end{array}
\text{for}\,\,\,  \sqrt{\frac{\eta_1}{\eta_0}} \le |\langle\Psi_0|\Psi_1\rangle|,
\end{eqnarray}

\begin{eqnarray}
\begin{array}{l}
E_0=\frac{1-\sqrt{\frac{\eta_1}{\eta_0}}|\langle \Psi_0|\Psi_1 \rangle|}{|\langle \Psi_1^\perp|\Psi_0 \rangle|^2}|\Psi_1^\perp\rangle\langle \Psi_1^\perp|\\
E_1=\frac{1-\sqrt{\frac{\eta_0}{\eta_1}}|\langle \Psi_0|\Psi_1 \rangle|}{|\langle \Psi_0^\perp|\Psi_1 \rangle|^2}|\Psi_0^\perp\rangle\langle \Psi_0^\perp|\\
E_?={\mathbb 1}-E_0 -E_1
\end{array}
\,\,\text{for}\,\,\, |\langle\Psi_0|\Psi_1\rangle| \le \sqrt{\frac{\eta_1}{\eta_0}} \le \frac{1}{|\langle\Psi_0|\Psi_1\rangle|},
\end{eqnarray}

\begin{eqnarray}
\begin{array}{l}
E_0=0\\
E_1=|\Psi_0^\perp \rangle \langle \Psi_0^\perp|\\
E_?=|\Psi_0 \rangle \langle \Psi_0|
\end{array}
\text{for}\,\,\, \frac{1}{|\langle\Psi_0|\Psi_1\rangle|} \le \sqrt{\frac{\eta_1}{\eta_0}}.
\end{eqnarray}


\section{Solution for n symmetric pure states}
Unambiguous discrimination can be consider for more than two states. The only requirement for an error-free discrimination is the linearly independence of the signal states as Chefles showed in 1998. An exact solutions can even be provided if the $n$ quantum states happen to be symmetric. Symmetric states are states that can be written in terms of a generator $|\Psi_0 \rangle$ and a unitary transformation $U$ such that $U^n={\mathbb 1}$. The complete set of symmetric states can be written as
\begin{eqnarray}
|\Psi_j\rangle&=&U |\Psi_{j-1}\rangle=U^j|\Psi_0\rangle,\,\,\, j=1,\dots,n-1\\
|\Psi_0\rangle&=&U|\Psi_{n-1}\rangle,\,\,\, U^n={\mathbb 1}.
\end{eqnarray}
Note that we choose the {\it a priori} probabilities to be equal in order not to break the symmetry. For such symmetric states, we can introduced a suitable orthonormal basis $\{|\gamma_k\rangle\}_k$ such that $|\Psi_j\rangle=\sum_{k=0}^{n-1} c_k e^{2 i \pi \frac{jk}{n}}|\gamma_k\rangle$ with $\sum_k |c_k|^2=1$ and $U=\sum_{k=0}^{n-1} e^{2 i \pi \frac{k}{n}} |\gamma_k\rangle\langle\gamma_k|$ \cite{chefles98a}. Note that the coefficients $c_k$ can be calculated thanks to the formula $|c_k|^2=\frac{1}{n^2}\sum_{j,j'} e^{2 i \pi k \frac{j-j'}{n}} \langle \Psi_{j'} | \Psi_j \rangle$. We define $c_{min}=min_k c_k$ and the optimal failure probabilities to unambiguously discriminate $n$ symmetric states is then given by
\begin{eqnarray}
Q^{\mathrm{opt}} = n |c_{min}|^2.
\end{eqnarray}

On the analytical side, some general properties of USD of mixed states were recently derived. We give here an overview of these results. First, there are the very general necessary and sufficient conditions for the optimality of a USD measurement derived by Eldar in \cite{eldar04a}. Unfortunately those conditions are pretty hard to solve. They can nevertheless be used to check the optimality of some USD POVM or, as we will do in chapter 5, to derive a new class of exact solutions. This class correspond to pairs of two Geometrically Uniform density matrices in four dimensions. Another general result on USD of two mixed states is the derivation of lower and upper bounds on the optimal failure probability. The lower bounds are expressed in terms of the fidelity. Therefore we first introduce this quantity. The upper bound is presented in term of the failure probabilities of some pure state case.

\section{Necessary and sufficient conditions for the optimality of a USD measurement}

Necessary and sufficient conditions for an optimal measurement that minimizes the probability of inconclusive result can be derived using argument of duality in vector space optimization \cite{eldar04a}. These conditions are valid for any number of mixed states. Let us now state the theorem.
\begin{theorem}
Let $\{ \rho_i \}$, $1 \le i \le n$ denote a set of density operators with their {\it a priori} probabilities $\{\eta_i\}$. Let denote $T_i$ and $\Delta_i$ two matrices such that $E_i=T_i \Delta_i T_i^\dagger$, $\Delta_i \ge 0$ and $T_i T_i^\dagger = \Pi_{{\cal S}_{E_i}}$, the projection onto the support of $E_i$, for all $1 \le i \le n$. Then {\it necessary and sufficient} conditions for a measurement $\{ E_k \}$, $k=?,1,\dots\,\,,n$ to be an optimal USD measurement are that there exists $Z \ge 0$ such that
\begin{eqnarray}
Z E_?&=&0\\
E_i (Z- \eta_i \rho_i) E_i &=&0,\,\,\, 1 \le i \le n\\
T_i^\perp(Z- \eta_i \rho_i)T_i^\perp &\ge& 0,\,\,\, 1 \le i \le n
\end{eqnarray}
\end{theorem}

We could rephrase this theorem for two mixed states only. The statement then is slightly simpler.

\begin{theorem}
Let $\rho_0$ and $\rho_1$ be two density matrices with {\it a priori} probabilities $\eta_0$ and $\eta_1$. We denote by $P_0^\perp$ and $P_1^\perp$, the projectors onto the kernel of $\rho_0$ and $\rho_1$. Then {\it necessary and sufficient} conditions for an optimal measurement $\{ E_k \}$, $k=?,0,1$ are that there exists $Z \ge 0$ such that
\begin{eqnarray}
Z E_?&=&0,\\
E_0 (Z- \eta_0 \rho_0) E_0 &=&0,\\
E_1 (Z- \eta_1 \rho_1) E_1 &=&0,\\
P_1^\perp(Z- \eta_0 \rho_0)P_1^\perp &\ge& 0,\\
P_0^\perp(Z- \eta_1 \rho_1)P_0^\perp &\ge& 0
\end{eqnarray}
\end{theorem}
One could try to find the general solution for unambiguously discriminating two mixed states by solving the above conditions. However, in the general case it appears difficult to find a positive semi-definite operator $Z$ fulfilling those conditions. Before ending this section, we can notice that
\begin{eqnarray}
\Tr(Z)=P_{success}^{\mathrm{opt}}.
\end{eqnarray}
Indeed Eqn.(2.19) is equivalent to $\sqrt{E_0} (Z- \eta_1 \rho_1) \sqrt{E_0} =0$. Its trace leads to $\Tr(Z E_0)=\eta_0 \Tr(\rho_0 E_0)$. Similarly Eqn.(2.20) yields $\Tr(Z E_1)=\eta_1 \Tr(\rho_1 E_1 )$ so that $\Tr(Z E_0)+ \Tr(Z E_1)=P_{success}^{\mathrm{opt}}$. The completeness relation ${\mathbb 1}=E_?+E_0+E_1$ together with Eqn.(2.18) gives $\Tr(Z)=P_{success}^{\mathrm{opt}}$.
Later in this thesis, we will use Eldar's necessary and sufficient conditions to derive a theorem about the rank of the POVM elements of an optimal USD measurement and a new class of exact solutions of USD.

\section{Bounds on the failure probability}

\subsection{Fidelity}
The fidelity $F(\rho_0,\,\, \rho_1)=\Tr(\sqrt{ \sqrt{\rho_0} \rho_1 \sqrt{\rho_0} })$ is a quantity to distinguish two mixed quantum states $\rho_0$ and $\rho_1$.\\

We can consider the two extreme cases $\rho_0=\rho_1$ and ${\cal S}_{\rho_0} \perp {\cal S}_{\rho_1}$. On one hand, if $\rho_0=\rho_1$ then $F(\rho_0,\,\, \rho_1)=1$. On the other hand, if $\rho_0$ and $\rho_1$ have orthogonal supports then $F(\rho_0,\,\, \rho_1)=0$. The fidelity takes value in $[0,1]$. when $F=1$, the two states are identical. When $F=0$, the two states have orthogonal supports. It is not obvious that the fidelity is a symmetric quantity in its two arguments, though it is as we will show here \cite{jozsa94a,uhlmann76a}. We can first consider the fidelity of two pure states.
\begin{eqnarray}
F(|\Psi_0 \rangle \langle \Psi_0 | ,\,\, | \Psi_1 \rangle \langle \Psi_1 |)&=& \Tr(\sqrt{|\Psi_0 \rangle \langle \Psi_0 | | \Psi_1 \rangle \langle \Psi_1 | |\Psi_0 \rangle \langle \Psi_0 |})\\ \nonumber
&=& |\langle \Psi_0 | \Psi_1 \rangle| \Tr(\sqrt{| \Psi_1 \rangle \langle \Psi_1 |})\\ \nonumber
&=& |\langle \Psi_0 | \Psi_1 \rangle|.
\end{eqnarray}
The fidelity of two pure states simply is the modulus of the overlap between those two pure states! The fidelity is here clearly symmetric. If we now consider mixed states, we can define the operators $F_0=\sqrt{\sqrt{\rho_0}\rho_1\sqrt{\rho_0}}$ and $F_1=\sqrt{\sqrt{\rho_1}\rho_0\sqrt{\rho_1}}$. They actually come from the polar decomposition
\begin{eqnarray}\label{PDD}
\sqrt{\rho_0} \sqrt{\rho_1}= F_0 V=V F_1.
\end{eqnarray}
As written in Eqn.(\ref{PDD}), the two operators $F_0$ and $F_1$ are unitary equivalent and their trace are equal. In other words,
\begin{eqnarray}
F(\rho_i,\,\, \rho_j)=\Tr(F_i)=\Tr(F_j)
\end{eqnarray}
and the fidelity is symmetric.
It might be sometimes difficult to work with the fidelity because of the three square roots involved in its definition and because of the noncommutativity of the density operators. For a review of its properties, the interested reader should look at Jozsa's 1994 paper \cite{jozsa94a} inspired by Uhlmann's {\it transition probability} \cite{uhlmann76a}. Let us however note here that in our work, the fidelity is given by $F(\rho_i,\,\, \rho_j)=\Tr(\sqrt{ \sqrt{\rho_i} \rho_j \sqrt{\rho_i} })$ and not by $F(\rho_i,\,\, \rho_j)= \{ \Tr(\sqrt{ \sqrt{\rho_i} \rho_j \sqrt{\rho_i} } ) \}^2$ \cite{jozsa94a} though the properties remain intact.

Actually one can construct a distance measure from the fidelity, the {\it Bures} distance $d_{\it Bures}^2(\rho_i,\,\, \rho_j)=2(1-F(\rho_i,\,\, \rho_j))$. It is well know that the problem of minimum error discrimination between two mixed states is linked to the {\it trace} distance as $P_{error}=\frac{1}{2}(1-\Tr(|\eta_0 \rho_0 - \eta_1 \rho1|)$. As we are going to see through this thesis, a link between Fidelity and the failure probability $Q$ in USD does exist. It is not as strong as the link between the {\it trace} distance as the error probability $P_{error}$ in MED.
In chapter 4, 5 and 6, we will intensively use the fidelity.

\subsection{Lower bound for the unambiguous discrimination of $n$ mixed states}
Y. Feng {\it et al.} obtained a very general lower bound for unambiguously discriminating $n$ mixed states $\{\rho_i\}$ with {\it a priori} probabilities $\{\eta_i\}$ \cite{feng04a}.
\begin{theorem}
Let $\{\rho_i\}$ be $n$ density matrices with their {\it a priori} probabilities $\eta_i$. We define the fidelity of two states $\rho_i$ and $\rho_j$ as $F(\rho_i,\,\, \rho_j)= \Tr(\sqrt{\sqrt{\rho_i}\rho_j\sqrt{\rho_i}})$. Then, for any USD measurement a lower bound on the failure probability $Q$ is
\begin{eqnarray}\label{first}
Q \ge \sqrt{ \frac{n}{n-1} \sum_{i \ne j}^{n} \eta_i \eta_j F^2(\rho_i,\,\, \rho_j) }.
\end{eqnarray}
\end{theorem}

Let us note here that another lower bound on the failure probability was derived by Y. Feng {\it et al.} (two of the three authors of Ref.~\cite{feng04a}) in an unpublished work \cite{feng04b}. Let us notice that this bound is given as an upper bound on the success probability.
\begin{theorem}
Let $\{\rho_i\}$ be $n$ density matrices with their {\it a priori} probabilities $\{\eta_i\}$. First we define the subspace $Mix(\rho_i)$ as $Mix(\rho_i)={\cal S}_{\rho_i} \cap \sum_{j \neq i} {\cal S}_{\rho_j}$. Second, we divide each $\rho_i$ in two parts, $\widetilde{\rho_i}$ and $\hat{\rho_i}$ such that ${\cal S}_{\hat{\rho_i}}=Mix(\rho_i)$ and ${\cal S}_{\widetilde{\rho_i}} \cap {\cal S}_{\hat{\rho_i}}={0}$. Finally we define the fidelity of two states $\rho_i$ and $\rho_j$ as $F(\rho_i,\,\, \rho_j)= \Tr(\sqrt{\sqrt{\rho_i}\rho_j\sqrt{\rho_i}})$. Then, for any USD measurement an upper bound on the success probability $P_{\textrm success}$ is
\begin{eqnarray}
P_{\textrm success} \le \sum_{i=1}^n \eta_i \Tr(\widetilde{\rho_i}) - \sqrt{ \frac{n}{n-1} \sum_{i \ne j}^{n} \eta_i \eta_j F^2(\widetilde{\rho_i},\,\, \widetilde{\rho_j}) }.
\end{eqnarray}
\end{theorem}
This last bound is tighter than the one in Theorem 5 since $\sum_{i=1}^n \eta_i \Tr(\widetilde{\rho_i}) \le 1$. The equality holds only if the density matrices $\rho_i$ do not have common subspaces. In that case, the two lower bounds in Eqn.(2.27) and Eqn.(2.28) are equal. We now focus on USD of two density matrices only. Rudolph {\it et al.} derived both lower and upper bounds on the failure probability to unambiguously discriminate two mixed states. This is the object of the last subsection of this chapter.

\subsection{Lower and upper bounds on the failure probability for the unambiguous discrimination of two mixed states}

\subsubsection{Lower bound}
\begin{figure}[h!]
  \centering
  \fbox{\includegraphics[width=14cm]{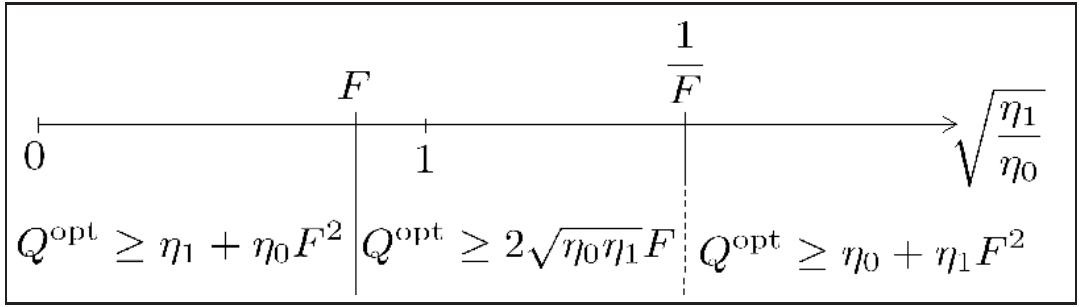}}
  \caption{Lower bounds on the optimal failure probability for USD of two density matrices}
  \label{usd2dmbound}
\end{figure}

In Ref.\cite{rudolph03a}, Rudolph {\it et al.} derived their lower bounds considering some purification of the two mixed states $\rho_0$ and $\rho_1$. Moreover, an interesting property of the fidelity is the following. Given two mixed states, we can consider all their possible purification and their overlap. In fact, the fidelity equals the maximum of the modulus of those overlaps. It is therefore not surprising that those lower bounds involve the optimal failure probability of two pure states where the overlap is replaced by the Fidelity (see Fig.~\ref{usd2dmbound}). More precisely, we end up with

\begin{theorem}
Let $\rho_0$ and $\rho_1$ be two density matrices with {\it a priori} probabilities $\eta_0$ and $\eta_1$. Let define the fidelity $F=\Tr(\sqrt{\sqrt{\rho_0}\rho_1\sqrt{\rho_0}})$ between these two mixed states. Then a lower bound on the failure probability of unambiguously discriminating $\rho_0$ and $\rho_1$ is
\begin{eqnarray}
Q^{\mathrm{opt}} &\ge& \eta_1+\eta_0 F^2  \,\,\, \mathrm{for} \,\,\, \sqrt{\frac{\eta_1}{\eta_0}} \le F,\\
Q^{\mathrm{opt}} &\ge& 2\sqrt{\eta_0\eta_1}F  \,\,\, \mathrm{for} \,\,\, F\le \sqrt{\frac{\eta_1}{\eta_0}} \le \frac{1}{F},\\
Q^{\mathrm{opt}} &\ge& \eta_0+\eta_1 F^2  \,\,\, \mathrm{for} \,\,\, \frac{1}{F} \le \sqrt{\frac{\eta_1}{\eta_0}}.
\end{eqnarray}
\end{theorem}


\subsubsection{Upper bound}

In the same paper \cite{rudolph03a}, the authors presented an upper bound on the optimal failure probability for unambiguous discrimination of two mixed states. This bound comes from considering several two dimensional USD problems rather that a global USD problem. The eigenbases for $E_0$ and $E_1$ here depend only on the supports of $\rho_0$ and $\rho_1$ and not on their eigenvalues. This leads naturally to an upper bound on the failure probability since the eigenvalues of $\rho_0$ and $\rho_1$ would allow to refine the measurement. The theorem presents a lower bound on the success probability instead of an upper bound on the failure probability. 

\begin{theorem}
Let $\rho_0$ and $\rho_1$ be two density matrices with {\it a priori} probabilities $\eta_0$ and $\eta_1$. We denote the dimension of their kernel ${\cal K}_0$ and ${\cal K}_1$ by $s_0$ and $s_1$ and assume that $s_0 \ge s_1$. There exist orthonormal bases $\{|k_b^j\rangle\}_{j=1}^{s_b}$ for ${\cal K}_b$ (b=0,1) such that for $1 \le j \le s_0$, $1 \le i \le s_1$,
\begin{eqnarray}
\langle k_0^j| k_1^i \rangle= Cos(\theta_j) \delta_{ij},
\end{eqnarray}
where the $\theta_j$ are the canonical angles between ${\cal K}_0$ and ${\cal K}_1$. In this case, a lower bound on the optimal success probability $P_{success}^{\mathrm{opt}}$ is 
\begin{eqnarray}
P_{success}^{\mathrm{opt}} \ge \sum_{j=1}^{s_1} P_{success}^{\mathrm{opt}}(|k_0^j\rangle,|k_1^j\rangle) + \sum_{j=s_1+1}^{s_0} \langle k_0^j | \rho_1 | k_0^j \rangle.
\end{eqnarray}
where
\begin{eqnarray}
P_{success}^{\mathrm{opt}}(|k_0^j\rangle,|k_1^j\rangle)=\Bigg\{ \begin{array}{cc}
&A_0^j+A_1^j-2 Cos(\theta_j) \sqrt{A_0^j A_1^j} \,\,\, \mathrm{for} \,\,\, Cos(\theta_j) < \sqrt{\frac{A_{min}^j}{A_{max}^j}} \\ 
&A_{max}^j   \,\,\, \mathrm{otherwise}\\ 
\end{array}
\end{eqnarray}
with $A_0^j=\eta_0 \langle k_1^j| \rho_0 | k_1^j \rangle$, $A_1^j=\eta_1 \langle k_0^j| \rho_1 | k_0^j \rangle$, $A_{min}^j=min\{A_0^j, A_1^j\}$ and $A_{max}^j=max\{A_0^j, A_1^j\}$.
\end{theorem}
Let us note that we will detail the construction of such orthogonal bases $\{|k_b^j\rangle\}_{j=1}^{s_b}$ in Chapter 3 when we will present the optimal unambiguous discrimination of two subspaces.\\

In the next chapter, we will find that any USD problem can be reduced to some standard situation. We will then see that some important tasks in Quantum Information Theory which are related to the USD of some mixed states can actually be reduced to some pure state case.

%% file: chapter4.tex
\chapter{A standard form} \label{standard_form}


We are searching for an optimal USD measurement to discriminate two arbitrary density matrices $\rho_0$ and $\rho_1$ with {\it a priori} probability $\eta_0$ and $\eta_1$ respectively. We find that this general problem can be reduced to a simpler standard situation thanks to three {\it reduction} theorems dealing with simple geometrical considerations. As their names indicate, the three {\it reduction} theorems allow to reduce the dimension of the USD problem. In fact, the reduction can also be applied to the case of more than two density matrices.\\

It is important to notice here that all the results on USD of mixed states known so far are reducible to some pure state scenarios. These cases are state filtering, unambiguous discrimination of two subspaces and unambiguous comparison of two pure states. Those three cases of USD of mixed states can be solved using some reduction theorem and the result of Jaeger and Shimony about USD of two pure states only. This underlines the fact that those cases were solved first because no new techniques were needed. In the following we will often refer to {\it non-reducible} mixed state case as generic USD problem. In the next chapters we are going to present two classes of exact solutions for such generic USD problems. But first of all, let us present, prove and use the three reduction theorems.\\

The first reduction theorem states that, if two density matrices share a common subspace (see Fig.~\ref{overlap}), no unambiguous discrimination is possible on it. Indeed any state vector in such a common subspace belongs to both $\rho_0$ and $\rho_1$ so that no conclusive result is possible. The failure probability restricted to this common subspace then equals unity. There is no optimization to perform onto this common subspace and we can focus our attention on the USD problem onto the orthogonal complement of this common subspace.\\

The second theorem is easy to understand, though the proof happens to be subtle. Let us consider the support ${\cal S}_{\rho_0}$ and ${\cal S}_{\rho_1}$ of two density matrices. Let us assume that there exists a subspace of ${\cal S}_{\rho_1}$ orthogonal to ${\cal S}_{\rho_0}$ (see Fig.~\ref{ortho}). This subspace can be equivalently denoted by ${\cal S}_{\rho_1} \cap {\cal S}_{\rho_0}^{\perp}$ or ${\cal S}_{\rho_1} \cap {\cal K}_{\rho_0}$. If we perform any measurement on that subspace, we can only detect $\rho_1$ but never $\rho_0$ since the measurement is orthogonal to ${\cal S}_{\rho_0}$. The difficulty step is to see that such a strategy is optimal. Here again, no optimization onto the subspace ${\cal S}_{\rho_1} \cap {\cal K}_{\rho_0}$ is needed. After splitting off ${\cal S}_{\rho_1} \cap {\cal K}_{\rho_0}$, we are left with a smaller USD problem. Of course, a similar reduction can be performed for the subspace ${\cal S}_{\rho_0} \cap {\cal K}_{\rho_1}$.\\

The last theorem refers to some block diagonal structure of the supports ${\cal S}_{\rho_0}$ and ${\cal S}_{\rho_1}$ of our two density matrices $\rho_0$ and $\rho_1$. If the supports ${\cal S}_{\rho_0}$ and ${\cal S}_{\rho_1}$ can be simultaneously decomposed into a direct sum of some subspaces, it seems reasonable that the optimal measurement can have the same property. Moreover we can choose the optimal measurement onto the total Hilbert space to be the direct sum of optimal measurements onto the smaller subspaces. In other words, we only have to look for optimality on each orthogonal subspace. This again simplifies the optimization task.\\

Let us now derive the three theorems.


\section{Overlapping supports}

In the first theorem, we will consider the situation where the supports of the two density matrices have a common subspace. This is the case whenever we find that
\begin{eqnarray}
{\rm dim}\left({\cal S}_{\rho_0}\right) + {\rm dim}\left({\cal S}_{\rho_1}\right) >
{\rm dim}\left({\cal H}\right). \;
\end{eqnarray}

\begin{figure}[h!]
  \centering
  \includegraphics[width=10cm]{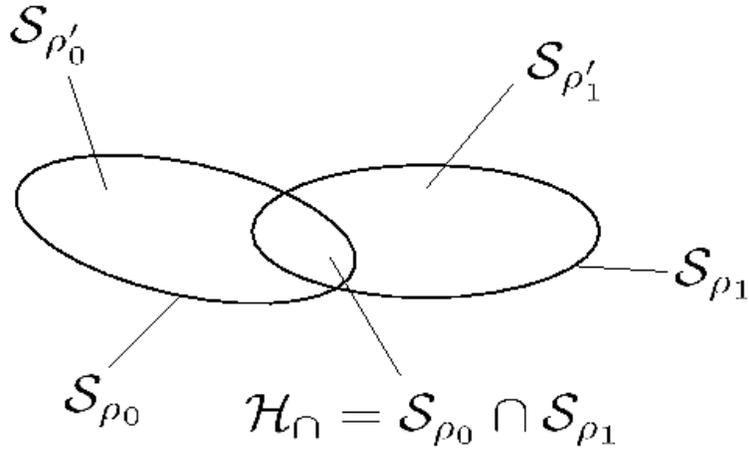}
  \caption{Illustration of a common subspace between $\rho_0$ and $\rho_1$}
  \label{overlap}
\end{figure}

Here ${\cal H}$ is the Hilbert space spanned by the two supports. In this case, it can be written as
\begin{eqnarray}
{\cal H} = {\cal H'}\oplus {\cal H_\cap}
\end{eqnarray}
where ${\cal H_\cap}= {\cal S}_{\rho_0} \cap {\cal S}_{\rho_1}$ is the common subspace of the two supports, and ${\cal H'}$, its  orthogonal complement  in ${\cal H}$ (see Fig.~\ref{overlap}). The first reduction theorem will eliminate the common subspace ${\cal H_\cap}$ from the problem. The intuitive reason is that in this subspace no unambiguous discrimination is possible, so the population of the two density matrices on it will contribute always only to the failure probability, never to the conclusive results. This is made precise in the following theorem. \\


\begin{theorem}
Reduction Theorem for a Common Subspace\\
\graybox{Suppose we are given two density matrices $\rho_0$ and $\rho_1$ in ${\cal H}$ with {\it a priori} probabilities $\eta_0$ and $\eta_1$ such that their respective supports ${\cal S}_{\rho_0}$ and ${\cal S}_{\rho_1}$ have a non-empty common subspace ${\cal H_\cap}$. We denote by ${\cal H'}$ the orthogonal complement of ${\cal H_\cap}$ in ${\cal H}$ while $\Pi_{\cal H_\cap}$ and $\Pi_{\cal H'}$ denote respectively the projector onto ${\cal H_{\cap}}$ and ${\cal H'}$. Then the optimal USD measurement is characterized by POVM elements of the form
\begin{eqnarray}
E^{opt}_0 & = & E^{'opt}_0 \\
E^{opt}_1 & = & E^{'opt}_1 \\
E^{opt}_? & = & E^{'opt}_? + \Pi_{\cal H_\cap}
\end{eqnarray}
\\
where the operators $E^{'opt}_0, E^{'opt}_1, E^{'opt}_?$ form a POVM $\{E_k^{'opt}\}$ with support on $\cal H'$ describing the OptUSDM of a reduced problem defined by
\begin{eqnarray}
\rho'_0=\frac{1}{N_0} \Pi_{\cal H'} \rho_0 \Pi_{\cal H'}, & \eta'_0= \frac{N_0 \eta_0}{N}, & N_0 = {\rm Tr}(\rho_0 \Pi_{\cal H'}) \\
\rho'_1=\frac{1}{N_1} \Pi_{\cal H'} \rho_1 \Pi_{\cal H'}, & \eta'_1= \frac{N_1 \eta_1}{N}, & N_1 = {\rm Tr}(\rho_1 \Pi_{\cal H'}) \\
N=N_0 \eta_0+ N_1 \eta_1\; .
\end{eqnarray}

And finally, the optimal failure probability $Q^{\mathrm{opt}}$ can be written in terms of $Q'^{\mathrm{opt}}$, the optimal failure probability of the reduced problem, as
\begin{eqnarray}
Q^{\mathrm{opt}} &=& 1-N + N Q'^{\mathrm{opt}}. \;
\end{eqnarray}}
\end{theorem}

\paragraph*{\bf Proof}
To prove the reduction theorem, we first need to recall that a USD measurement described by the POVM $\{E_k\}$ satisfies ${\rm Tr}(E_0\rho_1)=0$ and ${\rm Tr}(E_1\rho_0)=0$ by definition. It means, as a consequence of Lemma 1 given in the previous chapter, that $S_{E_0} \perp {S_{\rho_1}}$ and $S_{E_1} \perp {S_{\rho_0}}$. Since ${\cal H}_{\cap}$ is a subspace of ${S_{\rho_0}}$ and ${S_{\rho_1}}$, it follows that $S_{E_0} \perp {\cal H}_{\cap}$ and $S_{E_1} \perp {\cal H}_{\cap}$. Therefore, by writing the block-matrices in $\cal{H}= \cal{H}_\cap \oplus \cal{H}'$, we have

\begin{eqnarray}
E_0 = \left( \begin{array}{cc} 0 & 0 \\ 0 & E_0' \end{array} \right)\\
E_1 = \left( \begin{array}{cc} 0 & 0 \\ 0 & E_1' \end{array} \right)\;
\end{eqnarray}
The completeness relation on $\cal H$ implies firstly
\begin{eqnarray}
E_? = \left( \begin{array}{cc} {\mathbb 1}_{\cal {H}_\cap} & 0 \\ 0 & E_?' \end{array} \right) = \Pi_{\cal {H}_\cap}+ E'_?
\end{eqnarray}

and secondly by the completeness relation on the reduced subspace $\cal H'$
\begin{eqnarray}
\sum_k E_k' = {\mathbb 1}_{\cal{H'}}.
\end{eqnarray}
It follows also that the operators $E_k'$ ($k=0,1,?$) are positive semi-definite operators. Therefore, by definition,  $\{E_k'\}$ is a POVM on $\cal H'$. The fact that $E_?$ is equal to identity in the subspace $\cal H_{\cap}$ is here a direct consequence of the property of an USDM on $\cal H$. Next we will see that $\{E_k'\}$ is a POVM of a USD in $\cal H'$.\\

We define $\Pi_{\cal {H}_\cap}$ and $\Pi_{\cal{H'}}$ as the projector onto $\cal H_{\cap}$ and $\cal H'$ respectively. Thus  $\Pi_{\cal {H}_\cap} \oplus \Pi_{\cal{H'}}={\mathbb 1}_{\cal H}$. For any USDM, because of the diagonal block form of the POVM, we find for $Q$
\begin{eqnarray}
Q &=& \eta_0 {\rm Tr}(\rho_0 E_?)+\eta_1 {\rm Tr}(\rho_1 E_?) \nonumber \\
&=& (1-N_0) \eta_0 + (1-N_1) \eta_1\\
&+& (N_0\eta_0+N_1\eta_1) ( \eta_0' {\rm Tr}(\rho_0'E_?')+\eta_1' {\rm Tr}(\rho_1'E_?')) \nonumber \\
{\rm with} \,\, \rho'_0 &=& \frac{1}{{\rm Tr}(\rho_0 \Pi_{\cal H'})} \Pi_{\cal H'} \rho_0 \Pi_{\cal H'} \\
\rho'_1 &=& \frac{1}{{\rm Tr}(\rho_1 \Pi_{\cal H'})} \Pi_{\cal H'} \rho_1 \Pi_{\cal H'}. \;
\end{eqnarray}
Here $\eta'_i$ ($i=0,1$) is the {\it a priori} probability corresponding to the new density matrix $\rho'_i$ ($\eta'_0+\eta'_1=1$) 
\begin{eqnarray}
\eta'_0= \frac{N_0 \eta_0}{N_0 \eta_0+ N_1 \eta_1}, \, N_0 = {\rm Tr}(\rho_0 \Pi_{\cal H'})\\
\eta'_1= \frac{N_1 \eta_1}{N_0 \eta_0+ N_1 \eta_1}, \, N_1 = {\rm Tr}(\rho_1 \Pi_{\cal H'}).\;
\end{eqnarray}

We notice that ${\cal S}_{\rho'_0} \cap {\cal S}_{\rho'_1}=0$. Moreover, ${\rm Tr}(\rho_0 E_1)=0$ implies ${\rm Tr}(\rho'_0 E'_1)=0$ and ${\rm Tr}(\rho_1 E_0)=0$ implies ${\rm Tr}(\rho'_1 E'_0)=0$. Then $\{E_k'\}$ defines a POVM describing a USDM on $\{\rho_i',\,\, \eta_i'\}$ in $\cal H'$. The problem is now reduced to the subspace $\cal H'$. We now focus our attention on the optimality of the reduced USDM.\\

We can write $Q$ as
\begin{eqnarray}
Q&=&(1-N_0) \eta_0 + (1-N_1) \eta_1 + (N_0\eta_0+N_1\eta_1) Q'\\ \nonumber
&=&1-N+NQ'\;
\end{eqnarray}
where $Q'=\eta_0' {\rm Tr}(\rho_0'E_?')+\eta_1' {\rm Tr}(\rho_1'E_?')$ is, by definition, the failure probability of discriminating unambiguously $\rho_0'$ and $\rho_1'$ in $\cal H'$ with {\it a priori} probabilities $\eta_0'$, $\eta_1'$.

The previous equality implies that the failure probability $Q$ is minimal if and only if the failure probability $Q'$ is minimal. Thus we have that $\{E_k\}$ describes an optimal USDM on $\{\rho_i,\,\, \eta_i\}$ $\Leftrightarrow$ $Q$ is minimal $\Leftrightarrow$ $Q'$ is minimal $\Leftrightarrow$ $\{E'_k\}$ describes an optimal USDM on $\{\rho_i',\,\, \eta_i'\}$. This completes the proof. \hfill $\blacksquare$ \\

Let us note here that two subspaces that do not have a common subspace are not necessarily orthogonal. The formal statement is ${\cal S}_{\rho_0} \cap {\cal S}_{\rho_1}=\{0\} \nLeftrightarrow {\cal S}_{\rho_0} \perp {\cal S}_{\rho_1}$. Moreover we can give an easy way to know whether the two supports overlap of $\rho_0$ and $\rho_1$. In fact, it suffices to check whether the equation $dim({\cal H})=rank(\rho_0)+rank(\rho_1)=rank(\rho_0 + \rho_1)$ holds. Marsaglia and Styan proved that additivity of rank  of two matrices is related to the intersection of their column and row spaces in a simple way \cite{marsaglia72a}. Their result is given in the following theorem.

\begin{theorem}
Let A and B be two complex m{\rm x}n matrices. Let ${\cal C}_A$ and ${\cal C}_B$ be their column spaces and ${\cal R}_A$ and ${\cal R}_B$, their row spaces then\\

$rank(A+B)=rank(A)+rank(B)$ if and only if $dim({\cal C}_A \cap {\cal C}_B)=dim({\cal R}_A \cap {\cal R}_B)=\{0\}$.
\end{theorem}

In the more restricted case of two density matrices, which are Hermitian matrices, the column and row spaces simply are the support ${\cal C}_{\rho}={\cal R}_{\rho}={\cal S}_{\rho}$.


\section{Trivial orthogonal subspaces of the supports}
\begin{figure}[h!]
  \centering
  \includegraphics[width=10cm]{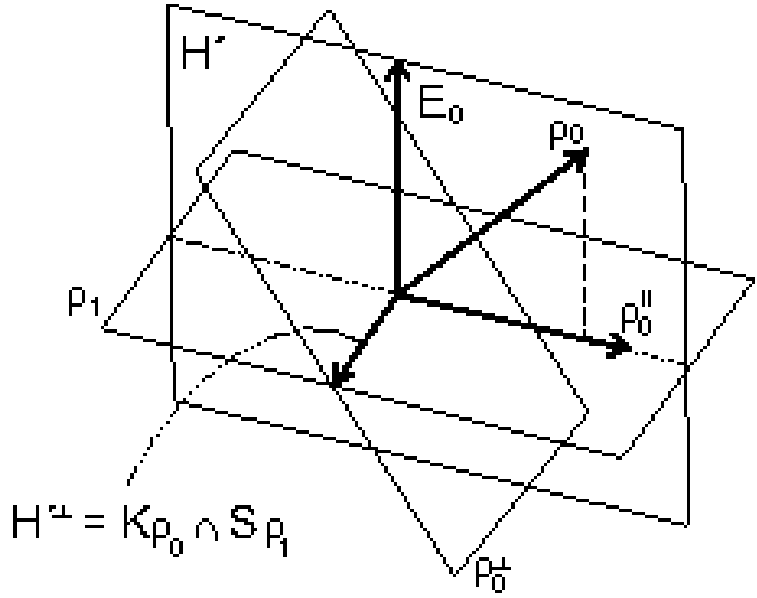}
  \caption{Illustration of the subspace ${\cal K}_{\rho_0} \cap {\cal  S}_{\rho_1}$}
  \label{ortho}
\end{figure}

We now consider the case where the supports of the two density matrices have
no common subspace. That can always be achieved thanks to the previous
reduction theorem for common subspace. If there is a part of ${\cal  S}_{\rho_1}$ orthogonal to ${\cal S}_{\rho_0}$, we can decompose ${\cal S}_{\rho_1}$ into this subspace and another one (see
Fig.~\ref{ortho}). It turns out that this subspace of ${\cal S}_{\rho_1}$ orthogonal to ${\cal S}_{\rho_0}$ can be split off and leads to an unambiguous discrimination without error. The same is true for ${\cal S}_{\rho_0}$.


\begin{theorem}
Reduction Theorem for Orthogonal Subspaces\\
\graybox{Suppose we are given two density matrices $\rho_0$ and $\rho_1$ in ${\cal H}$ with {\it a priori} probabilities $\eta_0$ and $\eta_1$. Assuming that their supports ${\cal S}_{\rho_0}$ and ${\cal S}_{\rho_1}$ have no common subspace, one can construct a decomposition
\begin{eqnarray}
{\cal H}={\cal H}' \oplus {\cal H}^{' \perp}
\end{eqnarray}
with ${\cal H}^{' \perp}=S_0^{\perp} \oplus S_1^{\perp}$, ${\cal S}_0^{\perp}={\cal K}_{\rho_0} \cap {\cal  S}_{\rho_1}$ and ${\cal S}_1^{\perp}={\cal K}_{\rho_1} \cap {\cal  S}_{\rho_0}$.

The solution of the optimal USDM problem can be given, with help of $\Pi_{{\cal S}^{\perp}_0}$ and $\Pi_{{\cal S}^{\perp}_1}$, the projection onto ${\cal S}_0^{\perp}$ and ${\cal S}_1^{\perp}$, respectively,  in ${\cal H}={\cal H}' \oplus {\cal H}^{' \perp}$, by
\begin{eqnarray}
E^{opt}_0 & = & E^{'opt}_0 + \Pi_{{\cal S}_1^{\perp}}\\
E^{opt}_1 & = & E^{'opt}_1 + \Pi_{{\cal S}_0^{\perp}}\\
E^{opt}_? & = & E^{'opt}_? .
\end{eqnarray}
The operators $E^{'opt}_0, E^{'opt}_1, E^{'opt}_?$ form a POVM $\{E_k^{'opt}\}$ with support on $\cal H'$ describing the OptUSDM of a reduced problem defined by
\begin{eqnarray}
\rho'_0=\frac{1}{N_0} \Pi_{\cal H'} \rho_0 \Pi_{\cal H'}, & \eta'_0= \frac{N_0 \eta_0}{N}, & N_0 = {\rm Tr}(\rho_0 \Pi_{\cal H'}) \\
\rho'_1=\frac{1}{N_1} \Pi_{\cal H'} \rho_1 \Pi_{\cal H'}, & \eta'_1= \frac{N_1 \eta_1}{N}, & N_1 = {\rm Tr}(\rho_1 \Pi_{\cal H'}) \\
N=N_0 \eta_0+ N_1 \eta_1.\;
\end{eqnarray}
And finally, the optimal failure probability $Q^{\mathrm{opt}}$ can be written in terms of $Q'^{\mathrm{opt}}$, the optimal failure probability of the reduced problem as
\begin{eqnarray}
Q^{\mathrm{opt}} &=& NQ'^{\mathrm{opt}}.\;
\end{eqnarray}}
\end{theorem}

\paragraph*{\bf Proof}
We translate the problem using a Naimark extension and a projection-valued measure (PVM). This idea is inspired by the first work of Sun {\it et al.} \cite{sun02a} where an extended Hilbert space has been used. Let us repeat the Naimark theorem.

Given a POVM $\{E_k\}$ on a Hilbert space $\cal H$, it exists an embedding of $\cal H$ into a larger Hilbert space $\cal R$ such that the measurement can be described by  projections onto orthogonal subspaces in $\cal R$. More precisely, there exist a Hilbert space $\cal R$, an embedding $\cal E$ such that ${\cal E }{\cal H} ={\cal R}$ and a PVM $\{R_k\}$ in $\cal R$ such that with P, the projection defined by $P{\cal R} ={\cal H}$, $E_k = PR_kP, \, \forall k$.

To the three POVM elements $E_k$ in $\cal H$ correspond three PVM elements $R_k$ in $\cal R$. The Hilbert space ${\cal R}$ can be decomposed into orthogonal subspaces
\begin{eqnarray}
{\cal R} = {\cal S}_{R_0} \oplus {\cal S}_{R_1} \oplus {\cal S}_{R_?}
\end{eqnarray}
which give raise to non-orthogonal subspaces in $\cal H$ as ${\cal S}_{E_k} = P{\cal S}_{R_k}P$. We can therefore translate properties of the USD POVM to the embedding of $\cal H$ into $\cal R$.

Next we take a look at the embedding of ${\cal S}_{\rho_0}$ and ${\cal S}_{\rho_1}$ into $\cal R$ and we translate the conditions for an USDM into the embedded language. We denote the embedded subspaces of $\cal R$ by the same symbol as the original subspace of $\cal H$. We can here introduce the projector $P^\perp$ onto the orthogonal complement ${\cal H}^\perp$ of ${\cal H}$ in ${\cal R}$ ($P+P^\perp={\mathbb 1}_{\cal R}$). Since ${\cal S}_{\rho_0} \in {\cal H}$, we have ${\rm Tr}(\rho_0 R_1)=\Tr(P\rho_0P R_1)=\Tr(\rho_0 E_1)=0$. This implies that ${\cal S}_{\rho_0}$ is orthogonal to ${\cal S}_{R_1}$. Similarly, we find that ${\cal S}_{\rho_1}$ is orthogonal to ${\cal S}_{R_0}$. Therefore, we can write
\begin{eqnarray}
{\cal S}_{\rho_0} \subset {\cal S}_{R_0} \oplus {\cal S}_{R_{?0}}\\
{\cal S}_{\rho_1} \subset {\cal S}_{R_1} \oplus {\cal S}_{R_{?1}}
\end{eqnarray}
where ${\cal S}_{R_{?0}}$ and ${\cal S}_{R_{?1}}$ are defined as subspaces of ${\cal S}_{R_?}$ with minimal dimension fulfilling the above decompositions in the sense that ${\cal S}_{R_{?i}}={\rm Support}(\Pi_{{\cal S}_{R_?}}{\cal S}_{\rho_i}\Pi_{{\cal S}_{R_?}})$ for $i=0,1$.

The optimality condition means in particular that no information should be obtained from the conditional states following an inconclusive result. If the two failure spaces ${\cal S}_{R_{?0}}$ and ${\cal S}_{R_{?1}}$ are different, it will be possible to distinguish the conditional states which arise from a projection onto ${\cal S}_{R_?}$ \cite{sun02a}. Indeed a detection in an orthogonal direction to one of the two subspaces will tell us which failure space was it or equivalently which state was sent. Therefore the optimality condition implies that ${\cal S}_{R_{?0}}={\cal S}_{R_{?1}}$ and then
\begin{eqnarray}
{\cal S}_{R_{?}}={\cal S}_{R_{?0}}={\cal S}_{R_{?1}}.
\end{eqnarray}
This is an important necessary condition for the optimality of a USD POVM. In the framework of the Naimark extension, this condition translates as follows. The equality of ${\cal S}_{R_{?0}}$ and ${\cal S}_{R_{?1}}$ implies that a subspace ${\cal S}_0^{\perp}={\cal K}_{\rho_0} \cap {\cal  S}_{\rho_1}$ satisfies ${\cal S}_0^{\perp} \subset {\cal S}_{R_1}$ in order to assure that the overlap between any state in ${\cal S}_0^{\perp}$ and any state in ${\cal S}_{\rho_0}$ will be zero. Similarly, ${\cal S}_1^{\perp}= {\cal K}_{\rho_1} \cap {\cal  S}_{\rho_0} \subset {\cal S}_{R_0}$.

Then there exist two subspaces ${\cal H}_1$ in ${\cal S}_{R_1}$ and ${\cal H}_0$ in ${\cal S}_{R_0}$ such that
\begin{eqnarray}
{\cal S}_{R_1}&=&{\cal S}_0^{\perp} \oplus {\cal H}_1\\
{\cal S}_{R_0}&=&{\cal S}_1^{\perp} \oplus {\cal H}_0.
\end{eqnarray}

The orthogonal projection $R_1$ then can be decomposed into a sum of orthogonal projectors as $\Pi_{{\cal S}^{\perp}_0} + \Pi_{{\cal H}_1}$, with $\Pi_{{\cal S}^{\perp}_0}\Pi_{{\cal H}_1}=0$, and the orthogonal projection $R_0$ as $\Pi_{{\cal S}^{\perp}_1} + \Pi_{{\cal H}_0}$, with $\Pi_{{\cal S}^{\perp}_1} \Pi_{{\cal H}_0}=0$. These projectors are mapped into $\cal H$ via the projection $P$. Since ${\cal S}_i^{\perp}$ is already in $\cal H$, we have $P\Pi_{{\cal S}_i^{\perp}}P = \Pi_{{\cal S}_i^{\perp}}$. We define $E'_i=P\Pi_{{\cal H}_i}P, \quad \forall i=0,1$ so that
\begin{eqnarray}
E_0 & = & E'_0 + \Pi_{{\cal S}_1^{\perp}}\\
E_1 & = & E'_1 + \Pi_{{\cal S}_0^{\perp}}.\;
\end{eqnarray}
Furthermore, the two supports ${\cal S}_{E_0'}$ and ${\cal S}_1^\perp$ are orthogonal since $\Pi_{{\cal H}_0} \Pi_{{\cal S}^{\perp}_1}=0$ implies $\Pi_{{\cal H}_0} P P \Pi_{{\cal S}^{\perp}_1}P=0$ so that $P\Pi_{{\cal H}_0} P P \Pi_{{\cal S}^{\perp}_1}P=E'_0 \Pi_{{\cal S}^{\perp}_1}=0$. Similarly the two supports ${\cal S}_{E_1'}$ and ${\cal S}_0^\perp$ are orthogonal too.

Moreover, ${\cal S}_{E_0} \perp {\cal S}_{\rho_1}$ and ${\cal S}_0^\perp \in {\cal S}_{\rho_1}$ so that ${\cal S}_{E_0} \perp {\cal S}_0^{\perp}$. Similarly, we have ${\cal S}_{E_1} \perp {\cal S}_1^{\perp}$. Then $E'_0$ and $E'_1$ have support on a subspace ${\cal H}'$, which is the complementary orthogonal subspace of ${\cal H}^{'\perp}={\cal S}_0^{\perp} \oplus {\cal S}_1^{\perp}$.

Therefore in ${\cal H}={\cal H}' \oplus {\cal S}_0^{\perp} \oplus {\cal S}_1^{\perp}={\cal H}'\oplus {\cal H}^{' \perp}$, we find
\begin{eqnarray}
E_0 = \left( \begin{array}{ccc} E'_0 & 0 & 0 \\ 0 & {\mathbb 1}_{{\cal S}_1^{\perp}} & 0 \\ 0 & 0 & 0 \end{array} \right)\\
E_1 = \left( \begin{array}{ccc} E'_1 & 0 & 0 \\ 0 & 0 & 0 \\ 0 & 0 & {\mathbb 1}_{{\cal S}_0^{\perp}} \end{array} \right).\;
\end{eqnarray}
From here, we will follow the same argumentation as we used in the proof of Theorem 9. The completeness relation on $\cal H$ implies firstly
\begin{eqnarray}
E_? = \left( \begin{array}{ccc} E'_? & 0 &0 \\ 0& 0 & 0 \\0 & 0 & 0 \end{array} \right)
\end{eqnarray}
and secondly the completeness relation on the reduced subspace $\cal H'$
\begin{eqnarray}
\sum_k E_k' = {\mathbb 1}_{\cal{H'}}.
\end{eqnarray}
It follows also that the $E_k'$ ($k=0,1,?$) are positive semi-definite operators. Therefore, by definition, $\{E_k'\}$ is a POVM on $\cal H'$. 

Let us note that ${\cal S}_{\rho_0} \subset {\cal S}_1^{\perp} \oplus {\cal H}_0 \oplus {\cal S}_{R_{?0}}$ and ${\cal S}_{\rho_1} \subset {\cal S}_0^{\perp} \oplus {\cal H}_1 \oplus {\cal S}_{R_{?1}}$. The fact that ${\cal S}_1^{\perp} \subset {\cal S}_{\rho_0}$ implies that
\begin{eqnarray}
{\cal S}_{\rho_0} = {\cal S}_1^{\perp} \oplus {\cal H}'_0,
\end{eqnarray}
with ${\cal H}'_0 \subset {\cal H}_0 \oplus {\cal S}_{R_{?0}}$. In the same way, with ${\cal H}'_1 \subset {\cal H}_1 \oplus {\cal S}_{R_{?1}}$,
\begin{eqnarray}
{\cal S}_{\rho_1} = {\cal S}_0^{\perp} \oplus {\cal H}'_1. 
\end{eqnarray}
Therefore, we can introduce a reduced problem onto ${\cal H}'$ defined such that ${\cal H}={\cal H}'\oplus {\cal S}_0^{\perp} \oplus {\cal S}_1^{\perp}$.

For any USDM, because of the diagonal block form of the POVM, we find for $Q$
\begin{eqnarray}
Q &=& \eta_0 {\rm Tr}(\rho_0E_?)+\eta_1 {\rm Tr}(\rho_1E_?) \\
&=& (N_0\eta_0+N_1\eta_1) ( \eta_0' {\rm Tr}(\rho_0'E'_?)+\eta_1' {\rm Tr}(\rho_1'E_?')) \nonumber \\
{\rm with}\,\, \rho'_0 &=& \frac{1}{{\rm Tr}(\rho_0 \Pi_{\cal H'})} \Pi_{\cal H'} \rho_0 \Pi_{\cal H'} \\
\rho'_1 &=& \frac{1}{{\rm Tr}(\rho_1 \Pi_{\cal H'})} \Pi_{\cal H'} \rho_1 \Pi_{\cal H'}. \;
\end{eqnarray}

Here $\eta'_i$ ($i=0,1$) is the {\it a priori} probability corresponding to the new density matrix $\rho'_i$ ($\eta'_0+\eta'_1=1$) 
\begin{eqnarray}
\eta'_0= \frac{N_0 \eta_0}{N_0 \eta_0+ N_1 \eta_1}, \, N_0 = {\rm Tr}(\rho_0 \Pi_{\cal H'})\\
\eta'_1= \frac{N_1 \eta_1}{N_0 \eta_0+ N_1 \eta_1}, \, N_1 = {\rm Tr}(\rho_1 \Pi_{\cal H'}).\;
\end{eqnarray}
Moreover, ${\rm Tr}(\rho_0 E_1)=0$ implies ${\rm Tr}(\rho'_0 E'_1)=0$ and ${\rm Tr}(\rho_1 E_0)=0$ implies ${\rm Tr}(\rho'_1 E'_0)=0$. Then $\{E_k'\}$ defines a POVM describing a USDM on $\{\rho_i'\}$ in $\cal H'$.\\

We can rewrite the failure probability $Q$ as
\begin{eqnarray}
Q=(N_0\eta_0+N_1\eta_1) Q'
\end{eqnarray}
where $Q'=\eta_0' {\rm Tr}(\rho_0'E_?')+\eta_1' {\rm Tr}(\rho_1'E_?')$ is, by definition, the failure probability of discriminating unambiguously $\rho_0'$ and $\rho_1'$ in $\cal H'$ with {\it a priori} probabilities $\eta'_0$ and $\eta'_1$, respectively.

And again, we have that $\{E_k\}$ describes an optimal USDM on $\{\rho_i,\,\, \eta_i\}$ $\Leftrightarrow$ $Q$ is minimal $\Leftrightarrow$ $Q'$ is minimal $\Leftrightarrow$ $\{E'_k\}$ describes an optimal USDM on $\{\rho_i',\,\, \eta_i'\}$. This completes the proof. \hfill $\blacksquare$ \\


\section{Block diagonal structure}
It is possible to state a last geometrical theorem which deals with two block diagonal density matrices $\rho_0$ and $\rho_1$. Schematically, $\rho_0$ and $\rho_1$ are then of the form

\begin{displaymath}
  \left(
    \begin{array}{ccc}
      \squarebox  &  0           & 0\\
          0        & \squarebox  &0 \\
             0     &     0        & \squarebox
    \end{array}
  \right).
\end{displaymath}
The problem of unambiguously discriminating such two density matrices can be reduced to smaller USD problems onto each one of the orthogonal subspaces. This is made more precise in the next theorem.

\begin{theorem}
Reduction Theorem for two block diagonal density matrices\\
\graybox{Suppose we are given two density matrices $\rho_0$ and $\rho_1$ in ${\cal H}$ with {\it a priori} probabilities $\eta_0$ and $\eta_1$. Suppose that $\rho_0$ and $\rho_1$ are block diagonal (in other words, it exists a set of orthogonal projectors $\{ \Pi_k\}$ such that $\sum_{k=1}^{n} \Pi_k={\mathbb 1}$ and $\rho_i=\sum_{k=1}^{n} \Pi_k\rho_i \Pi_k$, $i=0,1$). Then the optimal USD measurement can be chosen block diagonal where each block is optimal onto its restricted subspace.\\
More precisely, the optimal USD measurement is characterized by POVM elements of the form
\begin{eqnarray}
E^{opt}_i & = & \sum_k E^{k \,\, opt}_i.
\end{eqnarray}
For $k=1,...,n$, the operators $E^{k \,\, opt}_0, E^{k \,\, opt}_1, E^{k \,\, opt}_?$ form a POVM $\{E^{k \,\, opt}_j\}$ with support on ${\cal S}_{P_k}$ describing the OptUSDM of the reduced problem defined by
\begin{eqnarray}
\rho^k_0=\frac{1}{N^k_0} \Pi_k \rho_0 \Pi_k, & \eta^k_0= \frac{N^k_0 \eta_0}{N^k}, & N^k_0 = {\rm Tr}(\rho_0 \Pi_k) \\
\rho^k_1=\frac{1}{N^k_1} \Pi_k \rho_1 \Pi_k, & \eta^k_1= \frac{N^k_1 \eta_1}{N^k}, & N^k_1 = {\rm Tr}(\rho_1 \Pi_k) \\
N^k=N^k_0 \eta_0+ N^k_1 \eta_1\; .
\end{eqnarray}
And finally, the optimal failure probability can be written in terms of $Q^{k \,\, opt}$, the failure probability of the reduced problems, as
\begin{eqnarray}
Q^{\mathrm{opt}}=\sum_k N_k Q_k^{\mathrm{opt}}. \;
\end{eqnarray}}
\end{theorem}

\paragraph*{\bf Proof}
We start with two block diagonal mixed states $\rho_0$ and $\rho_1$ with {\it a priori} probabilities $\eta_0$ and $\eta_1$. In other words, we assume that it exists a set of orthogonal projectors $\{ \Pi_k\}$ such that $\sum_{k=1}^n \Pi_k={\mathbb 1}$ and $\rho_i=\sum_{k=1}^n \Pi_k\rho_i \Pi_k$, $i=0,1$. Next, we denote ${\cal S}_{\Pi_k}$, the support of the projector $\Pi_k$. We first show that only the restriction of the POVM to the $n$ orthogonal subspaces ${\cal S}_{\Pi_k}$ is relevant to the failure probability. Then we will show that optimality on each orthogonal subspace  ${\cal S}_{\Pi_k}$ leads to optimality on the total Hilbert space. Let us consider a USD POVM $\{E_j\}$ onto $\cal H$ and its failure probability $Q$ which can be written

\begin{eqnarray}
Q&=& \sum_i \eta_i \Tr(E_? \rho_i)\\ \nonumber
&=& \sum_i \eta_i \Tr(E_? (\sum_k \Pi_k \rho_i \Pi_k))\\ \nonumber
&=& \sum_k \sum_i \eta_i \Tr( \Pi_k E_? \Pi_k \Pi_k \rho_i \Pi_k)
\end{eqnarray}

We can obviously define $n$ reduced density matrices onto the $n$ subspaces ${\cal S}_{\Pi_k}$ as

\begin{eqnarray}
\rho_i^k&=&\frac{\Pi_k \rho_i \Pi_k}{N_i^k}\\
\eta_i^k&=&\frac{N_i^k \eta_i}{N^k}\\
N_i^k&=&\Tr(\Pi_k \rho_i)
\end{eqnarray}

with $N_k=\sum_i N_i^k \eta_i$. We can also consider the restrictions of the POVM elements $E_0, E_1$ and $E_?$ onto those $n$ subspaces. Thus

\begin{eqnarray}
E_0^k&=&\Pi_k E_0 \Pi_k\\ \nonumber
E_1^k&=&\Pi_k E_1 \Pi_k\\ \nonumber
E_?^k&=&\Pi_k E_? \Pi_k.
\end{eqnarray}

Obviously those operators $E_i^k$ ($i=0,1,?$) are positive semi-definite and add up to $\Pi_k$ since $\sum_i E_i= {\mathbb 1}$. Each restriction onto ${\cal S}_{\Pi_k}$ of a POVM $\{E_i\}$ then forms a POVM onto the subspace ${\cal S}_{\Pi_k}$.  Moreover $\Tr(E_i^k\rho_j^k)=\Tr(\Pi_k E_i \rho_j \Pi_k)=\Tr(\Pi_k E_i \rho_i \Pi_k)\delta_{ij}$ since $E_i \rho_j= E_i \rho_i \delta_{ij}$ for $i,j=0,1$, so that the $n$ POVMs are $n$ USD POVMs.

As a consequence, the failure probability for any two block diagonal density matrices can be expressed in terms of the failure probabilities $Q^k=\sum_i \eta_i^k \Tr(E_i^k \rho_i^k)$ of the $n$ reduced problems as 
\begin{eqnarray}
Q=\sum_k N_k Q^k.
\end{eqnarray}
We can now show that if each block is optimal then the block diagonal POVM onto $\cal H$ is optimal too.

To prove it, let us consider an optimal USD POVM onto each one of the $n$ orthogonal subspaces ${\cal S}_{\Pi_k}$. We denote $Q^{k \,\, opt}$ the optimal failure probability onto ${\cal S}_{\Pi_k}$. By definition of the optimal failure probability, $Q^k \ge Q^{k \,\, opt}$ for each subspace ${\cal S}_{\Pi_k}$. Since both $N_k$ and $Q^k$ are positive numbers, this yields
\begin{eqnarray}
Q \ge \sum_k N_k Q^{k \,\, opt}.
\end{eqnarray}
This bounds can be reached for $\{E_j\}$ being the direct sum of the $n$ optimal USD POVMs $\{E_j^k\}$ i.e.\ $E_j=\sum_{k=1}^n E^k_j$, $j=0,1,?$. The completes the proof. \hfill $\blacksquare$ \\


\section{A standard form of USD problem}

At this point, it is useful to introduce a notation to summarize our knowledge about the USD of two density matrices. We have ${\cal H}= {\cal S}_{\rho_0} + {\cal S}_{\rho_1}$ then ${\rm dim}\left({\cal H}\right) = {\rm dim}\left({\cal S}_{\rho_0}\right) + {\rm dim}\left({\cal S}_{\rho_1}\right) - {\rm dim}\left({\cal S}_{\rho_0} \cap {\cal S}_{\rho_1}\right)$. It implies, by denoting the dimension of the Hilbert space $\cal H$ as $d$, that the respective ranks $r_0$ and $r_1$ of the density matrices $\rho_0$ and $\rho_1$ satisfy
\begin{eqnarray}
r_0 + r_1 \ge d.
\end{eqnarray}
For example, the case of two density matrices of the same rank $(n-1)$ in an Hilbert space of dimension $n$ described by Rudolph {\it et al.} \cite{rudolph03a} can be written as ``$\left(n-1\right)+\left(n-1\right) > n$'' while the USD between one pure state and a mixed state described by Bergou {\it et al.} \cite{sun02a,bergou03a,herzog05b} can be characterized as the ``$1+n = (n+1)$'' case. We will see in the following section that important tasks in quantum information theory can be solved elegantly thanks to those three reduction theorems.\\

First of all, let us discuss some immediate consequences of the three above theorems. The first reduction theorem corresponds to the elimination of the common subspace. A common subspace is present when $r_0+r_1 >d$ holds. Its dimension is $d_{\cap}=r_0+r_1-d$. Therefore, after elimination of that subspace, we end up in the case $r'_0+r'_1=d'$ with $r'_0=r_0-d_{\cap}$ and similarly for $r'_1$ and $d'$. Then, we can reduce the Rudolph's case of discriminating unambiguously two density matrices of the same rank $(n-1)$ in an Hilbert space of dimension $n$ to the ``$1+1=2$'' case of two pure states because the common subspace is ($n-2$)-dimensional. Rudolph {\it et al.} \cite{rudolph03a} already noticed it in their paper. The reduction is constructive given $\rho_0$ and $\rho_1$.

The second reduction theorem corresponds to the elimination of the orthogonal part of one support with respect to the other, i.e., ${\cal K}_{\rho_0} \cap {\cal  S}_{\rho_1}$ and ${\cal K}_{\rho_1} \cap {\cal  S}_{\rho_0}$. The non-empty subspaces ${\cal K}_{\rho_0} \cap {\cal S}_{\rho_1}$  and ${\cal K}_{\rho_1} \cap {\cal S}_{\rho_0}$ can be found systematically. For example, ${\cal K}_{\rho_0} \cap {\cal S}_{\rho_1}$ can be found by projecting ${\cal S}_{\rho_0}$ onto ${\cal S}_{\rho_1}$ and then by taking the complementary orthogonal subspace in ${\cal S}_{\rho_2}$ of that projection. As a matter of fact, this assures that we can reduce a general USD problem always to that of two density matrices of the same rank $r$, $r \le \min(r_0, r_1)$, in a Hilbert space of $2r$ dimensions. Indeed, if after the first reduction, the rank of $\rho_1'$ is bigger than the rank of $\rho_0'$, then the subspace ${\cal K}_{\rho'_0} \cap {\cal S}_{\rho'_1}$ is at least of dimension $r'_1-r'_0$ and can be eliminated.
With the help of the first two reduction theorems, we can reduce any problem of discriminating unambiguously two density matrices $\rho_0$ and $\rho_1$, with rank $r_0$ and $r_1$ respectively, in a Hilbert space $\cal H$, into a problem of discriminating unambiguously two density matrices $\rho_0'$ and $\rho_1'$ with rank $r$ ($r \le \min (r_0, r_1)$) in $\cal H' \subset H$, a 2$r$-dimensional Hilbert space. The reduction is constructive. The first theorem allows us to split off the common subspace and the second theorem leads to the reduce problem of discriminating unambiguously two density matrices of the same rank. The third theorem tells us that if the two density matrices have a block diagonal structure, we can reduce the problem of unambiguously discriminating them to some smaller ones, each one corresponding to a block. In fact, the three reduction theorems allow us to define a {\it standard form} of USD problem as follows.

\begin{definition}
Standard form\\
\graybox{Any Unambiguous State Discrimination problem of two density matrices of rank $r_0$ and $r_1$ is reducible to that of two density matrices of the same rank $r \le \min (r_0, r_1)$) in a $2r$-dimensional Hilbert space without {\bf overlapping} supports, without {\bf trivial orthogonal} subspaces and without {\bf block diagonal} form. Such a problem is called a {\bf standard} Unambiguous State Discrimination problem.}
\end{definition}

The expression '{\bf trivial orthogonal} subspaces' stands for the subspaces ${\cal K}_{\rho_0} \cap {\cal  S}_{\rho_1}$ and ${\cal K}_{\rho_1} \cap {\cal  S}_{\rho_0}$. It is also interesting to note that the dimension of the failure space can not be greater than the lowest rank of the involved density matrices. In the language used in the proof of the second reduction theorem, we first have $E_? = P R_? P$ so that ${\rm dim}({\cal S}_{E_?}) \le {\rm dim}({\cal S}_{R_?})$. Second the dimension of ${\cal S}_{R_{?i}}$ can not be greater than $r_i$ because ${\cal S}_{R_{?i}}={\rm support}(R_?{\cal S}_{\rho_i}R_?)$, for $i=0,1$, and ${\cal S}_{R_{?}}={\cal S}_{R_{?1}}={\cal S}_{R_{?1}}$. Therefore ${\rm dim}{\cal S}_{E_?} \le \min_i {\rm dim}{\cal S}_{\rho_i}$ and we can define the maximum rank of $E_?$ as
\begin{eqnarray}
r_{E_?}^{max}=min (r_0,r_1).
\end{eqnarray}

This result looks natural considering that we can finally reduce any problem of discriminating two density matrices with rank $r_0$ and $r_1$, respectively, to the problem of discriminating two density matrices of the same rank $r$, $r \le \min_i r_i$.\\

Finally, a generalization to more than two density matrices can be achieved. Considering $n$ density matrices $\rho_k \,  (k=0...n-1)$ with {\it a priori} probabilities $\eta_k$, we can construct $n$ pairs of density matrices
\begin{eqnarray}
{\tilde \rho_0}=\rho_i, \,\,\,
i \in [0,..,n-1 ]
\end{eqnarray}
and
\begin{eqnarray}
{\tilde \rho_1}=\frac{\sum_{j=0, j \ne i}^{n-1} \eta_j\rho_j}{1-\eta_i}
\end{eqnarray}
with ${\tilde \eta}_0=\eta_i$, ${\tilde \eta}_1=1-\eta_i$, and apply the two reduction theorems to these two density matrices in the following sense (notice that ${\tilde \rho_1}$ has no physical meaning). As soon as a common subspace between any ${\cal S}_{\tilde \rho_0}$ and ${\cal S}_{\tilde \rho_1}$ exists, we can split it off from all the ${\cal S}\rho_i$'s because if we cannot discriminate unambiguously this part of the support of ${\tilde \rho_0}$ and ${\tilde \rho_1}$ then we can not discriminate unambiguously between this part of the support of all the $\rho_j$. The second theorem must be used more carefully. As soon as a subspace of ${\cal S}_{\tilde \rho_0}$ is orthogonal to ${\cal S}_{\tilde \rho_1}$ (${\cal K}_{\tilde \rho_1} \cap {\cal  S}_{\tilde\rho_0} \ne \{0\}$), we can eliminate it from the problem because it is orthogonal to the supports of all the $\rho_j$, $j \ne i$. However we cannot eliminate a subspace of ${\cal S}_{\tilde \rho_1}$ orthogonal to ${\cal S}_{\tilde \rho_0}$ (${\cal K}_{\tilde \rho_0} \cap {\cal  S}_{\tilde\rho_1} \ne \{0\}$) because we know nothing about the orthogonality of this subspace for all the states in $\tilde \rho_1$. In other words, we can only reduce the density matrix $\rho_i$ corresponding to ${\tilde \rho_0}$.\\

In the following section we are going to apply the reduction theorems to three important tasks in quantum information theory. Those tasks are State Filtering, Unambiguous Comparison of two subspaces and Unambiguous State Comparison of two pure states. We are going to see that those three tasks are reducible to some pure state case only.


\section{Applications of the reduction theorems}


\subsection{State Filtering}
Let us consider $n$ pure states $\{|\Psi_i\rangle\}$ with {\it a priori} probabilities $\{p_i\}$, $i=0,...,n-1$. We may want to group them in several sets and to unambiguously discriminate among these sets. This task is called {\it State Filtering} \cite{sun02a,herzog05b}. The simplest case deals with two sets only where the first set contains one pure state and the second set regroups the remaining $n-1$ states. This problem was studied in various papers by Bergou {\it et al.} \cite{sun02a,bergou03a,herzog05b} who gave the complete solution in  \cite{herzog05b}. We derive here this last result is an extremely simple way thanks to the second reduction theorem.\\

We have to unambiguously discriminate the two sets $\{|\Psi_0\rangle\}$ and $\{|\Psi_i\rangle\}_{i=1,...,n-1}$. We can consider the density matrices corresponding to these two sets as well as their {\it a priori} probabilities. The first density matrix obviously is $\rho_0=|\Psi_0\rangle \langle \Psi_0|$ with {\it a priori} probability $\eta_0=p_0$. The second mixed state can be written as
\begin{eqnarray}
\widetilde{\rho_1}=\sum_{i=1}^{n-1} p_i |\Psi_i\rangle \langle \Psi_i|.
\end{eqnarray}
This is not a proper density matrix since it is not normalized. We then must write $\rho_1=\frac{\sum_{i=1}^{n-1} p_i |\Psi_i\rangle \langle 
\Psi_i|}{\sum_{i=1}^{n-1} p_i}$. Its {\it a priori} probability simply is $\eta_1=\sum_{i=1}^{n-1} p_i=1-p_0$. State filtering finally is equivalent to unambiguously discriminate
\begin{eqnarray}
\rho_0=|\Psi_0\rangle \langle \Psi_0|
\end{eqnarray}
with {\it a priori} probability $\eta_0=p_0$ and
\begin{eqnarray}
\rho_1=\frac{\sum_{i=1}^{n-1} p_i |\Psi_i\rangle \langle \Psi_i|}{\eta_1}
\end{eqnarray}
with {\it a priori} probability $\eta_1=\sum_{i=1}^{n-1} p_i$.

After writing these two density matrices, the solution to the problem is trivial.

Indeed a consequence of Theorem 11 is that we can reduce the problem of USD between a pure state and a density matrix, a ``$1+n=(n+1)$'' case, to the problem of discriminating unambiguously two pure states, a ``$1+1=2$'' case, by splitting off ${\cal K}_{\rho_0} \cap {\cal S}_{\rho_1}$ of dimension $(n-1)$. The two reduced states are the original pure state $|\Psi_0\rangle$ and the unit vector corresponding to the projection of $\rho_0$ onto the support of the mixed state $\rho_1$. This unnormalized vector is given by $|\widetilde{\Psi_0''}\rangle=\Pi_1|\Psi_0\rangle$, where $\Pi_1$ is the projector onto the support of $\rho_1$. The corresponding unit vector simply is $|\Psi_0''\rangle=\frac{|\widetilde{\Psi_0''}\rangle}{||\widetilde{\Psi_0''}||}$.\\

Theorem 11 tells us that the optimal failure probability $Q^{opt}$ for State Filtering is given by
\begin{eqnarray}
Q^{opt} = NQ^{opt}(|\Psi_0\rangle,|\Psi_0''\rangle),
\end{eqnarray}
with
\begin{eqnarray}
\rho'_0=\frac{1}{N_0} \Pi_{\cal H'} \rho_0 \Pi_{\cal H'}, & \eta'_0= \frac{N_0 \eta_0}{N}, & N_0 = {\rm Tr}(\rho_0 \Pi_{\cal H'}) \\
\rho'_1=\frac{1}{N_1} \Pi_{\cal H'} \rho_1 \Pi_{\cal H'}, & \eta'_1= \frac{N_1 \eta_1}{N}, & N_1 = {\rm Tr}(\rho_1 \Pi_{\cal H'}) \\
N=N_0 \eta_0+ N_1 \eta_1,\\
{\cal H'}=\{|\Psi_0\rangle,|\Psi_0''\rangle\}.\;
\end{eqnarray}

Furthermore, the optimal failure probability for two pure states $|\Psi_0\rangle$ and $|\Psi_0''\rangle$ with {\it a priori} probabilities $\eta_0'$ and $\eta_1'$ is given by
\begin{eqnarray}
Q^{opt}(|\Psi_0\rangle,|\Psi_0''\rangle) = \eta_1' +\eta_0'|\langle\Psi_0|\Psi_0''\rangle|^2  \,\,\, \mathrm{for} \,\,\,  \sqrt{\frac{\eta_1'}{\eta_0'}} \le |\langle\Psi_0|\Psi_0''\rangle|,\\ \nonumber
\\
Q^{opt}(|\Psi_0\rangle,|\Psi_0''\rangle) = 2\sqrt{\eta_0'\eta_1'}|\langle\Psi_0|\Psi_0''\rangle| \,\,\, \mathrm{if} \,\,\, |\langle\Psi_0|\Psi_0''\rangle| \le \sqrt{\frac{\eta_1'}{\eta_0'}} \le \frac{1}{|\langle\Psi_0|\Psi_0''\rangle|},\\ \nonumber
\\
Q^{opt}(|\Psi_0\rangle,|\Psi_0''\rangle) = \eta_0'  +\eta_1' |\langle\Psi_0|\Psi_0''\rangle|^2 \, \,\,\, \mathrm{if} \,\,\,  \frac{1}{|\langle\Psi_0|\Psi_0''\rangle|} \le \sqrt{\frac{\eta_1'}{\eta_0'}}. \\ \nonumber
\end{eqnarray}

therefore the optimal failure probability $Q^{opt}$ of the non-reduced problem becomes

\begin{eqnarray}
Q^{\mathrm{opt}} = N(\eta_1' +\eta_0'|\langle\Psi_0|\Psi_0''\rangle|^2)  \,\,\, \mathrm{for} \,\,\,  \sqrt{\frac{\eta_1'}{\eta_0'}} \le |\langle\Psi_0|\Psi_0''\rangle|,\\ \nonumber
\\
Q^{\mathrm{opt}} = N(2\sqrt{\eta_0'\eta_1'}|\langle\Psi_0|\Psi_0''\rangle|) \,\,\, \mathrm{if} \,\,\, |\langle\Psi_0|\Psi_0''\rangle| \le \sqrt{\frac{\eta_1'}{\eta_0'}} \le \frac{1}{|\langle\Psi_0|\Psi_0''\rangle|},\\ \nonumber
\\
Q^{\mathrm{opt}} = N(\eta_0'  +\eta_1' |\langle\Psi_0|\Psi_0''\rangle|^2) \, \,\,\, \mathrm{if} \,\,\,  \frac{1}{|\langle\Psi_0|\Psi_0''\rangle|} \le \sqrt{\frac{\eta_1'}{\eta_0'}}. \\ \nonumber
\end{eqnarray}

If we denote $S=\sum_{j=1}^{n-1} p_j |\langle\Psi_0|\Psi_j\rangle|^2$, we find

\begin{eqnarray}
N_0&=&1\\ 
N_1&=&\frac{S}{\eta_1 ||\widetilde{\Psi_0''}||^2}\\ 
\eta_0'&=&\frac{\eta_0 N_0}{N}=\frac{p_0}{N}\\ 
\eta_1'&=&\frac{\eta_1 N_1}{N}=\frac{S}{N ||\widetilde{\Psi_0''}||^2}\\ 
|\langle \Psi_0|\Psi_0''\rangle|&=&||\widetilde{\Psi_0''}||.
\end{eqnarray}

We finally end up with

\begin{eqnarray}
Q^{\mathrm{opt}} &=& p_0 ||\widetilde{\Psi_0''}||^2  + \frac{S}{||\widetilde{\Psi_0''}||^2} \,\,\, \mathrm{if} \,\,\, \frac{S}{||\widetilde{\Psi_0''}||^4} \le p_0,\\
Q^{\mathrm{opt}} &=& 2\sqrt{p_0}\sqrt{S} \,\,\, \mathrm{if} \,\,\, S  \le p_0 \le \frac{S}{||\widetilde{\Psi_0''}||^4},\\
Q^{\mathrm{opt}} &=& p_0 + S \, \,\,\, \mathrm{if} \,\,\, p_0 \le S.
\end{eqnarray}

\subsection{Unambiguous Subspace Discrimination}

To unambiguously discriminate two subspaces, one has to unambiguously discriminate their respective bases. We can therefore consider the two ensembles corresponding to these two bases with a flat distribution because the basis vectors all possess the same probability of appearance. In fact we consider the projectors onto those respective bases as unnormalized mixed states and try to unambiguously discriminate them. In that sense, subspace discrimination is a special case of mixed state discrimination where the two density matrices are proportional to the projectors onto the respective subspaces.\\

There is a infinite amount of basis in which one can write a projector. Therefore the difficulty is to find a suitable basis of the space spanned by the two subspaces to discriminate. Such a suitable basis is given by the so-called {\it canonical bases} which allow us to write the two projectors in a block diagonal form, where each block is two-dimensional. This technique was used by Rudolph {\it et al.} for the derivation of the upper bound on the failure probability $Q$. Thus the unambiguous discrimination of two subspaces can be reduced to some pure state case and, because of that, be solved.\\

First, let us repeat that the first two reduction theorems permit us to focus our attention on the unambiguous discrimination of two subspaces $S_0$ and $S_1$ of rank $r$ in a $2r$-dimensional Hilbert space. Next we choose an orthogonal basis $\{|a_i \rangle\}$ of $S_0$ and an orthogonal basis $\{|b_j \rangle\}$ of $S_1$. The unambiguous discrimination between these two subspaces then corresponds to the unambiguous discrimination of $\rho_0=\frac{1}{r}\sum_i |a_i \rangle\langle a_i|$ and $\rho_1=\frac{1}{r}\sum_j |b_j \rangle \langle b_j|$.

Given two subspaces $S_0$ and $S_1$, it is always possible to find an orthonormal basis $\{|a_i \rangle\}$ of $S_0$ and an orthonormal basis $\{|b_j \rangle\}$ of $S_1$, called {\it canonical} or {\it principal bases} such that $\langle a_i|b_j \rangle = Cos (\theta_i)\delta_{ij}$, $Cos (\theta_i) \ge 0$. In such a basis, the projectors onto $S_0$ and $S_1$ are decomposed into a direct sum of $r$ two-dimensional subspaces. Thanks to theorem 12, the optimal solution to USD of two pure states is the only requirement for an optimal unambiguous discrimination of $S_0$ and $S_1$.

In fact, we can assume without loss of generality that $\langle a_i|b_j \rangle = Cos (\theta_i)\delta_{ij}$, $Cos (\theta_i) \ge 0$. Indeed, we can always construct the so-called {\it canonical bases} $\{|a_i \rangle\}$ and $\{|b_j \rangle\}$ for two subspaces if we follow Rudolph's technique \cite{rudolph03a}. Let $X_k$ be the (2$r$)x$r$ matrix whose columns span $S_k$. We then write a singular value decomposition of $X_0^\dagger X_1$,
\begin{eqnarray}
X_0^\dagger X_1=U_0 S U_1^\dagger,
\end{eqnarray}
where the $U_k$'s are two $r$x$r$ unitaries and $S$ is positive semi-definite and diagonal with $S_{ii}=Cos(\theta_i)$, ($\theta \in [0,2 \pi ]$). Let us define the vectors $|a_i \rangle$ as the $i^{th}$ column of $X_0 U_0$ and the vectors $|b_j \rangle$, the $j^{th}$ column of $X_1 U_1$. The set $\{|a_i \rangle\}$, respectively $\{|b_j \rangle\}$, forms an orthonormal basis of $S_0$, respectively $S_1$, since it is merely a rotation of a former basis. Moreover the vectors $|a_i \rangle$ and  $|b_i\rangle$ satisfy $\langle a_i|b_j \rangle = Cos (\theta_i)\delta_{ij}$. The angles $\theta_i$ are called the {\it canonical angles} and, the  vectors $|a_i \rangle$ and  $|b_i\rangle$, the {\it canonical vectors}. $|a_i \rangle$ and  $|b_i\rangle$ together span the total Hilbert space. The fundamental property $\langle a_i|b_j \rangle = Cos (\theta_i)\delta_{ij}$ allows us to write $\rho_0$ and $\rho_1$ in a block diagonal form, where each block is spanned by $\{|a_i \rangle, |b_i\rangle\}$. Indeed, in the basis $\{|a_1 \rangle\,|b_1 \rangle,|a_2 \rangle,|b_2 \rangle,\dots , |a_r \rangle,|b_r \rangle\}$, the two density matrices $\rho_0$ and $\rho_1$ takes the form

\begin{displaymath}
 \rho_k= \left(
    \begin{array}{ccc}
      \squarebox  &  0           & 0\\
          0        & \squarebox  &0 \\
             0     &     0        & \squarebox
    \end{array}
  \right)
\end{displaymath}
where, each block is a two-dimension subspace spanned by $\{|a_i \rangle\,|b_i \rangle\}$, orthogonal to the $r-1$ other two-dimensional subspaces $\{|a_k \rangle\,|b_k \rangle\}$, $k=1,\dots,i-1,i+1, \dots,n$.\\

Thanks to theorem 12 we can express the failure probability of unambiguously discriminating $S_0$ and $S_1$ as
\begin{eqnarray}
Q^{\mathrm{opt}}=\sum_k N^k Q^{k \,\, \mathrm{opt}},
\end{eqnarray}
where the $Q^{k \,\, \mathrm{opt}}$ are the optimal failure probabilities for unambiguously discriminating $|a_k \rangle$ and $|b_k \rangle$ with their corresponding {\it a priori} probabilities $\eta_0^k$ and $\eta_1^k$.\\

We can easily calculate all those quantities where $\Pi_k$ is the projector onto the two dimensional subspace spanned by $|a_k \rangle$ and $|b_k \rangle$. Thus
\begin{eqnarray}
N_i^k&=&\Tr(\Pi_k \rho_i)=\frac{1}{r}\\
N^k&=&\sum_i \eta_i N_i^k=\sum_i \eta_i \frac{1}{r}=\frac{1}{r}\\
\eta_i^k&=&\frac{\eta_i N_i^k}{N^k}=\eta_i.
\end{eqnarray}

Moreover, for each 2x2 subspace, the optimal failure probability between the two pure states $|a_k \rangle$ and $|b_k \rangle$ with {\it a priori} probabilities $\eta_0$ and $\eta_1$ is given by

\begin{eqnarray}
Q^{k \,\, \mathrm{opt}} &=& \eta_1 +\eta_0|\langle a_k|b_k\rangle|^2  \,\,\, \mathrm{for} \,\,\,  \sqrt{\frac{\eta_1}{\eta_0}} \le |\langle a_k|b_k\rangle|,\\ \nonumber
\\
Q^{k \,\, \mathrm{opt}} &=& 2\sqrt{\eta_0\eta_1}|\langle a_k|b_k\rangle| \,\,\, \mathrm{for} \,\,\, |\langle a_k|b_k\rangle| \le \sqrt{\frac{\eta_1}{\eta_0}} \le \frac{1}{|\langle a_k|b_k\rangle|},\\ \nonumber
\\
Q^{k \,\, \mathrm{opt}} &=& \eta_0  +\eta_1 |\langle a_k|b_k\rangle|^2 \, \,\,\, \mathrm{for} \,\,\,  \frac{1}{|\langle a_k| b_k\rangle|} \le \sqrt{\frac{\eta_1}{\eta_0}}. \\ \nonumber
\end{eqnarray}

In fact, the total failure probability can be expressed in terms of the {\it canonical angles} as
\begin{eqnarray}
Q^{\mathrm{opt}}=\frac{1}{r} \sum_i Q^{k \,\, \mathrm{opt}}
\end{eqnarray}
with for all i $\in [1,\dots,r]$,

\begin{eqnarray}
Q^{k \,\, \mathrm{opt}} = \eta_1 +\eta_0 Cos^2(\theta_k) \, \,\,\, \mathrm{for} \,\,\,  \frac{1}{Cos(\theta_k)} \le \sqrt{\frac{\eta_0}{\eta_1}},\\
Q^{k \,\, \mathrm{opt}} = 2\sqrt{\eta_0\eta_1}Cos(\theta_k) \,\,\, \mathrm{for} \,\,\, Cos(\theta_k) \le \sqrt{\frac{\eta_0}{\eta_1}} \le \frac{1}{Cos(\theta_k)},\\
Q^{k \,\, \mathrm{opt}} = \eta_0  +\eta_1 Cos^2(\theta_k) \,\,\, \mathrm{for} \,\,\,  \sqrt{\frac{\eta_0}{\eta_1}} \le Cos(\theta_k).
\end{eqnarray}

There are in conclusion numerous possible expressions (in principle $3^n$) of the optimal failure probability depending on the values of the canonical angles.


\subsection{Unambiguous State Comparison}
Let us consider a set of $n$ mixed quantum states $\{\sigma_i\}$ which occur with {\it a priori} probabilities $\{p_i\}$. We are given $m$ states out of that set and want to know with certainty whether all the $m$ states are identical or not. We name this task Unambiguous State Comparison '$m$ out of $n$', following the terminology introduced by Kleinmann {\it et al.} in \cite{kleinmann05}.

Such an unambiguous state comparison is a special case of unambiguous state discrimination. Indeed to decide with no errors whether the $m$ states are all identical or not, we have to unambiguously discriminate a first mixture of only identical states from a second mixture of non identical states. More precisely, we have to unambiguously discriminate the two density matrices
\begin{eqnarray}
\rho_0=\frac{1}{\eta_0} \sum_{i=1}^n (p_i \sigma_i)^{\otimes m}
\end{eqnarray}
and
\begin{eqnarray}
\rho_1=\frac{1}{\eta_1} \left( \sum_{i=1}^n p_i \sigma_i \right)^{\otimes m}-\frac{\eta_0}{\eta_1} \rho_0
\end{eqnarray}
where $\eta_0=\sum_{i=1}^n p_i^m$ and $\eta_1=1-\eta_0$ are introduced for normalization purpose.

In the next subsections, we are going to detail the unambiguous comparison of two pure states ('two out of two') and a special case of unambiguous comparison of $n$ pure states ('$n$ out of $n$'). We will see that those cases are reducible to some pure states scenarios and then analytically solvable.

\subsubsection{Unambiguous Comparison of two pure states}

The first case we study is the simplest situation of Unambiguous State Comparison. It involves only two pure states $|\Psi_+\rangle$ and $|\Psi_-\rangle$ with {\it a priori} probabilities $p_+$ and $p_-$. We know it is always possible to write two pure states in some suitable orthonormal basis $\{|0\rangle,\,\, |1\rangle\}$ as $|\Psi_\pm\rangle=\alpha |0\rangle \pm \beta |1\rangle$ where $\alpha$ and $\beta$ are real and such that $\alpha^2+\beta^2=1$. We can therefore denote by $\Theta$ the (real) overlap between $|\Psi_+\rangle$ and $|\Psi_-\rangle$ as $\Theta=\langle \Psi_+| \Psi_- \rangle=2\alpha^2-1$. First of all, we write the two density matrices to unambiguously discriminate. Thanks to Eqn.(3.98) and Eqn.(3.99), we can explicitly express them as
\begin{eqnarray}
\rho_0&=&\frac{1}{\eta_0} (p_+^2 |\Psi_+ \Psi_+ \rangle \langle \Psi_+ \Psi_+| + p_-^2 |\Psi_- \Psi_- \rangle \langle \Psi_- \Psi_-|),\\
\rho_1&=&\frac{1}{2} (|\Psi_+ \Psi_- \rangle \langle \Psi_+ \Psi_-| + |\Psi_- \Psi_+ \rangle \langle \Psi_- \Psi_+|).
\end{eqnarray}
with $\eta_0=p_+^2+p_-^2$ and $\eta_1=2 p_+ p_-$ so that $\eta_0 \ge \eta_1$ since $(p_+-p_-)^2 \ge 0$. Note that $|\Psi \Phi \rangle$ stands for $|\Psi \rangle \otimes |\Phi \rangle$. We will now show that these two mixed states are block diagonal.\\

In chapter 2, we have seen that their is a freedom on the state ensemble of a density matrix. More precisely, a mixed state is left unchanged under a unitary mixing of its state ensemble. Next we remark that the density matrix $\rho_1$ is left unchanged if one swaps $|\Psi_+ \Psi_-\rangle$ and $|\Psi_- \Psi_+\rangle$. Therefore, it seems natural to use a Discrete Fourier Transform to diagonalize $\rho_1$. That is why, we can consider for $\rho_1$ the two unnormalized vectors

\begin{eqnarray} 
\left(\begin{array}{c}
|\widetilde{b_+} \rangle\\
|\widetilde{b_-} \rangle
\end{array}
\right)
=\frac{1}{\sqrt{2}}
\left(\begin{array}{cc}
1&1\\
1&-1
\end{array}
\right)
\left(\begin{array}{c}
\frac{1}{\sqrt{2}} |\Psi_+ \Psi_- \rangle\\
\frac{1}{\sqrt{2}} |\Psi_- \Psi_+ \rangle
\end{array}
\right)
\end{eqnarray}
that is to say
\begin{eqnarray}
|\widetilde{b_\pm} \rangle =\frac{1}{2} \left( |\Psi_+ \Psi_- \rangle \pm |\Psi_- \Psi_+ \rangle \right).
\end{eqnarray}
This yields the new state ensemble $\{\frac{1}{\sqrt{2 (1 \pm \Theta^2)}}, |b_\pm \rangle \}$
where
\begin{eqnarray}
|b_\pm \rangle =\frac{1}{\sqrt{2 (1 \pm \Theta^2)}} \left( |b_+ b_- \rangle \pm |b_- b_+ \rangle \right).
\end{eqnarray}
We finally end up with
\begin{eqnarray}
\rho_1=\frac{1}{2} ( (1 + \Theta^2) |b_+ \rangle \langle b_+| + (1 - \Theta^2) |b_- \rangle \langle b_-| ).
\end{eqnarray}

It is worth noticing that, since $\langle b_+|b_-\rangle=0$, the state vectors $|b_\pm \rangle$ are the eigenvectors of $\rho_1$ with eigenvalues $b_\pm = \frac{1}{2}(1 \pm \Theta^2)$.

In that form, it appears obvious that $\rho_0$ and $\rho_1$ are block-diagonal. To convince ourself, we simply write the different overlaps involved here.
\begin{eqnarray}
\langle \Psi_+ \Psi_+ | b_+ \rangle &=& \frac{2 \Theta}{\sqrt{2(1+\Theta^2)}}\,\,\,,\\
\langle \Psi_- \Psi_- | b_+ \rangle &=& \frac{2 \Theta}{\sqrt{2(1+\Theta^2)}},\\
\langle \Psi_+ \Psi_+ | b_- \rangle &=& 0,\\
\langle \Psi_- \Psi_- | b_- \rangle &=& 0.
\end{eqnarray}

It remains to give the optimal failure probability to unambiguously discriminate $\rho_0$ and $\rho_1$ or equivalently the failure probability to unambiguously compare two pure states $|\Psi_\pm \rangle$.

In fact $|b_-\rangle$ is orthogonal to $\rho_0$ and to $|b_+\rangle$ or in other words $|b_-\rangle \in {\cal S}_{\rho_1} \cap {\cal K}_{\rho_0}$. Thanks to Theorem 11, we know that this direction $| b_- \rangle$ can be perfectly discriminated. This direction does not contribute to the failure probability for unambiguously comparing $|\Psi_+ \rangle$ and $|\Psi_- \rangle$. We are left with the three dimensional subspace spanned by $\rho_0$ and $| b_+ \rangle$. Since $\rho_0$ is two dimensional, Theorem 11 can again be used. It tells us that we can reduce this USD problem further and only consider the problem of two pure states $| b_+ \rangle $ and $| b_+'' \rangle$ with proper {\it a priori} probabilities.

We introduce here the projection $|\widetilde{ b_+''} \rangle $ of $|b_+ \rangle$ onto the support of $\rho_0$. The corresponding unit vector is $| b_+'' \rangle=\frac{|\widetilde{b_+''}\rangle}{||\widetilde{b_+''}||}$ cited above. We proceed as we did for the case of state filtering where here $\Pi_k$ is the projector onto the two dimensional subspace spanned by $| b_+ \rangle $ and $| b_+'' \rangle $.

Theorem 11 tells us that the optimal failure probability $Q^{opt}$ is given by
\begin{eqnarray}
Q^{opt} = NQ^{opt}(|b_+\rangle,|b_+''\rangle),
\end{eqnarray}
with

\begin{eqnarray}
\rho'_ 0=\frac{1}{N_0} \Pi_{\cal H'} \rho_0 \Pi_{\cal H'}, & \eta'_0= \frac{N_0 \eta_0}{N}, & N_0 = {\rm Tr}(\rho_0 \Pi_{\cal H'}) \\
\rho'_1=\frac{1}{N_1} \Pi_{\cal H'} \rho_1 \Pi_{\cal H'}, & \eta'_1= \frac{N_1 \eta_1}{N}, & N_1 = {\rm Tr}(\rho_1 \Pi_{\cal H'}) \\
N=N_0 \eta_0+ N_1 \eta_1,\\
{\cal H'}=\{|b_+\rangle,|b_+''\rangle\}.\;
\end{eqnarray}

Let us calculate the relevant quantities $N_1$, $N_0$ and $\langle b_+''|b_+ \rangle$. Since $|b_+ \rangle$ is an eigenvector of $\rho_1$, $N_1$ simply is its eigenvalue. Thus
\begin{eqnarray}
N_1=\frac{1+\Theta^2}{2}.
\end{eqnarray}

To find $N_0$ and $\langle b_+''|b_+ \rangle$ we first have to calculate $|\widetilde{b_+''} \rangle$ and $|b_+'' \rangle$. We can express $|\widetilde{b_+''} \rangle$ in the non-orthogonal basis $\{|\Psi_+ \Psi_+ \rangle, \,\,\, |\Psi_- \Psi_- \rangle \}$ of ${\cal S}_{\rho_0}$ so that
\begin{eqnarray}
|\widetilde{b_+''} \rangle &=& \langle \Psi_+ \Psi_+|b_+ \rangle |\Psi_+ \Psi_+ \rangle\\ \nonumber
&+& \left( \frac{\langle  \Psi_- \Psi_-|b_+ \rangle - \Theta^2 \langle  \Psi_+ \Psi_+|b_+ \rangle }{1-\Theta^4} \right)  \left( |\Psi_- \Psi_-\rangle - \Theta^2 |\Psi_+ \Psi_+\rangle \right)\\ \nonumber
&=& \frac{2 \Theta}{(1+\Theta^2)\sqrt{2(1+\Theta^2)}}(|\Psi_+ \Psi_+\rangle +  |\Psi_- \Psi_-\rangle).
\end{eqnarray}
The norm of this vector therefor is
\begin{eqnarray}
\sqrt{\langle \widetilde{b_+''}|\widetilde{b_+''} \rangle}=\frac{2 \Theta}{1+\Theta^2}
\end{eqnarray}
which yields
\begin{eqnarray}
|b_+'' \rangle = \frac{1}{\sqrt{2(1+\Theta^2)}}(|\Psi_+ \Psi_+\rangle +  |\Psi_- \Psi_-\rangle).
\end{eqnarray}
Since $N_0=\Tr(\Pi_{\cal H'} \rho_0 \Pi_{\cal H'})=\langle b_+''|\rho_0|b_+'' \rangle$ we simply obtain
\begin{eqnarray}
N_0=\frac{1+\Theta^2}{2}=N_1.
\end{eqnarray}
Finally, the last relevant quantity simply is
\begin{eqnarray}
\langle b_+''|b_+ \rangle=||\widetilde{b_+''}||=\frac{2 \Theta}{1+\Theta^2}.
\end{eqnarray}

Considering the three possible regimes of the optimal failure probability for two pure states, we end up with $Q^{opt}$, the failure probability of unambiguously comparing the two pure states $|\Psi_\pm \rangle$, expressed as

\begin{eqnarray}Q^{\mathrm{opt}} &=& \eta_0 \frac{1+\Theta^2}{2} + \eta_1 \frac{ 2 \Theta^2}{1+\Theta^2} \, \,\,\, \mathrm{for} \,\,\,  \frac{1+\Theta^2}{2 \Theta^2} \le \sqrt{\frac{\eta_1}{\eta_0}}. \\ \nonumber
\\
Q^{\mathrm{opt}} &=& 2 \sqrt{\eta_0 \eta_1} \Theta \,\,\,\,\,\,\, \mathrm{for} \,\,\, \frac{2 \Theta^2}{1+\Theta^2} \le \sqrt{\frac{\eta_1}{\eta_0}} \le \frac{1+\Theta^2}{2 \Theta^2} ,\\ \nonumber
\\
Q^{\mathrm{opt}} &=& \eta_0 \frac{ 2 \Theta^2}{1+\Theta^2} + \eta_1\frac{1+\Theta^2}{2}  \,\,\, \mathrm{for} \,\,\,  \sqrt{\frac{\eta_1}{\eta_0}} \le \frac{2 \Theta^2}{1+\Theta^2},\\ \nonumber
\end{eqnarray}

Let us note here, that we could derive the last expressions of $Q^{\mathrm{opt}}$ using the result derived for State Filtering (Eqn.(3.78) to (3.85)) with the following correspondences
\begin{eqnarray}
p_0 &=& \eta_1\frac{1+\Theta^2}{2}, \\
p_j &=& p_\pm ,\\
S &=& \eta_0 \frac{2\Theta^2}{1+\Theta^2},\\
||\widetilde{\Psi_0''}|| &=& ||\widetilde{b_+''}||=\frac{2 \Theta}{1+\Theta^2}.
\end{eqnarray}

In the next application of our reduction theorems to state comparison, we will use more properties of the Discrete Fourier Transform.

\subsubsection{Unambiguous Comparison of $n$ pure states with a simple symmetry}

We propose to study the problem of comparing $n$ linearly independent pure states $|\Psi_i\rangle$ with equal {\it a priori} probabilities $p_i=\frac{1}{n}$ and equal real overlaps $\Theta=\langle \Psi_i| \Psi_j \rangle$, $\forall i,j=1,\dots,n$.

Related to this comparison task, Eqn.(3.98) and (3.99) tell us that there is a USD problem that involves two density matrices $\rho_0$ and $\rho_1$ and their {\it a priori} probabilities expressed as
\begin{eqnarray}
\rho_0&=&\frac{1}{n} \sum_{i=1}^n |\Psi_i \dots \Psi_i \rangle \langle \Psi_i \dots \Psi_i |\\
\eta_0&=&\frac{1}{n^{n-1}}
\end{eqnarray}
and
\begin{eqnarray}
\rho_1&=&\frac{n^{n-1}}{n^{n-1}-1}\xi^{\otimes n}-\frac{1}{n^{n-1}-1} \rho_0\\
\eta_1&=&\frac{n^{n-1}-1}{n^{n-1}}
\end{eqnarray}
where
\begin{eqnarray}
\xi=\frac{1}{n} \sum_{i=1}^n |\Psi_i\rangle \langle \Psi_i |.
\end{eqnarray}
Note that $\xi$ is not a projector since the vectors $|\Psi_i\rangle$ are in general not orthogonal.
We will now show that these two density matrices are block diagonal and that their unambiguous discrimination can be reduced to $n$ two pure states USD problems only.\\

Actually we can consider the cyclic permutation $C$ that maps $|\Psi_i \rangle$ to $|\Psi_{i+1} \rangle$ for $i=0,n-1$ and $|\Psi_n \rangle$ to $|\Psi_0 \rangle$ and the Discrete Fourier Transform. From now on, all the indexes are given modulo $n$ to simplify the notations. In fact, it is pretty clear that both $\rho_0$ and $\rho_1$ are invariant under the cyclic permutation $C^{\otimes n}$. We can therefore, as we have already done for the comparison of two pure states, use the Discrete Fourier Transform to change the state ensemble of $\rho_0$ and $\rho_1$. If we do so, we will see that both $\rho_0$ and $\rho_1$ are block diagonal where each block is an eigenspace of $C^{\otimes n}$. The main reason for that is that the permutation operator $C$ is diagonalized by the Discrete Fourier Transform. Importantly, the $n$ vectors states of $\rho_0$ are $n$ eigenvectors of $C^{\otimes n}$ with distinct eigenvalues (i.e.\ the $n$ roots of unity). Therefore, the $n$ vectors states of $\rho_0$ are in different eigenspaces of $C^{\otimes n}$. As a matter of fact, $\rho_0$ and $\rho_1$ are block diagonal where only one vector state of $\rho_0$ is in each eigenspace of $C^{\otimes n}$. Thanks to theorem 11 and 12, the USD of $\rho_0$ and $\rho_1$ is reducible to $n$ two pure states cases.\\

Now that the flow of the argumentation is clear, let us first that $\rho_0$ and $\rho_1$ are invariant under $C^{\otimes n}$.

First, we examine the action of $C^{\otimes n}$ on $\rho_0$.
\begin{eqnarray}
C^{\otimes n}\rho_0{C^\dagger}^{\otimes n}&=&C^{\otimes n}\frac{1}{n} \sum_{i=1}^n (|\Psi_i\rangle \langle \Psi_i |)^{\otimes n}{C^\dagger}^{\otimes n}\\
&=&\frac{1}{n} \sum_{i=1}^n C^{\otimes n}(|\Psi_i\rangle \langle \Psi_i |)^{\otimes n}{C^\dagger}^{\otimes n}\\
&=&\frac{1}{n} \sum_{i=1}^n (C|\Psi_i\rangle \langle \Psi_i |C^\dagger)^{\otimes n}\\
&=&\frac{1}{n} \sum_{i=1}^n (|\Psi_{i+1}\rangle \langle \Psi_{i+1}|)^{\otimes n}\\
&=&\frac{1}{n} \sum_{i'=1}^n (|\Psi_{i'}\rangle \langle \Psi_{i'}|)^{\otimes n}\\
&=&\rho_0
\end{eqnarray}
where the index $n+1$ equals $1$ since the indexes are given modulo $n$. We can also investigate the action of $C$ the operator $\xi=\frac{1}{n} \sum_{i=1}^n |\Psi_i\rangle \langle \Psi_i |$.
\begin{eqnarray}
C\xi C^\dagger&=& C \left( \frac{1}{n}  \sum_{i=1}^n |\Psi_i\rangle \langle \Psi_i | \right) C^\dagger\\
&=&\frac{1}{n}  \sum_{i=1}^n C|\Psi_i\rangle \langle \Psi_i |C^\dagger\\
&=&\frac{1}{n} \sum_{i'=1}^n |\Psi_{i'}\rangle \langle \Psi_{i'}|\\
&=&\xi.
\end{eqnarray}
Since $\xi$ is invariant under $C$, $\xi^{\otimes n}$ is invariant under $C^{\otimes n}$. $\rho_1=\frac{n^{n-1}}{n^{n-1}-1}\xi^{\otimes n}-\frac{1}{n^{n-1}-1} \rho_0$ where both $\xi^{\otimes n}$ and $\rho_0$ are invariant under $C^{\otimes n}$, the immediate consequence is that $\rho_1$ is invariant under $C^{\otimes n}$ too.

The Discrete Fourier Transform is the main tool of the next calculations. The matrix elements of $U$ are given by
\begin{eqnarray}
U_{jk}=\frac{1}{\sqrt{n}}e^{2i\pi \frac{(j-1)(k-1)}{n}},\,\,\, k=1,\dots,n.
\end{eqnarray}
The eigenvalues of $C$ simply are the $n$ roots of unity which can be expressed as
\begin{eqnarray}
\lambda_j=e^{-2i\pi \frac{k-1}{n}},\,\,\, k=1,\dots,n.
\end{eqnarray}
Let us briefly derive this result. In a tensor representation, $C_{qk}=\delta_{(q+1)k}$ therefore
\begin{eqnarray}
(UCU^\dagger)_{pj}&=&\sum_{qk} U_{pq} C_{qk} U_{kj}^\dagger\\
&=&\sum_{qk} \frac{1}{\sqrt{n}}e^{2i\pi \frac{(p-1)(q-1)}{n}} \delta_{(q+1)k} \frac{1}{\sqrt{n}}e^{-2i\pi \frac{(k-1)(j-1)}{n}}\\
&=&\frac{1}{n}\sum_{q} e^{2i\pi \frac{(p-1)(q-1)}{n}} e^{-2i\pi \frac{q(j-1)}{n}}\\
&=& \frac{1}{n} \sum_{q} e^{2i\pi q \frac{(p-1)-(j-1)}{n}} e^{-2i\pi \frac{(p-1)}{n}}\\
&=& e^{-2i\pi \frac{(p-1)}{n}} \frac{1}{n} \sum_{q} e^{2i\pi q \frac{(p-j)}{n}} \\
&=& e^{-2i\pi \frac{p-1}{n}} \delta_{pj}
\end{eqnarray}
where we used the relation
\begin{eqnarray}
\frac{1}{n} \sum_{q} e^{2i\pi \frac{q(p-j)}{n}}=\delta_{pj}.
\end{eqnarray}

The unitary freedom in the ensemble of a density matrix allows us to write any density matrix $\rho=\sum_i \mu_i |\mu_i\rangle\langle\mu_i|$ as $\sum_i \nu_i |\nu_i\rangle\langle\nu_i|$ where
\begin{eqnarray}\label{new}
\sqrt{\nu_i}|\nu_i\rangle=\sum_j U_{ij}\sqrt{\mu_j}|\mu_j\rangle.
\end{eqnarray}
We now change the set of state ensemble of both $\rho_0$ and $\rho_1$. In the former case, we use the Discrete Fourier Transform U, a ($n$x$n$) matrix acting on $n$ non normalized vectors $\frac{1}{\sqrt{n}}|\Psi_j \dots \Psi_j \rangle$. In the later case, we use the unitary transformation $U$ on $n$ non normalized vectors $\frac{1}{\sqrt{n}} |\Psi_{j}\rangle$ to change the state ensemble of $\xi$ and therefore to change the state ensemble of $\rho_1$ too.

We begin with the state ensemble of $\rho_0$ and its new {\it a priori} probabilities $\nu_i$ thanks to Eqn.(1.6).
\begin{eqnarray}
\nu_i&=&\frac{1}{n}\sum_{kj} \langle \Psi_k \dots \Psi_k|U_{ik}^*U_{ij}|\Psi_j \dots \Psi_j \rangle\\
&=&\frac{1}{n}\sum_{kj} U_{ik}^*U_{ij}\langle \Psi_k \dots \Psi_k|\Psi_j \dots \Psi_j \rangle\\
&=&\frac{1}{n}  (\sum_k  U_{ik}^* \sum_{j \ne k}U_{ij} \langle \Psi_k \dots \Psi_k|\Psi_j \dots \Psi_j \rangle  + \sum_k  U_{ik}^* U_{ik} \langle \Psi_k \dots \Psi_k|\Psi_j \dots \Psi_j \rangle )\\
&=&\frac{1}{n}  ( \Theta^n \sum_k  U_{ik}^* \sum_{j \ne k}U_{ij} + \sum_k  |U_{ik}|^2 ).
\end{eqnarray}
At that point of the calculation, two cases must be considered. On one hand there is the case where $i=1$ and on the other hand, $i\ne 1$. Two properties of the Discrete Fourier Transform are important here. They can be summarized as
\begin{eqnarray}
\sum_{j=1}^n U_{ij}&=&\left\{
\begin{array}{c}
\sqrt{n} \,\,\, {\rm if}\,\,\, i=1\\
0 \,\,\, {\rm if}\,\,\, i \ne 1
\end{array},
\right.\\
\sum_{j=1}^n |U_{ij}|^2&=&1 \,\,\, \forall i.
\end{eqnarray}
The above calculation of the new {\it a priori} probabilities $\nu_i$ for $i=1$ then leads to
\begin{eqnarray}
\nu_1&=&\frac{1}{n}  ( \Theta^n \sum_k  U_{1k}^* \sum_{j \ne k}U_{1j} + \sum_k  |U_{1k}|^2 )\\
&=&\frac{1}{n}  ( \Theta^n \sum_k  U_{1k}^* (\sqrt{n}-U_{1k}) + 1 )\\
&=&\frac{1}{n}  ( \Theta^n ( \sqrt{n} \sum_k  U_{1k}^* - \sum_k  |U_{1k}|^2) + 1 )\\
&=&\frac{1}{n}  ( \Theta^n ( n - 1) + 1 ).
\end{eqnarray}
A similar calculation for $i \ne 1$ gives
\begin{eqnarray}
\nu_i&=&\frac{1}{n}  ( \Theta^n \sum_k  U_{ik}^* \sum_{j \ne k}U_{ij} + \sum_k  |U_{ik}|^2 )\\
&=&\frac{1}{n}  ( \Theta^n \sum_k  U_{ik}^* (0-U_{ik}) + 1 )\\
&=&\frac{1}{n}  ( -\Theta^n \sum_k  |U_{ik}|^2 + 1 )\\
&=&\frac{1}{n}  ( - \Theta^n + 1 ).
\end{eqnarray}
Finally, $\rho_0$ takes the form
\begin{eqnarray}
\rho_0=\frac{1+(n-1)\Theta^n}{n}|\Phi_1 \rangle \langle \Phi_1| + \frac{1-\Theta^n}{n}\sum_k |\Phi_k \rangle \langle \Phi_k|\\
\end{eqnarray}
with
\begin{eqnarray}
|\Phi_1 \rangle=\frac{1}{\sqrt{1+(n-1)\Theta^n}}\sum_j |\Psi_j\dots \Psi_j\rangle,\\
|\Phi_k \rangle=\frac{1}{\sqrt{1-\Theta^n}}\sum_j e^{2i\pi \frac{(k-1)(j-1)}{n}} |\Psi_j\dots \Psi_j\rangle \,\,\, {\rm for}\,\,\, i \ne 1.
\end{eqnarray}
The fundamental property of those states vector $|\Phi_j\rangle$, $i=1,\dots,n$ is that they are eigenvectors of $C^{\otimes n}$ with $n$ distinct eigenvalues. Note here that $C^{\otimes n}$ has the same eigenvalues than $C$ because this eigenvalues are roots of unity. In other words,
\begin{eqnarray}
C^{\otimes n} |\Phi_j \rangle= \lambda_j |\Phi_j \rangle,
\end{eqnarray}
with $\lambda_j=e^{-2i\pi \frac{k-1}{n}}$, $k=1,\dots,n$.
Indeed the operator $C^{\otimes n}$ acts on the vector $|\Phi_k \rangle$ as
\begin{eqnarray}
C^{\otimes n}|\Phi_k \rangle&=&C^{\otimes n}\frac{1}{\sqrt{1-\Theta^n}}\sum_j e^{2i\pi \frac{(k-1)(j-1)}{n}} |\Psi_j\dots \Psi_j\rangle\\
&=&\frac{1}{\sqrt{1-\Theta^n}}\sum_j e^{2i\pi \frac{(k-1)(j-1)}{n}} C^{\otimes n} |\Psi_j\dots \Psi_j\rangle\\
&=&\frac{1}{\sqrt{1-\Theta^n}}\sum_j e^{2i\pi \frac{(k-1)(j-1)}{n}} C|\Psi_j\rangle \otimes \dots \otimes C |\Psi_j\rangle\\
&=&\frac{1}{\sqrt{1+(n-1)\Theta^n}}\sum_{j} e^{2i\pi \frac{(k-1)(j-1)}{n}} |\Psi_{j+1}\dots \Psi_{j+1}\rangle\\
&=&\frac{1}{\sqrt{1+(n-1)\Theta^n}}\sum_{j} e^{2i\pi \frac{(k-1)(j+1-1)}{n}} e^{-2i\pi \frac{k-1}{n}} |\Psi_{j+1}\dots \Psi_{j+1}\rangle\\
&=& e^{-2i\pi \frac{k-1}{n}} \frac{1}{\sqrt{1+(n-1)\Theta^n}}\sum_{j'} e^{2i\pi \frac{(k-1)(j'-1)}{n}} |\Psi_{j'}\dots \Psi_{j'}\rangle\\
&=& e^{-2i\pi \frac{k-1}{n}}|\Phi_k \rangle \\
&=& \lambda_k |\Phi_k \rangle.
\end{eqnarray}
By definition, $\rho_0$ can be written in a block diagonal form where each block is an eigenspace of $C^{\otimes n}$. 

We follow the same technique to change the state ensemble of $\rho_1$. Since $\rho_1=\frac{n^{n-1}}{n^{n-1}-1}\xi^{\otimes n}-\frac{1}{n^{n-1}-1} \rho_0$, we focus our interest on the matrix $\xi$. We use the Discrete Fourier Transform $U$ acting on the $n$ unnormalized vectors $\frac{1}{\sqrt{n}}|\Psi_{j}\rangle$ to change the state ensemble of $\xi$ and, as a consequence, of $\rho_1$.

We calculate the new {\it a priori} probabilities $\upsilon_i$ of the new state ensemble of $\xi$.
\begin{eqnarray}
\upsilon_i&=&\frac{1}{n}\sum_{kj} \langle \Psi_k k|U_{ik}^*U_{ij}|\Psi_j \rangle\\
&=&\frac{1}{n}\sum_{kj} U_{ik}^*U_{ij}\langle \Psi_k | \Psi_j \rangle\\
&=&\frac{1}{n}  (\sum_k  U_{ik}^* \sum_{j \ne k}U_{ij} \langle \Psi_k | \Psi_j \rangle  + \sum_k  U_{ik}^* U_{ik} \langle \Psi_k \dots \Psi_k|\Psi_j \dots \Psi_j \rangle )\\
&=&\frac{1}{n}  ( \Theta \sum_k  U_{ik}^* \sum_{j \ne k}U_{ij} + \sum_k  |U_{ik}|^2 ).
\end{eqnarray}
This calculation is similar to $\rho_0$'s case. Only the quantity $\Theta^n$ is changed to $\Theta$. Therefore, we end up with
\begin{eqnarray}
\upsilon_i=\left\{
\begin{array}{c}
\frac{1}{n}(1+(n-1)\Theta),\,\,\, i=1\\
\frac{1}{n}(1-\Theta),\,\,\, \forall i \ne 1.
\end{array}\right.
\end{eqnarray}

Finally, $\xi$ takes the form
\begin{eqnarray}
\xi=\frac{1+(n-1)\Theta}{n}|\Upsilon_1 \rangle \langle \Upsilon_1| + \frac{1-\Theta}{n}\sum_k |\Upsilon_k \rangle \langle \Upsilon_k|
\end{eqnarray}
with
\begin{eqnarray}
|\Upsilon_1 \rangle&=&\frac{1}{\sqrt{1+(n-1)\Theta}}\sum_j |\Psi_j\rangle,\\
|\Upsilon_k \rangle&=&\frac{1}{\sqrt{1-\Theta}}\sum_j e^{2i\pi \frac{(k-1)(j-1)}{n}} |\Psi_j\rangle \,\,\, {\rm for}\,\,\, i \ne 1.
\end{eqnarray}
The immediate consequence is that
\begin{eqnarray}
\rho_1&=&\frac{n^{n-1}}{n^{n-1}-1}\left( \frac{1+(n-1)\Theta}{n}|\Upsilon_1 \rangle \langle \Upsilon_1| + \frac{1-\Theta}{n}\sum_k |\Upsilon_k \rangle \langle \Upsilon_k|\right)^{\otimes n}\\
&-&\frac{1}{n^{n-1}-1} \frac{1+(n-1)\Theta^n}{n}|\Phi_1 \rangle \langle \Phi_1| + \frac{1-\Theta^n}{n}\sum_k |\Phi_k \rangle \langle \Phi_k|
\end{eqnarray}
Moreover, the state vectors $|\Upsilon_{j}\rangle$ of $\xi$ are eigenvectors of $C$ therefore the state vectors $|\Upsilon_{i1} \dots \Upsilon_{in}\rangle$ of $\xi^{\otimes n}$ are eigenvectors of $C^{\otimes n}$. A short calculation can verify this claim.
\begin{eqnarray}
C |\Phi_{j}\rangle &=& C \sum_j U_{jk} |\Psi_{k}\rangle\\
&=& \sum_k U_{jk} C |\Psi_{k}\rangle\\
&=& \sum_k U_{jk} |\Psi_{k+1}\rangle\\
&=& \sum_{k} e^{2i\pi \frac{(j-1)(k-1)}{n}} |\Psi_{k+1}\rangle\\
&=& \sum_{k} e^{2i\pi \frac{(j+1-1)(k-1)}{n}} e^{-2i\pi \frac{(k-1)}{n}} |\Psi_{k+1}\rangle\\
&=& e^{-2i\pi \frac{(j-1)}{n}} \sum_{k} e^{2i\pi \frac{(j-1)(k+1-1)}{n}} |\Psi_{k+1}\rangle\\
&=& e^{-2i\pi \frac{(j-1)}{n}} \sum_{k'} e^{2i\pi \frac{(j+1-1)(k'-1)}{n}} |\Psi_{k'}\rangle\\
&=& e^{-2i\pi \frac{(j-1)}{n}} |\Phi_{j}\rangle\\
&=& \lambda_j |\Phi_{j}\rangle.
\end{eqnarray}
This implies that
\begin{eqnarray}
C \otimes \dots \otimes C |\Phi_{i1} \dots \Phi_{i{n}} \rangle &=& C|\Phi_{i1} \rangle \otimes \dots \otimes C | \Phi_{i{n}} \rangle\\
&=&\lambda_{i1}|\Phi_{i1} \rangle \otimes \dots \otimes \lambda_{in}| \Phi_{i{n}} \rangle\\
&=&\lambda_{i1}\dots\lambda_{in}|\Phi_{i1} \dots \Phi_{i{n}} \rangle
\end{eqnarray}

Since the state vectors of $\xi^{\otimes n}$ are eigenvectors of $C^{\otimes n}$, $\xi^{\otimes n}$, like $\rho_0$, is block diagonal, where each block in an eigenspace of $C^{\otimes n}$. The immediate consequence is that $\rho_1$, linear combination of $\xi^{\otimes n}$ and $\rho_0$ is block diagonal, too.\\

Let us denote $S_k$, the eigenspace associated with the eigenvalues $\lambda_k$ of $C^{\otimes n}$ and $\Pi_k$ the orthogonal projector onto $S_k$. We have
\begin{eqnarray}
\rho_0&=&\sum_k \Pi_k \rho_0 \Pi_k,\\
\rho_1&=&\sum_k \Pi_k \rho_1 \Pi_k.\\
\end{eqnarray}
Therefore, Theorem 12 tells us to focus our attention onto the $n$ reduced problem defined by the two density matrices $\rho_0^k=\frac{\Pi_k \rho_0 \Pi_k}{\Tr(\Pi_k \rho_0)}$ and $\rho_1^k=\frac{\Pi_k \rho_0 \Pi_k}{\Tr(\Pi_k \rho_1)}$. Moreover the reduced density matrix $\rho_0^k$ simply is a pure state
\begin{eqnarray}
\Pi_k \rho_0 \Pi_k=|\phi_k \rangle \langle \phi_k|
\end{eqnarray}
since the $n$ state vectors of $\rho_0$ are eigenvectors of $C^{\otimes n}$ with distinct eigenvalues. By means of Theorem 11, we can reduce the USD problem of unambiguously discriminating $\rho_0^k$ and $\rho_1^k$ to the one of two pure states only.\\

Finally the unambiguous discrimination of $\rho_0$ and $\rho_1$ or, equivalently, the unambiguous comparison of $n$ linearly independent pure states $|\Psi_i\rangle$ with equal {\it a priori} probabilities $p_i=\frac{1}{n}$ and equal real overlaps $\Theta=\langle \Psi_i| \Psi_j \rangle$, $\forall i,j=1,\dots,n$ is reducible to $n$ two pure states cases.

The goal of this section was to show that the unambiguous comparison of $n$ pure states with equal {\it a priori} probabilities and equal and real overlaps is reducible to some pure state case. As we have already indicated in the introduction, the question to know whether any unambiguous comparison of pure states is always reducible to some pure state cases remains opened. However, as expected, the unambiguous comparison of mixed states is generally not reducible to some pure states case \cite{kleinmann05}. 

This concludes this chapter. In the next chapter, we will derive the first class of exact solutions for a generic USD problem.


%% file: chapter5.tex
\chapter{First class of exact solutions} \label{first_class}

The structure of this chapter is the following. In the section \ref{bounds}, we derive three lower bounds on the failure probability  to unambiguously discriminate two density matrices in three regimes of the ratio between the two {\it a priori} probabilities. Our derivation uses the Cauchy-Schwarz inequality and  allows us to look for necessary and sufficient conditions to reach the lower bound in each regime of the {\em a priori} probabilities. In section \ref{sec:parallel}, we report the notion of {\em parallel addition} that leads to some useful relations for USD in connection with our first reduction theorem. In section 4.3, we finally derive the main result of this chapter as a theorem: a necessary and sufficient set of two conditions for the failure probability to reach the bounds are given. We also give the corresponding optimal POVM.

With that result, we give the optimal USD POVM of a wide class of pairs of mixed states. This class corresponds to pairs of mixed states for which the lower bounds (one for each of the three regimes depending on the ratio between the {\it a priori} probabilities) on the failure probability $Q$ are saturated. This class in nonempty since it contains some pairs of generic mixed states as well as any pair of pure states. For those pairs, we provide the first analytical solutions for unambiguous discrimination of generic mixed states. This goes beyond known results which are all reducible to some pure state case as we have seen in chapter 2 and 3. 

\section{Lower bounds on the failure probability}
\label{bounds}
	 
The failure probability $Q$ of a USD strategy is given by $Q=\sum_i Q_i$, where $Q_i=\eta_i \Tr (E_? \rho_i)$. From this definition we immediately see that $Q_i \le \eta_i$. In this chapter, we consider the USD of two signal states $\rho_0$ and $\rho_1$ that are mixed states with {\it a priori} probabilities $\eta_0$ and $\eta_1$. Accordingly, our POVM contains three elements $\{E_0,E_1,E_?\}$ which correspond respectively to the conclusive detection of $\rho_0$, to the conclusive detection of $\rho_1$ and to an inconclusive result. The failure probability then equals $Q=Q_0+Q_1$.

Our interest is first focused on the product $Q_0Q_1$. We can give a lower bound expressed in terms of the fidelity $F = \Tr(\sqrt{\sqrt{\rho_0}\rho_1\sqrt{\rho_0}})$ of the two states $\rho_0$ and $\rho_1$. The bounds, formulated in the following theorem, are tighter than those given in chapter 2. Moreover, we pay additional attention to the condition under which the bounds can be reached.

\begin{theorem} Lower bound on the product $Q_0Q_1$\\
\graybox{Let $\rho_0$ and $\rho_1$ be two density matrices with {\it a priori} probabilities $\eta_0$ and $\eta_1$. We define the fidelity of the two states $\rho_0$ and $\rho_1$ as $F= \Tr(\sqrt{\sqrt{\rho_0}\rho_1\sqrt{\rho_0}})$. Then, for any USD measurement with inconclusive outcome $E_?$, the product of the two probabilities $Q_0$ and $Q_1$ to fail to identify respectively the state $\rho_0$ and $\rho_1$ is such that
\begin{eqnarray}
Q_0 Q_1\ge \eta_0 \eta_1F^2.
\end{eqnarray}
The equality holds if and only if the unitary operator $V$ arising from a polar decomposition
\begin{eqnarray}
\sqrt{\rho_0}\sqrt{\rho_1}=\sqrt{\sqrt{\rho_0}\rho_1\sqrt{\rho_0}}\,\, V
\end{eqnarray}
satisfies
\begin{equation} \label{condition}
V^\dagger \sqrt{\rho_0}\sqrt{E_?}=\alpha \sqrt{\rho_1} \sqrt{E_?}\\
\end{equation}
for some $\alpha \in \mathbb{R}^+$.}
\end{theorem}

Before we turn to the proof of this theorem, note that relation (\ref{condition}) implies a condition on the optimality of a USD POVM \cite{sun02a,bergou03a, raynal03a}. It is clear that optimality of a specific USD measurement implies that the conditional states after the inconclusive results do not allow further USD measurements as we already discussed it in chapter 3. This condition is satisfied, for example, when the supports of the conditional states coincide. We find a stronger property whenever equality holds in  Theorem 13. Indeed, if  we have $V^\dagger \sqrt{\rho_0}\sqrt{E_?}= \alpha \sqrt{\rho_1}\sqrt{E_?}$  with $\alpha \in \mathbb{R}^+$, then it follows immediately that  $\sqrt{E_?} \rho_0 \sqrt{E_?} = \alpha^2 \sqrt{E_?} \rho_1 \sqrt{E_?}$. This means that the conditional states corresponding to inconclusive results must be identical up to normalization. Therefore no information whatsoever about the signal state can be extracted from these conditional states.

\paragraph*{\bf Proof of Theorem 13}
This theorem was stimulated by the proof of the {\it nonbroadcasting} theorem \cite{barnum96a}. The basic ingredient for the derivation of the bound is the Cauchy-Schwarz inequality:

\begin{theorem} \cite{lancaster85}
Cauchy-Schwarz inequality\\
If x and y are members of a unitary space then
$\|x\| \|y\|\ge |(x,y)|$.\\
The equality holds if and only if $x=\alpha \, y$ for some $\alpha$ in $\mathbb C$.
\end{theorem}
A unitary space is  a complex linear space $\mathcal{S}$ together with an inner product from $\mathcal{S} \times \mathcal{S}$ to $\mathbb C$. Therefore the complex space of bounded operators acting on a Hilbert space is a complete unitary space (i.e.\ every Cauchy sequence converge) if we consider for two elements $A$ and $B$ the inner product $\Tr(A B^{\dagger})$. The Cauchy-Schwarz inequality then takes the form $\sqrt{\Tr(AA^\dagger)}\sqrt{\Tr(BB^\dagger)}\ge|\Tr(AB^\dagger)|$ where equality holds for $A=\alpha B$, $\alpha$ in $\mathbb C$.

Let us now consider a POVM element $E_k$ and two density matrices $\rho_0$ and $\rho_1$. We can decompose these three operators as $\rho_1=\sqrt{\rho_1}\sqrt{\rho_1}$ and $E_k=\sqrt{E_k}\sqrt{E_k}$ and $\rho_0=\sqrt{\rho_0}U U^\dagger \sqrt{\rho_0}$ where $U$ is an arbitrary unitary transformation coming from the freedom in the decomposition of a positive semi-definite operator.
Hence we obtain from the Cauchy-Schwarz inequality with $A=U^\dagger\sqrt{\rho_0}\sqrt{E_k}$ and $B=\sqrt{\rho_1}\sqrt{E_k}$
\begin{eqnarray}
\sqrt{\Tr(E_k \rho_0)} \sqrt{ \Tr(E_k \rho_1)}\ge |{\Tr(U^\dagger \sqrt{\rho_0} \sqrt{E_k} \sqrt{E_k} \sqrt{\rho_1})|}=|{\Tr(U^\dagger \sqrt{\rho_0} E_k \sqrt{\rho_1})|}.
\end{eqnarray}
By Theorem 14, the equality holds if and only if $U^\dagger \sqrt{\rho_0}\sqrt{E_k}=\alpha \sqrt{\rho_1} \sqrt{E_k}$, for some $\alpha \in \mathbb{C}$.
\\
We now consider a USD POVM $\{E_k\}_{k=0,1,?}$. Using the fact that $\Tr(E_0 \rho_1)=\Tr(E_1 \rho_0)=0$, we find for $E_0$ and $E_1$

\begin{eqnarray}
0=\sqrt{\Tr(E_0 \rho_0)} \sqrt{\Tr(E_0 \rho_1)} \ge |\Tr(U^\dagger \sqrt{\rho_0} E_0 \sqrt{\rho_1})|,\\
0=\sqrt{\Tr(E_1 \rho_0)} \sqrt{\Tr(E_1 \rho_1)} \ge |\Tr(U^\dagger \sqrt{\rho_0} E_1 \sqrt{\rho_1})|.
\end{eqnarray}
This simply means that $\Tr(U^\dagger \sqrt{\rho_0} E_0 \sqrt{\rho_1})=\Tr(U^\dagger \sqrt{\rho_0} E_1 \sqrt{\rho_1})=0$.
For $E_?$, we obtain
\begin{eqnarray}
\sqrt{\Tr(E_? \rho_0)} \sqrt{\Tr(E_? \rho_1)}\ge |\Tr(U^\dagger \sqrt{\rho_0} E_? \sqrt{\rho_1})|.
\end{eqnarray}
From this it follows that  we can write
\begin{eqnarray} \label{CS}
\sqrt{\Tr(E_? \rho_0)} \sqrt{\Tr(E_? \rho_1)}\ge |\Tr(U^\dagger \sqrt{\rho_0} E_? \sqrt{\rho_1})+0+0|=|{\Tr(U^\dagger \sqrt{\rho_0}\sqrt{\rho_1})|}\;,
\end{eqnarray}
where we used the relation $\sum_k E_k = {\mathbb 1}$. Furthermore, the inequality (\ref{CS}) must hold for any unitary matrix $U$ so that we find
\begin{equation} \label{maxx}
\sqrt{\Tr(E_? \rho_0)} \sqrt{\Tr(E_? \rho_1)}\ge \max_U |\Tr(U^\dagger \sqrt{\rho_0} \sqrt{\rho_1})|.
\end{equation} Here, again, the equality holds if and only if a unitary operator $U_{\mathrm{max}}$ which maximizes the right hand side satisfies
\begin{eqnarray}\label{Umax}
U^\dagger_{\mathrm{max}} \sqrt{\rho_0}\sqrt{E_?}=\alpha \sqrt{\rho_1} \sqrt{E_?}
\end{eqnarray}
for some $\alpha \in \mathbb{C}$.
To find the unitary matrices $U_{\mathrm{max}}$ that maximize  $|\Tr(U^\dagger \sqrt{\rho_0} \sqrt{\rho_1})|$  we use the following lemma:

\begin{lemma}
\label{maxUlemma}
For any operator $A$ in the space $M_n$ of $n\times n$ matrices we find
\begin{eqnarray}
\max_W |\Tr(AW)|=Tr(|A|)
\end{eqnarray}
where the maximum is taken over all unitary matrices.  The maximum is reached for any unitary operator $W$ that can be written as $W =V^{\dagger}e^{\imath \phi}$. Here $e^{\imath \phi}$ is an arbitrary phase while the unitary operator $V$ is defined via a polar decomposition
\begin{eqnarray}
A=|A| \, V
\end{eqnarray}
with $|A|=\sqrt{A A^\dagger}=V \, \sqrt{A^\dagger A}\, V^{\dagger}$. (See proof in Appendix B.)
\end{lemma}
Let us introduce the operators  $F_0:=|\sqrt{\rho_0}\sqrt{\rho_1}|=\sqrt{\sqrt{\rho_0}\rho_1\sqrt{\rho_0}}$ and $F_1=V^\dagger F_0 V=\sqrt{\sqrt{\rho_1}\rho_0\sqrt{\rho_1}}$, which are motivated by the polar decomposition
\begin{eqnarray}\label{PD}
\sqrt{\rho_0} \sqrt{\rho_1}= F_0 V=V F_1.
\end{eqnarray}
These operators are related to the fidelity of the two density matrices through the relation $F=\Tr(\sqrt{\sqrt{\rho_1}\rho_0\sqrt{\rho_1}})=\Tr(F_0)=\Tr(F_1)$ \cite{jozsa94a}. Thanks to lemma \ref{maxUlemma}, Eqn. (\ref{maxx}) implies
\begin{eqnarray}
\sqrt{\Tr(E_? \rho_0)} \sqrt{\Tr(E_? \rho_1)}\ge |{\Tr(|\sqrt{\rho_0}\sqrt{\rho_1}|)|}= \Tr(|\sqrt{\rho_0}\sqrt{\rho_1}|)\; 
\end{eqnarray}
where equality now holds if and only if $U_{max}$ in (\ref{Umax}) arises from a polar decomposition of $\sqrt{\rho_0}\sqrt{\rho_1}$. In other words, we have
\begin{eqnarray} \label{condmax}
V^{\dagger}e^{\imath \phi} \sqrt{\rho_0}\sqrt{E_?}=\alpha \sqrt{\rho_1} \sqrt{E_?}
\end{eqnarray}
for some $\alpha \in \mathbb{C}$.  

Next we use the definitions of the partial failure probabilities  $Q_i=\eta_i \Tr(E_? \rho_i)$ and  choose the phase $e^{\imath \phi}$ to be the same as the phase of $\alpha$ in (\ref{condmax})  to obtain the desired inequality $Q_0 Q_1\ge \eta_0 \eta_1 F^2$. Equality in the previous equation then holds if and only if $V^\dagger \sqrt{\rho_0}\sqrt{E_?}=\alpha \sqrt{\rho_1} \sqrt{E_?}$, for some $\alpha \in \mathbb{R}^+$. This completes the proof. \hfill $\blacksquare$ \\

We can now derive the bounds in the different regimes of the ratio $\frac{\eta_1}{\eta_0}$ between the two {\it a priori} probabilities. Actually, the procedure is to find the minimum of the failure probability $Q=Q_0+Q_1$ under the constraints of the previous derived inequality $Q_0 Q_1\ge \eta_0 \eta_1 F^2$. According to Theorem 13, we can provide the necessary and sufficient condition for equality. 

\begin{theorem}\label{3_bounds_theorem}Lower bounds on the failure probability\\
\graybox{Let $\rho_0$ and $\rho_1$ be two density matrices with {\it a priori} probabilities $\eta_0$ and $\eta_1$. We define the fidelity $F$ of the two states $\rho_0$ and $\rho_1$ as $\Tr(\sqrt{\sqrt{\rho_0}\rho_1\sqrt{\rho_0}})$. We denote by $P_0$ and $P_1$, the projectors onto the support of $\rho_0$ and $\rho_1$. Then, for any USD measurement with inconclusive outcome $E_?$, the failure probability $Q$ obeys
\begin{eqnarray} \label{3_regimes}
Q \ge \eta_1 \frac{F^2}{\Tr(P_1 \rho_0)}+\eta_0 \Tr(P_1 \rho_0) \,\,\, &\textrm{for}& \,\,\, \sqrt{\frac{\eta_1}{\eta_0}} \le \frac{\Tr(P_1 \rho_0)}{F}\\
Q \ge 2\sqrt{\eta_0\eta_1}F \,\,\, &\textrm{for}& \,\,\, \frac{\Tr(P_1 \rho_0)}{F}\le \sqrt{\frac{\eta_1}{\eta_0}} \le \frac{F}{\Tr(P_0 \rho_1)}\\
Q \ge \eta_0 \frac{F^2}{\Tr(P_0 \rho_1)}+\eta_1 \Tr(P_0 \rho_1) \,\,\, &\textrm{for}& \,\,\, \frac{F}{\Tr(P_0 \rho_1)} \le \sqrt{\frac{\eta_1}{\eta_0}}.
\end{eqnarray}
Equality holds if and only if the unitary operator $V$ arising from a polar decomposition $\sqrt{\rho_0}\sqrt{\rho_1}=\sqrt{\sqrt{\rho_0}\rho_1\sqrt{\rho_0}}\,\, V$ satisfies  $V^\dagger \sqrt{\rho_0}\sqrt{E_?}= \alpha \sqrt{\rho_1} \sqrt{E_?}$, with $\alpha=\frac{\Tr(P_1 \rho_0)}{F}$, $\alpha =\sqrt{\frac{\eta_1}{\eta_0}}$  and $\alpha = \frac{F}{\Tr(P_0 \rho_1)}$ in the the first, second and third regime, respectively.}
\end{theorem}

\paragraph*{\bf Proof}
First of all, according to Theorem 13, we know that for any USD measurement the inequality $Q_0Q_1 \ge \eta_0 \eta_1F^2$ i.e.\ $Q_1 \ge \frac{\eta_0 \eta_1F^2}{Q_0}$ holds. It follows that the failure probability is lower bounded as

\begin{equation}
\label{optfunction}
Q \ge Q_0+\frac{\eta_0 \eta_1F^2}{Q_0}\; .
\end{equation}

Since we are interested in a lower bound on $Q$, let us consider the case where equality holds in Eqn.~(\ref{optfunction}). In this case, we have
\begin{eqnarray}\label{Q1}
Q_0 Q_1 = \eta_0 \eta_1F^2 \\
Q = Q_0+\frac{\eta_0 \eta_1F^2}{Q_0}\;
\end{eqnarray}
From Theorem 13 we know that Eqn.(\ref{Q1}) holds if and only if $V^\dagger \sqrt{\rho_0}\sqrt{E_?}= \alpha \sqrt{\rho_1} \sqrt{E_?}$, for some $\alpha \in {\mathbb R}^+$. We will now connect $\alpha$ to the other quantities. The previous relation implies, via the respective definitions, that 
\begin{equation}\label{Q2}
Q_0=\alpha^2 \frac{\eta_0}{\eta_1} Q_1 \; .
\end{equation}
The former relationship corresponds to the proportionality between two vectors of the vector space of bounded operators while the latter relationship corresponds to the proportionality between their norms. We can combine the two equations (\ref{Q1}) and (\ref{Q2}) to
\begin{eqnarray}\label{alphadef}
Q_0&=&\alpha \eta_0 F\\
Q_1&=&\frac {1}{\alpha} \eta_1 F \; .
\end{eqnarray}
 So the final statement is that $Q=Q_0+\frac{\eta_0 \eta_1F^2}{Q_0}$ if and only if $V^\dagger \sqrt{\rho_0}\sqrt{E_?}= \alpha \sqrt{\rho_1} \sqrt{E_?}$, where $\alpha$ now is explicitly related to the other parameters as  $Q_0 = \alpha \eta_0 F$ and $Q_1 = \frac {1}{\alpha} \eta_1 F$.

Second, we have to derive a range constraint on $Q_0$ and $Q_1$.  We know already that $Q_i \le \eta_i$. Moreover, from work by Herzog and Bergou in \cite{herzog05a}, we learn that $\eta_0 \Tr(P_1 \rho_0) \le Q_0$ and $\eta_1 \Tr(P_0 \rho_1) \le Q_1$. Indeed, from the structure of the USD POVM elements, we have $E_0+E_1+E_?={\mathbb 1}$ with $\mathcal{S}_{E_0} \subset \mathcal{K}_{\rho_1}$ and $\mathcal{S}_{E_1} \subset \mathcal{K}_{\rho_0}$. We consider only the non-trivial case where the supports of $\rho_0$ and $\rho_1$ are not identical. Then the structure must be such that $E_1+E_?=P_1 + R$ where $P_1$ is the projection onto the support of $\rho_1$ and $R$ is a positive semi-definite operator with support $\mathcal{S}_{R} \subset \mathcal{K}_{\rho_1}$ which satisfies $E_0+R = P_1^\perp$ otherwise $\Tr(E_0\rho_1)\ne 0$. Then it follows that the partial success probability $P^{s}_0$ is $P^{s}_0=\eta_0 \Tr(E_0 \rho_0)=\eta_0 \Tr(P^\perp_1 \rho_0) - \eta_0 \Tr(R \rho_0)$. In our non-trivial case we will have $\Tr(R \rho_0)>0$ as soon as $R \neq 0$. This yields $P^{s}_0 \le \eta_0 \Tr(P^\perp_1 \rho_0)$ or equivalently $Q_0 \ge \eta_0 \Tr(P_1 \rho_0)$. In the same way, one can find $Q_1 \ge \eta_1 \Tr(P_0 \rho_1)$. We then have
\begin{eqnarray}
\label{constraintQ0}
\eta_0 \Tr(P_1 \rho_0) \le Q_0 \le \eta_0, \\
\label{constraintQ1}
\eta_1 \Tr(P_0 \rho_1) \le Q_1 \le \eta_1. 
\end{eqnarray}
These two constraints can be combined in
\begin{eqnarray}\label{range}
\eta_0 \Tr(P_1 \rho_0) \le Q_0 \le \eta_0 \frac{F^2}{\Tr(P_0 \rho_1)}.
\end{eqnarray}
This can be seen as follows. Since $Q_1=\frac{\eta_0 \eta_1F^2}{Q_0}$, the constraint (\ref{constraintQ1}) on $Q_1$ takes the form
\begin{eqnarray}
\label{constraintQ0bis}
\eta_0 F^2\le Q_0 \le \eta_0 \frac{F^2}{\Tr(P_0 \rho_1)}.
\end{eqnarray}
We now have two lower bounds and two upper bounds on $Q_0$ ((\ref{constraintQ0}) and (\ref{constraintQ0bis})) and we want to find the tighter ones. To do that, let us consider the USD POVM given by $\{E_?=P_1,E_0=P_1^\perp,E_1=0\}$. Thank to Theorem 13, we find $\eta_0 \eta_1 F^2 \le \eta_0 \eta_1 \Tr(P_1 \rho_0) \Tr(P_1 \rho_1)$ or in other words $\eta_0 F^2 \le \eta_0 \Tr(P_1 \rho_0)$. We can also consider the USD POVM given by $\{E_?=P_0,E_0=0,E_1=P_0^\perp\}$ and with Theorem 13, we have $\eta_0\frac{F^2}{\Tr(P_0 \rho_1)} \le \eta_0$.  Finally, we obtain $\eta_0 \Tr(P_1 \rho_0) \le Q_0 \le \eta_0 \frac{F^2}{\Tr(P_0 \rho_1)}$.

Next, we define the function $q(Q_0)=Q_0+\frac{\eta_0 \eta_1F^2}{Q_0}$  and minimize it under the constraint $\eta_0 \Tr(P_1 \rho_0) \le Q_0 \le \eta_0 \frac{F^2}{\Tr(P_0 \rho_1)}$. The resulting minimum will constitute a lower bound for $Q$. The function $q(Q_0)$ is convex ($\frac{d^2 q}{dQ_0^2}(Q_0) \ge 0$) and, therefore, it takes its minimum at the point $Q_0^{\text{min}}$ where the derivative vanishes ($\frac{d q}{dQ_0}(Q_0) = 0$ yielding $Q_0^{\text{min}}=\sqrt{\eta_0 \eta_1} F$) or at the limits of the constraint interval ($Q_0^{\text{min}}=\eta_0 \Tr(P_1 \rho_0)$ and $Q_0^{\text{min}}=\eta_0 \frac{F^2}{\Tr(P_0 \rho_1)}$). That gives us the minimum of the function $q(Q_0)$ in three different regimes. In the first regime we have  $q_{\text{min}}(Q_0)=\eta_0 \Tr(P_1 \rho_0) + \eta_1 \frac{F^2}{\Tr(P_1 \rho_0)}$ and $Q_0^{\text{min}}=\eta_0 \Tr(P_1 \rho_0)$ if $\sqrt{\eta_0 \eta_1} F \le \eta_0 \Tr(P_1 \rho_0)$ that is to say if $\sqrt{\frac{\eta_1}{\eta_0}} \le \frac{\Tr(P_1 \rho_0)}{F}$. In the second regime we have  $q_{\text{min}}(Q_0)=2\sqrt{\eta_0\eta_1}F$ and $Q_0^{\text{min}}=\sqrt{\eta_0\eta_1}F$ if $\frac{\Tr(P_1 \rho_0)}{F} \le \sqrt{\frac{\eta_1}{\eta_0}} \le \frac{F}{\Tr(P_0 \rho_1)}$. The third regime gives  $q_{\text{min}}(Q_0)=\eta_0 \frac{F^2}{\Tr(P_0 \rho_1)}+\eta_1 \Tr(P_0 \rho_1)$ and $Q_0^{\text{min}}=\eta_0 \frac{F^2}{\Tr(P_0 \rho_1)}$ if $\frac{F}{\Tr(P_0 \rho_1)} \le \sqrt{\frac{\eta_1}{\eta_0}}$.

As a result we obtain lower bounds for the failure probability $Q$ in three regimes as given in Eqn.~(\ref{3_regimes}). For each regime, the value of $Q_0$ which minimized $q(Q_0)$ is given and via Eqn.~(\ref{alphadef}) we find the corresponding value that $\alpha$ has to take. We read off the values as  $\alpha=\frac{\Tr(P_1 \rho_0)}{F}$, $\alpha=\sqrt{\frac{\eta_1}{\eta_0}}$ and $\alpha=\frac{F}{\Tr(P_0 \rho_1)}$ for the first, second and third regime, respectively. This completes the proof. \hfill $\blacksquare$ \\

Let us note that, by construction, those bounds are tighter than the ones in chapter 2 \cite{rudolph03a}. Indeed, one could recover the three bounds in \cite{rudolph03a} by looking for the minimum of the function $q(Q_0)$ under the weaker constraints $\eta_0 F^2 \le Q_0 \le \eta_0$ as we will show in the last section of this chapter.

\section{Parallel addition $\rho_0 \Sigma^{-1}\rho_1$}
\label{sec:parallel}

Before deriving the central theorem of this chapter and then provide the first class of exact solution for USD of two generic mixed states, we will first recall some useful results of linear algebra. We denote by $M^{-1}$ the pseudo-inverse of a matrix $M$, which has not necessarily full rank. The pseudo-inverse can be defined via the singular-value decomposition of $M=UDV$ as $M^{-1}=UD^{-1}V$, where $U$ and $V$ are unitaries and $D$ is a positive semi-definite and diagonal matrix. Whenever $M$ is of full rank, the pseudo-inverse coincides with the inverse.  In general, it is not known how to express the pseudo inverse of a sum $(A+B)^{-1}$  in terms of the pseudo inverses $A^{-1}$ and $B^{-1}$ \cite{anderson69a,fill98a}. However, a related operation $A(A+B)^{-1}B$, called {\it parallel addition} and denoted by $A:B$ has been defined and studied in 1969 by Anderson and Duffin and will turn out useful in our context. 
\begin{lemma}\label{parallel}
{\rm\cite{anderson69a}} Let $A$ and $B$ be two positive semi-definite matrices in $M_n$, then the support ${\mathcal S}_{A:B}$ of $A:B$ is given in terms of the supports of $A$ and $B$ as
\begin{eqnarray}
{\mathcal S}_{A:B}={\mathcal S}_A \cap {\mathcal S}_B.
\end{eqnarray}
(See proof in Appendix C.)
\end{lemma}

Next let us recall the first reduction theorem for USD of mixed states (Theorem 9).  We consider the problem of discriminating unambiguously two density matrices $\rho_0$ and $\rho_1$ with {\it a priori} probabilities $\eta_0$ and $\eta_1$. We denote by $r_0$ the rank of $\rho_0$ and by $r_1$ the rank of $\rho_1$. A general USD problem can satisfy $r_0+r_1 \ge d$, where $d$ is the dimension of the Hilbert space ${\mathcal H}$ spanned by the two states. This means in particular that the two supports can overlap.

In the first reduction theorem it has been shown that any such USD problem can always be reduced to the one of discriminating $\rho'_0$ and $\rho'_1$, two density matrices of rank $r'_0$ and $r'_1$ with {\it a priori} probabilities $\eta'_0$ and $\eta'_1$, spanning the same Hilbert space ${\mathcal H}$ of dimension $d=r'_0+r'_1$. Indeed we can split off any common subspace of the supports ${\mathcal S}_{\rho_0} \cap {\mathcal S}_{\rho_1}$ to end up with ${\mathcal S}_{\rho'_0} \cap {\mathcal S}_{\rho'_1}=\{0\}$. As we have already seen, two supports do not overlap if and only if $rank(\rho'_0)+rank(\rho'_1)=rank(\rho'_0 + \rho'_1)$ holds. In such a reduced case, Lemma~\ref{parallel} implies ${\mathcal S}_{\rho'_0:\rho'_1}=0$ that is to say $\rho'_0:\rho'_1=0$.\\

We defining $\Sigma:=\rho'_0+\rho'_1$ to write the parallel addition as  $\rho'_0 \Sigma^{-1} \rho'_1$. Since $rank(\rho'_0 + \rho'_1)=\text{dim}({\mathcal H})$, we end up with $\Sigma$ having full rank and $\Sigma \Sigma^{-1}={\mathbb 1}_{\mathcal H}$. We therefore  have the following corollary to Lemma~\ref{parallel},
\begin{corollary}
Let $\rho_0$ and $\rho_1$ be two density matrices spanning a Hilbert space ${\mathcal H}$. Let $\Sigma$ be the full rank operator defined as the sum of these two density matrices. $${\rm If}\,\,\, {\mathcal S}_{\rho_0} \cap {\mathcal S}_{\rho_1}=\{0\}\,\,\, {\rm then}\,\,\, \rho_0\Sigma^{-1}\rho_1=0.$$
\end{corollary}

According to the first reduction theorem we can, without loss of generality, consider only USD problems of two density matrices without overlap of their supports. In the following, we consider two density matrices $\rho_0$ and $\rho_1$ (which are positive semi-definite matrices) such that ${\mathcal S}_{\rho_0} \cap {\mathcal S}_{\rho_1}=\{0\}$ or equivalently $rank(\rho_0 + \rho_1)=rank(\rho_0)+rank(\rho_1)=\text{dim}({\mathcal H})$. As explained above, for such a problem, $\rho_0\Sigma^{-1}\rho_1=0$, with $\Sigma = \rho_0 + \rho_1$ having full rank. This leads to
\begin{eqnarray}
\rho_i&=&\rho_i\Sigma^{-1}\rho_i,\,\,\, i=0,1
\end{eqnarray}
since $\Sigma \Sigma^{-1}={\mathbb 1}_{\mathcal H}$. The projectors onto the supports of $\rho_i$, $i=0,1$, can then be written as
\begin{eqnarray}
P_i=\sqrt{\rho_i}\Sigma^{-1}\sqrt{\rho_i},\,\,\, i=0,1
\end{eqnarray}

To finish, let us precise that the two density matrices involved in a standard USD problem fulfill all the above properties since they do not overlap.

\section{Necessary and sufficient conditions - first class of exact solutions}

We are now ready to derive the main result of this chapter. The first part of this result gives compact necessary and sufficient conditions for a pair of mixed states to saturate the bounds of the failure probability $Q$. The second part gives  the corresponding POVMs in an explicit form.

\begin{theorem}\label{boundsaturation}Necessary and sufficient conditions to saturate the bounds on the failure probability\\
\graybox{Consider a USD problem defined by the two density matrices $\rho_0$ and $\rho_1$ and their respective {\it a priori} probabilities $\eta_0$ and $\eta_1$ such that their supports satisfy ${\mathcal S}_{\rho_0} \cap {\mathcal S}_{\rho_1}=\{0\}$ (Any USD problem of two density matrices can be reduced to such a form according to Theorem 9). Let $F_0$ and $F_1$ be the two operators $\sqrt{\sqrt{\rho_0}\rho_1\sqrt{\rho_0}}$ and $\sqrt{\sqrt{\rho_1}\rho_0\sqrt{\rho_1}}$. The fidelity $F$ of the two states $\rho_0$ and $\rho_1$ is then given by $F=\Tr(F_0)=\Tr(F_1)$. We denote by $P_0$ and $P_1$, the projectors onto the support of $\rho_0$ and $\rho_1$. The optimal failure probability $Q^{\textrm{opt}}$ for USD then satisfies 
\begin{eqnarray} \label{theocentral}
Q^{\mathrm{opt}} &=& \eta_1 \frac{F^2}{\Tr(P_1 \rho_0)}+\eta_0 \Tr(P_1 \rho_0) \, \nonumber \Leftrightarrow  \,
\left\{\begin{array}{c}
\rho_0-\frac{\Tr(P_1 \rho_0)}{F} F_0 \ge 0 \\ 
\rho_1-\frac{F}{\Tr(P_1 \rho_0)}F_1 \ge 0 \\ 
\end{array}\right. \,\,\, \mathrm{for} \,\,\, \sqrt{\frac{\eta_1}{\eta_0}} \le \frac{\Tr(P_1 \rho_0)}{F}\\ \nonumber
Q^{\mathrm{opt}} &=& 2\sqrt{\eta_0\eta_1}F   \Leftrightarrow  \, 
\left\{\begin{array}{cc}
\rho_0-\sqrt{\frac{\eta_1}{\eta_0}}F_0 \ge 0 \\ 
\rho_1-\sqrt{\frac{\eta_0}{\eta_1}}F_1 \ge 0 \\ 
\end{array}\right. \,\,\, \mathrm{for} \,\,\, \frac{\Tr(P_1 \rho_0)}{F}\le \sqrt{\frac{\eta_1}{\eta_0}} \le \frac{F}{\Tr(P_0 \rho_1)}\\ \nonumber
Q^{\mathrm{opt}} &=& \eta_0 \frac{F^2}{\Tr(P_0 \rho_1)}+\eta_1 \Tr(P_0 \rho_1)  \Leftrightarrow  \, 
\left\{\begin{array}{cc}
\rho_0-\frac{F}{\Tr(P_0 \rho_1)}F_0 \ge 0 \\ 
\rho_1-\frac{\Tr(P_0 \rho_1)}{F} F_1 \ge 0 \\ 
\end{array}\right. \,\,\, \mathrm{for} \,\,\, \frac{F}{\Tr(P_0 \rho_1)} \le \sqrt{\frac{\eta_1}{\eta_0}}\\ 
\end{eqnarray}
The POVM elements that realize these optimal failure probabilities, if the corresponding conditions are fulfilled,  are given by
\begin{eqnarray}
E_0&=&\Sigma^{-1} \sqrt{\rho_0} \left(\rho_0-\alpha F_0 \right) \sqrt{\rho_0}\Sigma^{-1} \\ \nonumber
E_1&=&\Sigma^{-1} \sqrt{\rho_1} \left(\rho_1-\frac{1}{\alpha} F_1\right)\sqrt{\rho_1}\Sigma^{-1} \\ \nonumber
E_?&=&\Sigma^{-1} \left(\sqrt{\alpha} \sqrt{\rho_0}+\frac{1}{\sqrt{\alpha}} \sqrt{\rho_1} V^\dagger\right) F_0 \left(\sqrt{\alpha} \sqrt{\rho_0}+\frac{1}{\sqrt{\alpha}}V \sqrt{\rho_1}\right)\Sigma^{-1} 
\end{eqnarray}
with $\alpha=\frac{\Tr(P_1 \rho_0)}{F}$ for the first regime, $\alpha=\sqrt{\frac{\eta_1}{\eta_0}}$ for the second regime and $\alpha=\frac{F}{\Tr(P_0 \rho_1)}$ for the third regime and where the unitary operator $V$ arises from a polar decomposition $\sqrt{\rho_0}\sqrt{\rho_1}=\sqrt{\sqrt{\rho_0}\rho_1\sqrt{\rho_0}}\,\, V$.}
\end{theorem}

\paragraph*{\bf Proof of Theorem 16}

First, we give a proof for the necessary conditions.\\

\paragraph*{ Proof for the necessary conditions}

From Theorem \ref{3_bounds_theorem} we know that the bounds on the failure probability are satisfied whenever  $V^\dagger \sqrt{\rho_0}E_?=\alpha \sqrt{\rho_1} E_?$ with $\alpha=\frac{\Tr(P_1 \rho_0)}{F}$, $\alpha=\sqrt{\frac{\eta_1}{\eta_0}}$ and $\alpha=\frac{F}{\Tr(P_0 \rho_1)}$ for the three regimes, respectively. 

We replace $E_?$ by ${\mathbb 1}-E_0-E_1$, multiply on the left by $V$ and on the right by $\sqrt{\rho_0}$. This leads us to
\begin{eqnarray}\label{pos}
\rho_0-\alpha F_0=\sqrt{\rho_0}E_0\sqrt{\rho_0}
\end{eqnarray}
where we used the relation (\ref{PD}) $\sqrt{\rho_0} \sqrt{\rho_1}= F_0 V$ and the fact that the support of $\rho_i$ and $E_j$ are orthogonal for $i\neq j$. Indeed, in Lemma 1, we have seen that $\Tr(E_i\rho_j)=0 \Leftrightarrow E_i\rho_j=0$ because $E_i$ and $\rho_j$ are positive semi-definite operators. The right hand side in (\ref{pos}) is positive semi-definite because of the form $AA^\dagger$ with $A=\sqrt{\rho_0}\sqrt{E_0}$. Thus $\rho_0-\alpha F_0$ must be positive semi-definite as well. A similar calculation where we multiply on the right by $\sqrt{\rho_1}$ instead of by $\sqrt{\rho_0}$ leads us to
\begin{eqnarray}
\rho_1-\frac{1}{\alpha} F_1=\sqrt{\rho_1}E_1\sqrt{\rho_1}
\end{eqnarray}
which is again a positive semi-definite operator.\\

With that we have proved that if equality holds in the bounds of Theorem \ref{3_bounds_theorem} then we have
\begin{eqnarray}\label{posop}
\left\{\begin{array}{c}
\rho_0-\alpha F_0 \geq 0\\
\rho_1-\frac{1}{\alpha} F_1\geq 0
\end{array}
\right.
\end{eqnarray}which form, therefore, necessary conditions for equality in the bounds of Theorem \ref{3_bounds_theorem}. 

\paragraph*{ Proof for the sufficient conditions}
Now we start with the assumption that the conditions (\ref{posop}) are fulfilled. Then we can construct an explicit POVM saturating the bound, therefore providing that the conditions are sufficient. Let us define the following POVM elements : 
\begin{eqnarray} \label{POVM}
E_0&=&\Sigma^{-1} \sqrt{\rho_0} \left(\rho_0-\alpha F_0 \right)\sqrt{\rho_0}\Sigma^{-1} \\ \nonumber
E_1&=&\Sigma^{-1} \sqrt{\rho_1} \left(\rho_1-\frac{1}{\alpha} F_1 \right)\sqrt{\rho_1}\Sigma^{-1} \\ \nonumber
E_?&=&\Sigma^{-1} \left(\sqrt{\alpha} \sqrt{\rho_0}+\frac{1}{\sqrt{\alpha}} \sqrt{\rho_1} V^{\dagger} \right) F_0 \left(\sqrt{\alpha} \sqrt{\rho_0}+\frac{1}{\sqrt{\alpha}}V \sqrt{\rho_1}\right)\Sigma^{-1}
\end{eqnarray}

First, let us verify that this is indeed a valid POVM. The three operators are positive semi-definite since they are of the form $A^\dagger M A$ where $M$ is a positive semi-definite operator. In the first two cases this is true because of the conditions (\ref{posop}), in the third case it follows from the positivity of $F_0$. The three operators sum to identity,  $E_0+E_1+E_? = {\mathbb 1}$, as can be checked by straightforward though lengthy calculation, making use of Eqn.~(\ref{PD}).  
Next, we have to check that the given POVM is a valid USD POVM, that is, $\Tr(\rho_0 E_1)=\Tr(\rho_1 E_0)=0$. This relation holds since the supports of $\rho_0$ and $\rho_1$ do not  overlap. Therefore,  corollary 3 applies and we have  $\rho_0 \Sigma^{-1} \rho_1=0$ from which follows that   $\sqrt{\rho_0} \Sigma^{-1} \rho_1=0$ and $\sqrt{\rho_1} \Sigma^{-1} \rho_0=0$. Finally, one can check in a straightforward though lengthy calculation, exploiting the properties used in the previous checks that this  POVM leads to the three desired failure probabilities. This completes the proof. \hfill $\blacksquare$ \\

Let us first note that the assumption about the non-overlapping supports was only used to prove the sufficiency of the conditions. Their necessity does not require this assumption.\\

Moreover given a pair of two density matrices with their {\it a priori} probabilities, the middle regime does not always exists. A necessary condition for its existence is
\begin{eqnarray}
\Tr(P_1 \rho_0) \Tr(P_0 \rho_1) \le F^2
\end{eqnarray}
as pointed out by Ulrike Herzog in \cite{herzog05a}.

To conclude the presentation of our first class of exact solutions, we would like to repeat that only the first reduction theorem is needed to derive Theorem 16. In chapter 6, we will provide pairs of density matrices that fall in this class as well as pairs of density matrices that are not included in it. It means that this class contains pairs of density matrices but does not cover all pairs.

\section{The two pure states case revisited}\label{revisited}

It is possible to use Theorem 16 for two pure states $|\Psi_\pm\rangle$. We change here the label of the two states from '$0/1$' to '$+/-$' since one can always write two pure states $|\Psi_\pm\rangle=\alpha |0\rangle \pm \beta |1\rangle$ where $\alpha$ and $\beta$ are real and such that $\alpha^2+\beta^2=1$ in some suitable orthonormal basis $\{|0\rangle,\,\, |1\rangle\}$. For two pure states, the operators $F_\pm$ are easy to explicit. Indeed $F_+=F|\Psi_+\rangle \langle \Psi_+|$ and $F_-=F|\Psi_-\rangle \langle \Psi_-|$ with $F=|\langle \Psi_+| \Psi_- \rangle|=|2\alpha^2-1|$. Moreover one has the simple relation $\Tr(P_+\rho_-)=\Tr(P_- \rho_+)=F^2$.

The conditions in Theorem 16 then take the following form:

\begin{eqnarray}
\begin{array}{c}
(1-F^2)\rho_+ \ge 0
\end{array} \,\,\, \mathrm{for} \,\,\, \sqrt{\frac{\eta_-}{\eta_+}} \le F\\ \nonumber
\left\{\begin{array}{c}
(1-\sqrt{\frac{\eta_-}{\eta_+}}F) \rho_+ \ge 0 \\ 
(1-\sqrt{\frac{\eta_+}{\eta_-}}F) \rho_- \ge 0 
\end{array}\right. \,\,\, \mathrm{for} \,\,\, F \le \sqrt{\frac{\eta_-}{\eta_+}} \le \frac{1}{F}\\ \nonumber
\begin{array}{c}
(1-F^2)\rho_-\ge 0\\ 
\end{array} \,\,\, \mathrm{for} \,\,\, \frac{1}{F} \le \sqrt{\frac{\eta_-}{\eta_+}}\\ \nonumber
\end{eqnarray}

Since $(1-\sqrt{\frac{\eta_-}{\eta_+}}F)$ and $(1-\sqrt{\frac{\eta_+}{\eta_-}}F)$ for $\frac{1}{F} \le \sqrt{\frac{\eta_-}{\eta_+}} \le F$ range between $0$ and $F^2$, the constraints above are always fulfilled and our result reduces to that of Shimony and Jaeger. Moreover we can give the POVM elements in a compact form thanks to the operator $\Sigma^{-1}$. The choice of our basis yields
\begin{eqnarray}
\rho_\pm=\left(
\begin{array}{cc}
\alpha^2 & \pm \alpha \beta\\
\pm \alpha \beta & \beta^2\end{array}
\right)
\end{eqnarray}
such that
\begin{eqnarray}
\Sigma^{-1}=\frac{1}{2}\left(
\begin{array}{cc}
\alpha^{-2} & 0\\
0 & \beta^{-2}\end{array}
\right).
\end{eqnarray}
It is therefore easy to write the optimal USD POVM as follows
\begin{eqnarray}
E_+=\frac{(1-\alpha F)}{4} \left(
\begin{array}{cc}
\alpha^{-2} & \frac{ 1}{\alpha \beta}\\
\frac{ 1}{\alpha \beta} & \beta^{-2}\end{array}
\right),
\end{eqnarray}
\begin{eqnarray}
E_-=\frac{(1-\frac {F}{\alpha})}{4}\left(
\begin{array}{cc}
\alpha^{-2} & \frac{- 1}{\alpha \beta}\\
\frac{- 1}{\alpha \beta} & \beta^{-2}\end{array}
\right)
\end{eqnarray}
and 
\begin{eqnarray}
E_?={\mathbb 1} - E_+ - E_-
\end{eqnarray}
with $\alpha=F$ for the first regime, $\alpha=\sqrt{\frac{\eta_-}{\eta_+}}$ for the second regime and $\alpha=\frac{1}{F}$ for the third regime. This expression of $E_\pm$ leads naturally to the desired failure probability $Q^{\mathrm{opt}}=F(\alpha \eta_+ + \frac{\eta_-}{\alpha})$ with the respective $\alpha$s.

We can go beyond this remark and investigate under which conditions our bounds reduce to those given in chapter 2. The bounds derived by Rudolph {\it et al.} in \cite{rudolph03a} take the form
\begin{eqnarray}
Q^{\mathrm{opt}} & \ge & \eta_1+\eta_0 F^2  \,\,\, \mathrm{for} \,\,\, \sqrt{\frac{\eta_1}{\eta_0}} \le F,\\ \nonumber
Q^{\mathrm{opt}} & \ge & 2\sqrt{\eta_0\eta_1}F  \,\,\, \mathrm{for} \,\,\, F\le \sqrt{\frac{\eta_1}{\eta_0}} \le \frac{1}{F},\\ \nonumber
Q^{\mathrm{opt}} & \ge & \eta_0+\eta_1 F^2  \,\,\, \mathrm{for} \,\,\, \frac{1}{F} \le \sqrt{\frac{\eta_1}{\eta_0}}.
\end{eqnarray}

Actually one can find Rudolph's bounds following the argumentation in the proof of Theorem 15 but using the weaker constraint $\eta_0 F^2 \le Q_0 \le \eta_0$. This means in particular that our bounds are tighter. To convince ourself, we can nevertheless consider our bounds and Rudolph's bounds in the five regimes of the ratio $\sqrt{\frac{\eta_1}{\eta_0}}$ given by
\begin{eqnarray}
0 \le F \le \frac{\Tr(P_1 \rho_0)}{F} \le \frac{F}{\Tr(P_0 \rho_1)} \le \frac{1}{F}.
\end{eqnarray}
Note that this ordering is due to the Theorem 13 that tells us that $F \le \frac{\Tr(P_1 \rho_0)}{F}$ since $F^2 \le \Tr(P_1 \rho_0)$ and $\frac{F}{\Tr(P_0 \rho_1)} \le \frac{1}{F}$ since $F^2 \le \Tr(P_0 \rho_1)$. On the other hand, the inequality $\frac{\Tr(P_1 \rho_0)}{F} \le \frac{F}{\Tr(P_0 \rho_1)}$ is not always fulfilled as we already discussed (see Eqn.~(4.38)). We can now compare the two bounds in each regime.\\

In the middle regime given by $\frac{\Tr(P_1 \rho_0)}{F} \le \sqrt{\frac{\eta_1}{\eta_0}} \le \Tr(P_0 \rho_1)$, the two bounds are equal.
In the second regime given by $F \le \sqrt{\frac{\eta_1}{\eta_0}} \le \frac{\Tr(P_1 \rho_0)}{F}$, Rudolph's bound still equals the overall lower bound $2\sqrt{\eta_0 \eta_1} F$ and is therefore less or equal than our bound. In the third regime given by $\frac{F}{\Tr(P_0 \rho_1)} \le \sqrt{\frac{\eta_1}{\eta_0}} \le \frac{1}{F}$, a similar argument holds: Rudolph's bound still equals the overall lower bound $2\sqrt{\eta_0 \eta_1} F$ and is therefore less or equal than our bound.\\

In the outer regimes, things are a bit more subtle. We must again consider the function $q(Q_0)= Q_0+ \frac{\eta_0 \eta_1 F^2}{Q_0}$. This function decreases for $0 \le Q_0 \le \sqrt{\eta_0 \eta_1} F$ and increases for $\sqrt{\eta_0 \eta_1} F \le Q_0$.\\
In the first regime, we have by definition $\sqrt{\frac{\eta_1}{\eta_0}} \le F \le \frac{\Tr(P_1 \rho_0)}{F}$ (See Eqn.~(4.38)). We can multiply this inequality by $\eta_0 F$ to get $\sqrt{\eta_0 \eta_1} F \le \eta_0 F^2 \le \eta_0 \Tr(P_1 \rho_0)$. For that range, $q(Q_0)$ increases so that $Q(\eta_0 F^2) \le Q(\eta_0 \Tr(P_1 \rho_0))$ or in other words: $\eta_0 F^2 + \eta_1 \le \eta_0 \Tr(P_1 \rho_0)+\eta_1 \frac{F^2}{\Tr(P_1 \rho_0)}$.\\
In the fifth regime, we have $\frac{F}{\Tr(P_0 \rho_1)} \le \frac{1}{F} \le \sqrt{\frac{\eta_1}{\eta_0}}$ (See Eqn.(4.50)). We can again multiply this inequality by $\eta_0 F$ to get $\eta_0 \frac{F^2}{\Tr(P_0 \rho_1)} \le \eta_0 \le \sqrt{\eta_0 \eta_1} F$. For that range, $q(Q_0)$ decreases so that $Q(\eta_0 \frac{F^2}{\Tr(P_0 \rho_1)}) \ge Q(\eta_0)$ or in other words: $\eta_0 \frac{F^2}{\Tr(P_0 \rho_1)} + \eta_1 \Tr(P_0 \rho_1) \le \eta_0 + \eta_1 F^2$.\\

Since our bounds are tighter, Rudolph's bounds are reached if and only if, first, the conditions in Theorem 16 are fulfilled and, second, the equalities $\Tr(P_0\rho_1)=\Tr(P_1 \rho_0)=F^2$ hold like in the pure state case. Let us now state the corresponding theorem and give the only part of the proof that changes with respect to theorems 15 and 16.

\begin{theorem} Necessary and sufficient conditions to saturate the bounds in \cite{rudolph03a}\\
\graybox{Consider a USD problem defined by the two density matrices $\rho_0$ and $\rho_1$ and their respective {\it a priori} probabilities $\eta_0$ and $\eta_1$ such that their supports satisfy ${\mathcal S}_{\rho_0} \cap {\mathcal S}_{\rho_1}=\{0\}$ (Any USD problem of two density matrices can be reduced to such a form according to Theorem 9). Let $F_0$ and $F_1$ be the two operators $\sqrt{\sqrt{\rho_0}\rho_1\sqrt{\rho_0}}$ and $\sqrt{\sqrt{\rho_1}\rho_0\sqrt{\rho_1}}$. The fidelity $F$ of the two states $\rho_0$ and $\rho_1$ is then given by $F=\Tr(F_0)=\Tr(F_1)$. We denote by $P_0$ and $P_1$, the projectors onto the support of $\rho_0$ and $\rho_1$. The optimal failure probability $Q^{\textrm{opt}}$ for USD then satisfies
\begin{eqnarray} \label{3_regimes_2}
Q^{\mathrm{opt}} = \eta_1+\eta_0 F^2 \, & \Leftrightarrow & \,
\left\{\begin{array}{c}
\rho_0-F F_0 \ge 0 \\ 
\rho_1-\frac{1}{F}F_1 = 0 \\ 
\end{array}\right. \,\,\, \mathrm{for} \,\,\, \sqrt{\frac{\eta_1}{\eta_0}} \le F\\ \nonumber
Q^{\mathrm{opt}} = 2\sqrt{\eta_0\eta_1}F  & \Leftrightarrow & \, 
\left\{\begin{array}{cc}
\rho_0-\sqrt{\frac{\eta_1}{\eta_0}}F_0 \ge 0 \\ 
\rho_1-\sqrt{\frac{\eta_0}{\eta_1}}F_1 \ge 0 \\ 
\end{array}\right. \,\,\, \mathrm{for} \,\,\, F\le \sqrt{\frac{\eta_1}{\eta_0}} \le \frac{1}{F}\\ \nonumber
Q^{\mathrm{opt}} = \eta_0+\eta_1 F^2 & \Leftrightarrow & \, 
\left\{\begin{array}{cc}
\rho_0-\frac{1}{F}F_0 = 0 \\ 
\rho_1-F F_1 \ge 0 \\ 
\end{array}\right. \,\,\, \mathrm{for} \,\,\, \frac{1}{F} \le \sqrt{\frac{\eta_1}{\eta_0}}\\ \nonumber
\end{eqnarray}
The POVM elements that realize these optimal failure probabilities, if the corresponding conditions are fulfilled,  are given by
\begin{eqnarray}
E_0&=&\Sigma^{-1} \sqrt{\rho_0} \left(\rho_0-\alpha F_0 \right) \sqrt{\rho_0}\Sigma^{-1} \\ \nonumber
E_1&=&\Sigma^{-1} \sqrt{\rho_1} \left(\rho_1-\frac{1}{\alpha} F_1\right)\sqrt{\rho_1}\Sigma^{-1} \\ \nonumber
E_?&=&\Sigma^{-1} \left(\sqrt{\alpha} \sqrt{\rho_0}+\frac{1}{\sqrt{\alpha}} \sqrt{\rho_1} V^\dagger\right) F_0 \left(\sqrt{\alpha} \sqrt{\rho_0}+\frac{1}{\sqrt{\alpha}}V \sqrt{\rho_1}\right)\Sigma^{-1} 
\end{eqnarray}
with $\alpha=F$ for the first regime, $\alpha=\sqrt{\frac{\eta_1}{\eta_0}}$ for the second regime and $\alpha=\frac{1}{F}$ for the third regime and where the unitary operator $V$ arises from a polar decomposition $\sqrt{\rho_0}\sqrt{\rho_1}=\sqrt{\sqrt{\rho_0}\rho_1\sqrt{\rho_0}}\,\, V$.}
\end{theorem}

In the first regime $\alpha=F$ implies that $E_1 = 0$. The resulting POVM has to be a projective measurement with projections onto the support of $\rho_1$ and onto its orthogonal complement, i.e.\ $E_0=P^\perp_1$, $E_1=0$ and $E_?=P_1$. A direct proof from the explicit expressions in Eqn.~(4.48) is difficult, however a simple reasoning allows to verify this statement. We consider only the non-trivial case where the supports of $\rho_0$ and $\rho_1$ are not identical. Of course, a two-element USD POVM satisfies $E_0+E_?={\mathbb 1}$ with $\mathcal{S}_{E_0} \subset \mathcal{S}_{\rho_1}$. Then its structure must be such that $E_?=P_1 + R$ where $P_1$ is the projection onto the support of $\rho_1$ and $R$ is an operator with support $\mathcal{S}_{R} \subset \mathcal{K}_{\rho_1}$ which satisfies $E_0+R = P_1^\perp$. Then it follows that $Q=\eta_1 + \eta_0 \Tr(P_1 \rho_0) + \eta_0 \Tr(R \rho_0)$. In our non-trivial case we will have $\Tr(R \rho_0)>0$ as soon as $R \neq 0$. Therefore we find as an optimal solution within this class of two-element USD POVM, the POVM with $R=0$ leading to $E_?=P_1$ and $E_0=P_1^\perp$. We can actually write the failure probability as $Q^{\mathrm{opt}} = \eta_1+\eta_0 F^2$. Indeed $\rho_1=\frac{1}{F}F_1$ then $\rho^2_1=\frac{1}{F^2} \sqrt{\rho_1}\rho_0 \sqrt{\rho_1}$. This implies $F^2\rho_1=P_1\rho_0P_1$ and finally $\Tr(P_1 \rho_0)=F^2$. This is consistent with the results derived above and gives the correct failure probability. In the third regime, we have $\alpha=\frac{1}{F}$. Therefore $E_0=0$ and the corresponding POVM is a projective measurement with $E_0=0$, $E_1=P^\perp_0$, $E_?=P_0$.

\paragraph*{\bf Proof of Theorem 17} 

We will  only derive the three minima of the function $q(Q_0)=Q_0+\frac{\eta_0 \eta_1F^2}{Q_0}$ since the remaining part of the proof does not change (the proof correspond to Theorem 15 where the bounds are derived). Here we consider weaker range constraints on $Q_0$ and $Q_1$: $0 \le Q_0 \le \eta_0$ and $0 \le Q_1 \le \eta_1$. We then minimize $q(Q_0)$ under the constraint $\eta_0 F^2 \le Q_0 \le \eta_0$. Again, the function $q(Q_0)$ is convex ($\frac{d^2 q}{dQ_0^2}(Q_0) \ge 0$) and, therefore, it takes its minimum at the point $Q_0^{\text{min}}$ where the derivative vanishes ($\frac{d q}{dQ_0}(Q_0) = 0$ yielding $Q_0^{\text{min}}=\sqrt{\eta_0 \eta_1} F$) or at the limits of the constraint interval ($Q_0^{\text{min}}=\eta_0 F^2$ and $Q_0^{\text{min}}=\eta_0$). That gives us the minimum of the function $q(Q_0)$ in three different regimes. In the first regime we have  $q_{\text{min}}(Q_0)=\eta_0 F^2 + \eta_1 $ and $Q_0^{\text{min}}=\eta_0 F^2$ if $\sqrt{\eta_0 \eta_1} F \le \eta_0 F^2$ that is to say if $\sqrt{\frac{\eta_1}{\eta_0}} \le F$. In the second regime we have  $q_{\text{min}}(Q_0)=2\sqrt{\eta_0\eta_1}F$ and $Q_0^{\text{min}}=\sqrt{\eta_0\eta_1}F$ if $F \le \sqrt{\frac{\eta_1}{\eta_0}} \le \frac{1}{F}$. The third regime gives  $q_{\text{min}}(Q_0)=\eta_0 + \eta_1 F^2$ and $Q_0^{\text{min}}=\eta_0$ if $\frac{1}{F} \le \sqrt{\frac{\eta_1}{\eta_0}}$.

As a result we obtain lower bounds for the failure probability $Q$ in three regimes as given in Eqn.~(\ref{3_regimes_2}). Since $Q_0=\alpha \eta_0 F$, we read off the values of $\alpha$ as  $\alpha=F$, $\alpha=\sqrt{\frac{\eta_1}{\eta_0}}$ and $\alpha=\frac{1}{F}$ for the first, second and third regime, respectively. This completes the proof. \hfill $\blacksquare$ \\

In the next chapter, we will derive a second class of exact solutions. This class is concerned with pairs of geometrically uniform states in four dimensions.

%% file: chapter6_correction.tex
\chapter{Second class of exact solutions} \label{second_class}

In this chapter, we derive three important results. First we derive a theorem concerned with the rank of an optimal USD measurement. Next, we propose a corollary which is interested in the spectrum of an optimal USD measurement. Finally we give the main result of this chapter, a second class of exact solutions. This class corresponds to any pair of {\it geometrically uniform} states in four dimensions. To be proved, this result requires most of the theorems previously derived in this thesis.

\section{Overall lower bound and rank of the POVM elements}
The maximum rank $r^{max}_{E_i}$ of a USD POVM element $E_i$, $i=0,1$ is
\begin{eqnarray}
r_{E_0}^{max}&=&dim({\cal K}_{\rho_1}),\\ 
r_{E_1}^{max}&=&dim({\cal K}_{\rho_0}).\\
\end{eqnarray}
In the case where ${\mathcal S}_{\rho_0} \cap {\mathcal S}_{\rho_1}=\{0\}$, the {\it maximum rank} of the USD POVM elements $E_i$, $i=0,1$ is $dim({\cal S}_{\rho_i})$, the rank of the mixed states $\rho_i$. Indeed $E_i$ has support in ${\cal K}_{\rho_j}$, $i,j=0,1$, $j\neq i$ and therefore, if ${\mathcal S}_{\rho_0} \cap {\mathcal S}_{\rho_1}=\{0\}$,
\begin{eqnarray}
rank(E_i) & \le & dim({\cal K}_{\rho_j})\\
& \le & dim({\cal H})- dim({\cal S}_{\rho_j})\\
& \le & dim({\cal S}_{\rho_0})+ dim({\cal S}_{\rho_1}) - dim({\cal S}_{\rho_j})\\
& \le &  dim({\cal S}_{\rho_i}),\,\,\, i=0,1.
\end{eqnarray}
Note that in Chapter 3, we already proved that
\begin{eqnarray}
r_{E_?}^{max}=min(dim({\cal S}_{\rho_0}),dim({\cal S}_{\rho_1})).
\end{eqnarray}

The first theorem of this chapter states that the two POVM elements $E_0$ and $E_1$ of an optimal USDM both have {\it maximum rank} only if the two operators $\rho_0-\sqrt{\frac{\eta_1}{\eta_0}}F_0$ and $\rho_1-\sqrt{\frac{\eta_0}{\eta_1}}F_1$ are positive semi-definite. The attentive reader can recognize the two operators involved in the middle regime of Theorem 16.


\begin{theorem}Rank of $E_0$ and $E_1$\\
\graybox{Consider a USD problem defined by two density matrices $\rho_0$ and $\rho_1$ and their respective {\it a priori} probabilities $\eta_0$ and $\eta_1$ such that their supports satisfy ${\mathcal S}_{\rho_0} \cap {\mathcal S}_{\rho_1}=\{0\}$ (Any USD problem of two density matrices can be reduced to such a form according to Theorem 9). Consider also an optimal measurement $\{E_0^{opt},E_1^{opt}, E_?^{opt}\}$ to that problem. Let $F_0$ and $F_1$ be the two operators $\sqrt{\sqrt{\rho_0}\rho_1\sqrt{\rho_0}}$ and $\sqrt{\sqrt{\rho_1}\rho_0\sqrt{\rho_1}}$. The fidelity $F$ of the two states $\rho_0$ and $\rho_1$ is then given by $F=\Tr(F_0)=\Tr(F_1)$.\\

If the two POVM elements $E_0^{opt}$ and $E_1^{opt}$ have maximal rank $dim({\cal S}_{\rho_0})$ and $dim({\cal S}_{\rho_1})$, respectively, then
\begin{eqnarray}
\left\{
\begin{array}{c}
\rho_0-\sqrt{\frac{\eta_1}{\eta_0}} F_0 \geq 0 \\
\rho_1-\sqrt{\frac{\eta_0}{\eta_1}} F_1 \geq 0.
\end{array}
\right.
\end{eqnarray}}
\end{theorem}

\paragraph*{\bf Proof}
We consider an optimal measurement for unambiguously discriminating two mixed states $\rho_0$ and $\rho_1$. We can therefore use the necessary and sufficient conditions derived by Eldar \cite{eldar04a}. We recall them here.
Necessary and sufficient conditions for a measurement $\{ E_k \}$, $k=?,0,1$ to be optimal are that there exists $Z \ge 0$ such that
\begin{eqnarray}
Z E_?=0,\\
E_0 (Z- \eta_0 \rho_0) E_0 =0,\\
E_1 (Z- \eta_1 \rho_1) E_1 =0,\\
P_1^\perp(Z- \eta_0 \rho_0)P_1^\perp \ge 0,\\
P_0^\perp(Z- \eta_1 \rho_1)P_0^\perp \ge 0.
\end{eqnarray}
If $E_0$ and $E_1$ have {\it maximum rank} and Eqn.~(5.11) and Eqn.~(5.12) are fulfilled then the two Hermitian operators $P_1^\perp(Z-\eta_0 \rho_0)P_1^\perp$ and $P_0^\perp(Z-\eta_1 \rho_1)P_0^\perp$ must vanish. Indeed the situation is the following. We consider two positive operators $A$ and $B$, with $A$ full rank and $ABA^\dagger=0$. We can see this relation as of the form $CC^\dagger=0$ with $C=A\sqrt{B}$. Moreover, such an equation $CC^\dagger=0$ is equivalent to $C=0$ for any matrix $C$ (See Appendix A for a proof of this statement). Consequently, $ABA^\dagger=0$ is equivalent to $A\sqrt{B}=0$. Finally, since $A$ is full rank $A^{-1}$ exists and $B$ must vanish.\\

In Eqn.~(5.11) and (5.13), we have $A=E_0$ and $B=P_1^\perp(Z-\eta_0 \rho_0)P_1^\perp$. In Eqn.~(5.12) and (5.14), we have $A=E_1$ and $B=P_0^\perp(Z-\eta_1 \rho_1)P_0^\perp$. As a result, $P_1^\perp(Z-\eta_0 \rho_0)P_1^\perp$ and $P_0^\perp(Z-\eta_1 \rho_1)P_0^\perp$ must vanish if $E_0$ and $E_1$ have maximum rank. Finally to prove the statement of the theorem we can show the following equivalence:
\begin{eqnarray}
\exists Z\ge 0 \,\,\,\textrm{such that}\,\,\,
\left\{\begin{array}{c}
ZE_?=0\\
P_0^\perp(Z-\eta_1 \rho_1)P_0^\perp=0\\
P_1^\perp(Z-\eta_0 \rho_0)P_1^\perp=0\\
\end{array}
\right.
\Leftrightarrow
\left\{
\begin{array}{c}
\rho_0-\sqrt{\frac{\eta_1}{\eta_0}} F_0 \geq 0 \\
\rho_1-\sqrt{\frac{\eta_0}{\eta_1}} F_1 \geq 0
\end{array}
\right.
\end{eqnarray}
where $P_0^\perp(Z-\eta_1 \rho_1)P_0^\perp$ and $P_1^\perp(Z-\eta_0 \rho_0)P_1^\perp$ are positive semi-definite operators.
To prove this statement, we proceed by equivalence.\\

Since the two supports do not overlap, we can make use of the full rank operator $\Sigma^{-1}=(\rho_0 + \rho_1)^{-1}$ introduced in chapter 4. Let us repeat here that its main property is
\begin{eqnarray}
\rho_i \Sigma^{-1} \rho_j= \rho_i \delta_{ij},\,\,\,i=0,1.
\end{eqnarray}
As a consequence, we get the interesting relations
\begin{eqnarray}
\rho_0 \Sigma^{-1}&=&\rho_0 \Sigma^{-1} P_1^\perp,\\
P_1^\perp \rho_0 \Sigma^{-1}&=&P_1^\perp.
\end{eqnarray}
Indeed $\rho_0 \Sigma^{-1}=\rho_0 \Sigma^{-1}(P_1+P_1^\perp)=\rho_0 \Sigma^{-1}\rho_1 \rho_1^{-1} + \rho_0 \Sigma^{-1} P_1^\perp)=\rho_0 \Sigma^{-1} P_1^\perp$. Moreover, $P_1^\perp=P_1^\perp {\mathbb 1}=P_1^\perp(\rho_0 +\rho_1) \Sigma^{-1}=P_1^\perp \rho_0 \Sigma^{-1}$. The same relations are of course true when we swap $0$ and $1$.
\begin{eqnarray}
\rho_1 \Sigma^{-1}&=&\rho_1 \Sigma^{-1} P_0^\perp,\\
P_0^\perp \rho_1 \Sigma^{-1}&=&P_0^\perp.
\end{eqnarray}
It follows that the two equalities $P_1^\perp(Z-\eta_0 \rho_0)P_1^\perp=0$ and $P_0^\perp(Z-\eta_1 \rho_1)P_0^\perp=0$ are equivalent to $\rho_0 \Sigma^{-1}(Z-\eta_0 \rho_0)\Sigma^{-1}\rho_0=0$ and $\rho_1 \Sigma^{-1} (Z-\eta_1 \rho_1)\Sigma^{-1} \rho_1=0$. Hence the assertion\\

\begin{eqnarray}
\exists Z\ge 0 \,\,\,\textrm{such that}\,\,\,
\left\{\begin{array}{c}
ZE_?=0\\
P_0^\perp(Z-\eta_1 \rho_1)P_0^\perp=0\\
P_1^\perp(Z-\eta_0 \rho_0)P_1^\perp=0\\
\end{array}
\right.
\end{eqnarray}
 can be replaced by
\begin{eqnarray}\label{Z1}
\exists Z\ge 0 \,\,\,\textrm{such that}\,\,\,
\left\{\begin{array}{c}
ZE_?=0\\
\rho_i\Sigma ^{-1} Z \Sigma^{-1} \rho_i=\eta_i \rho_i \,\textrm{, for}\,\,\, i=0,1.
\end{array}
\right.
\end{eqnarray}
Since the operator $Z$ is positive, we know it exists an operator $Y$ such that $Z=Y Y^\dagger$. We can insert it in $\rho_i\Sigma ^{-1} Z \Sigma^{-1} \rho_i=\eta_i \rho_i$ and find that it exists $W_i$, a unitary transformation such that
\begin{eqnarray}\label{Z2}
W_i^\dagger Y^\dagger \Sigma^{-1}\rho_i=\sqrt{\eta_i}\sqrt{\rho_i},\,\,\, i=0,1.
\end{eqnarray}
Moreover, $\Sigma$ is full rank. As a result we can decompose $Z$ as $Z=\rho_0 \Sigma^{-1} Z \Sigma^{-1} \rho_0 + \rho_0 \Sigma^{-1} Z \Sigma^{-1} \rho_1 +\rho_1 \Sigma^{-1} Z \Sigma^{-1} \rho_0 + \rho_1 \Sigma^{-1} Z \Sigma^{-1} \rho_1$. This directly yields
\begin{eqnarray}
Z&=&\eta_0 \rho_0 + \eta_1 \rho_1 + \sqrt{\eta_0 \eta_1} \sqrt{\rho_0} W_0^\dagger W_1 \sqrt{\rho_1} + \sqrt{\eta_0 \eta_1} \sqrt{\rho_1} W_1^\dagger W_0 \sqrt{\rho_0}\\ \nonumber
&=&(\sqrt{\eta_0} \sqrt{\rho_0} W_0^\dagger W_1 +\sqrt{\eta_1} \sqrt{\rho_1}) (\sqrt{\eta_0}W_1^\dagger W_0 \sqrt{\rho_0} +\sqrt{\eta_1} \sqrt{\rho_1})
\end{eqnarray}
We finally read off $Y^\dagger$ as
\begin{eqnarray}
Y^\dagger=\sqrt{\eta_0}W^\dagger \sqrt{\rho_0} +\sqrt{\eta_1} \sqrt{\rho_1}
\end{eqnarray}
where $W^\dagger=W_1^\dagger W_0.$\\

We now make use of the relation $ZE_?=0$ which is equivalent to $Y^\dagger E_?=0$ since $AA^\dagger=0 \Leftrightarrow A=0$ for any matrix $A$. We can explicitly write $Y^\dagger E_?=0$ with $Y^\dagger=\sqrt{\eta_0}W^\dagger \sqrt{\rho_0} +\sqrt{\eta_1} \sqrt{\rho_1}$ and $W=W_0^\dagger W_1$. This leads to the statement
\begin{eqnarray}\label{Z3}
\exists Y,W \,\,\,\textrm{such that}\,\,\,
\left\{\begin{array}{c}
WW^\dagger={\mathbb 1},\\
YY^\dagger=Z,\\
-\sqrt{\eta_0}W^\dagger \sqrt{\rho_0} E_?=\sqrt{\eta_1} \sqrt{\rho_1}E_?.
\end{array}
\right.
\end{eqnarray}
In fact, this relation $-\sqrt{\eta_0}W^\dagger \sqrt{\rho_0} E_?=\sqrt{\eta_1} \sqrt{\rho_1}E_?$ is only possible when $-W$ is a unitary transformation coming from a polar decomposition of $\sqrt{\rho_0}\sqrt{\rho_1}$ otherwise theorem 13 in chapter 4 is violated. Indeed theorem 13 tells us that the product between $Q_0$ and $Q_1$ is lower bounded as
\begin{equation}
Q_0 Q_1\ge \eta_0 \eta_1F^2
\end{equation} where the equality holds if and only if a unitary operator $V$ arising from a polar decomposition
\begin{eqnarray}
\sqrt{\rho_0}\sqrt{\rho_1}=\sqrt{\sqrt{\rho_0}\rho_1\sqrt{\rho_0}}\,\, V
\end{eqnarray}
satisfies
\begin{equation}
V^\dagger \sqrt{\rho_0}\sqrt{E_?}=\alpha \sqrt{\rho_1} \sqrt{E_?}\\
\end{equation}
for some $\alpha \in \mathbb{R}^+$.
Moreover $F=\max_U |\Tr(U^\dagger \sqrt{\rho_0} \sqrt{\rho_1})|$ is reached only for unitaries $U$ coming from a polar decomposition of $\sqrt{\rho_0}\sqrt{\rho_1}$. For any unitary $V$ which does not come from a polar decomposition, we then have the strict inequality $F > |\Tr(V^\dagger \sqrt{\rho_0} \sqrt{\rho_1})|$. In other words, if $V$ does not come from a polar decomposition then 
\begin{equation}
\eta_0 \eta_1 F^2 > \eta_0 \eta_1 |\Tr(V^\dagger \sqrt{\rho_0} \sqrt{\rho_1})|.
\end{equation}
Moreover, since $V^\dagger \sqrt{\rho_0}\sqrt{E_?}=\alpha \sqrt{\rho_1} \sqrt{E_?}$, the Cauchy-Schwarz (in)equality tells us that
\begin{eqnarray}
Q_0Q_1&=&\eta_0 \eta_1Tr(E_?\rho_0)Tr(E_? \rho_1)\\
&=&\eta_0 \eta_1|Tr(V^\dagger \sqrt{\rho_0} E_? \sqrt{\rho_1})|\\
&=&\eta_0 \eta_1|Tr(V^\dagger \sqrt{\rho_0} \sqrt{\rho_1})|.
\end{eqnarray}
Consequently $\eta_0 \eta_1 F^2 > Q_0Q_1$ and the theorem 13 is violated. This implies that $-W$ comes from a polar decomposition of $\sqrt{\rho_0}\sqrt{\rho_1}$.
At that point, we simply use the equivalence derived in chapter 4 
\begin{eqnarray}
-\sqrt{\eta_0}W^\dagger \sqrt{\rho_0} E_?=\sqrt{\eta_1} \sqrt{\rho_1}E_? \Leftrightarrow 
\left\{
\begin{array}{c}
\rho_0-\sqrt{\frac{\eta_1}{\eta_0}} F_0 \geq 0 \\
\rho_1-\sqrt{\frac{\eta_0}{\eta_1}} F_1 \geq 0.
\end{array}
\right.
\end{eqnarray}
Indeed Theorem 13 tells us that, for any $-W$ coming from a polar decomposition of $\sqrt{\rho_0}\sqrt{\rho_1}$,
\begin{eqnarray}
-\sqrt{\eta_0}W^\dagger \sqrt{\rho_0} E_?=\sqrt{\eta_1} \sqrt{\rho_1}E_? \Leftrightarrow Q^{\mathrm{opt}} = 2\sqrt{\eta_0\eta_1}F.
\end{eqnarray}
And Theorem 16 says that, for any $-W$ coming from a polar decomposition of $\sqrt{\rho_0}\sqrt{\rho_1}$,
\begin{eqnarray}
Q^{\mathrm{opt}} = 2\sqrt{\eta_0\eta_1}F \Leftrightarrow
\left\{\begin{array}{c}
\rho_0-\sqrt{\frac{\eta_1}{\eta_0}} F_0 \geq 0 \\
\rho_1-\sqrt{\frac{\eta_0}{\eta_1}} F_1 \geq 0.
\end{array}
\right.
\end{eqnarray}This completes the proof. \hfill $\blacksquare$\\

There are at least three consequences to the theorem above. First, it indicates that an optimal POVM is, in general, unlikely to have its elements $E_0$ and $E_1$ of maximum rank. This comes from the fact that the positivity of two operators $\rho_0-\sqrt{\frac{\eta_1}{\eta_0}} F_0$ and $\rho_1-\sqrt{\frac{\eta_0}{\eta_1}} F_1 $ is only possible the middle regime defined by $\frac{\Tr(P_1 \rho_0)}{F}\le \sqrt{\frac{\eta_1}{\eta_0}} \le \frac{F}{\Tr(P_0 \rho_1)}$. Second, we can use Theorem 18 to investigate further the spectrum of an optimal USDM. Last but not least, we can derive a new class of exact solutions for the problem of unambiguously discriminating two mixed states.\\

\section{Maximum rank and {\it a priori} probabilities}
Theorem 18 can be rephrased as
\begin{eqnarray}\label{conditions}
\textrm{If}\,\,\, 
\left\{
\begin{array}{c}
\rho_0-\sqrt{\frac{\eta_1}{\eta_0}} F_0 \geq 0 \\
\rho_1-\sqrt{\frac{\eta_0}{\eta_1}} F_1 \geq 0
\end{array}
\right.
\textrm{is violated then}
\left\{\begin{array}{c}
rank(E_0)<dim({\cal S}_{\rho_0})\\
\textrm{or}\\
rank(E_1)<dim({\cal S}_{\rho_1}).
\end{array}
\right.
\end{eqnarray}
In this section, we discuss why Theorem 18 suggests that $E_0$ and $E_1$ have maximum rank only in a small regime of the ratio between the two {\it a priori} probabilities around 1.\\

We already know that the positivity conditions in (5.37) are quite restrictive since they are reachable only in the middle regime of the ratio $\sqrt{\frac{\eta_1}{\eta_0}}$. Indeed we repeat here that
\begin{eqnarray}
Q^{\mathrm{opt}} = 2\sqrt{\eta_0\eta_1}F  & \Leftrightarrow & \, 
\left\{\begin{array}{cc}
\rho_0-\sqrt{\frac{\eta_1}{\eta_0}}F_0 \ge 0 \\ \nonumber
\rho_1-\sqrt{\frac{\eta_0}{\eta_1}}F_1 \ge 0 \\ 
\end{array}\right. \,\,\, \mathrm{for} \,\,\, \frac{\Tr(P_1 \rho_0)}{F}\le \sqrt{\frac{\eta_1}{\eta_0}} \le \frac{F}{\Tr(P_0 \rho_1)}\\
\end{eqnarray}
where $Q^{\mathrm{opt}}=2 \sqrt{\eta_0 \eta_1} F$ is an overall lower bound on the failure probability that cannot be reached in the two outer regimes.\\

Second, the boundaries of this middle regime can actually be made tighter. Indeed the three regimes of the ratio $\sqrt{\frac{\eta_1}{\eta_0}}$ where built considering some constraints on $Q_0$ and $Q_1$. Stronger constraints means tighter boundaries and the constraints on $Q_0$ and $Q_1$ could in principle be made stronger if more knowledge on the two density matrices $\rho_0$ and $\rho_1$ is provided.\\

Let us give such an example of stronger constraints on $Q_0$ for, say, a POVM having the symmetry $E_1=UE_0 U$ where $U$ is a unitary transformation\footnote{We will see in the next section that such a symmetry is possible for USD of two geometrically uniform states.}.

Since $E_0 \subset {\mathcal K}_{\rho_1}$, there exists $R \ge 0$ in ${\cal K}_{\rho_1}$ such that $E_1+E_?=P_1+R$. Moreover the POVM element $E_?$ is invariant under $U$ since $U E_? U =U ({\mathbb 1} -E_0 -E_1) U= ({\mathbb 1} -E_1 -E_0)=E_?$. Hence, $E_0+E_?=U(E_1+E_?)U = P_0 + URU$. We therefore obtain the trace equality
\begin{eqnarray}\label{R}
\Tr(E_?)=2\Tr(R).
\end{eqnarray}
Indeed $\Tr(E_1+E_?)= \Tr(P_1) + \Tr(R)$ and $\Tr(E_0+E_?)= \Tr(P_0) + \Tr(R)$ so that $\Tr({\mathbb 1})+\Tr(E_?)=\Tr(P_0)+\Tr(P_1)+2\Tr(R)$. And, for a standard USD problem, the equality $\Tr({\mathbb 1})=\Tr(P_0)+\Tr(P_1)$ holds.

We can now consider $Q_0$. $E_1+E_?=P_1+R$ and $\Tr(E_1 \rho_0)=0$, we can consequently write
\begin{eqnarray}
Q_0&=&\eta_0 \Tr(E_?\rho_0)\\
&=&\eta_0 \Tr(E_? \rho_0) + \eta_0 \Tr(E_1 \rho_0)\\
&=&\eta_0 \Tr(P_1 \rho_0) + \eta_0 \Tr(R \rho_0).
\end{eqnarray}
The operator $P_1^\perp \rho_0 P_1^\perp$ is a positive semi-definite operator so that its eigenvalues are all positive or equal to $0$. We can here introduce $\lambda_{min}$, its smallest non vanishing eigenvalue. It follows that $Q_0 \ge \eta_0 \Tr(P_1 \rho_0)+ \eta_0 \Tr(R) \lambda_{min}$. Together with Eqn.(\ref{R}) this yields
\begin{eqnarray}
Q_0 &\ge & \eta_0 \Tr(P_1 \rho_0)+ \frac{\eta_0 \lambda_{min}}{2} \Tr(E_?)\\
& \ge & \eta_0 \Tr(P_1 \rho_0)+ \frac{\eta_0 \lambda_{min}}{2} \Tr(E_? \rho_0).
\end{eqnarray}
In other words, for any USD POVM such that $E_1=UE_0 U$ where $U$ is a unitary transformation,
\begin{eqnarray}
Q_0 \ge \frac{\eta_0 \Tr(P_1 \rho_0)}{1-\lambda_{min}/2}
\end{eqnarray}
where $\lambda_{min}=min \{Spec(P_1^\perp \rho_0 P_1^\perp)\}$. It becomes clear that with more knowledge on the mixed states $\rho_0$ and $\rho_1$, we could make the boundaries of the middle regime tighter. The extreme case would be a middle regime reduced to $\sqrt{\frac{\eta_1}{\eta_0}}=1$. These considerations might indicate that, in general, $E_0$ and $E_1$ have {\it maximum rank} only for some range of the ratio between the {\it a priori} probabilities around $\eta_1=\eta_0=1/2$.

\section{A fourth, incomplete, reduction theorem}

In the case where ${\mathcal S}_{\rho_0} \cap {\mathcal S}_{\rho_1}=\{0\}$, the {\it maximum rank} of the USD POVM elements $E_i$, $i=0,1$ is $r_i$, the rank of the mixed states $\rho_i$. Moreover if not only ${\mathcal S}_{\rho_0} \cap {\mathcal S}_{\rho_1}=\{0\}$ but also ${\mathcal K}_{\rho_0} \cap {\mathcal S}_{\rho_1}=\{0\}$ and ${\mathcal K}_{\rho_1} \cap {\mathcal S}_{\rho_0}=\{0\}$ then $\rho_0$ and $\rho_1$ have the same rank $r$ in a $2r$-dimensional Hilbert space and we end up with
\begin{eqnarray}
r_{E_i}^{max}=r, i=0,1,?
\end{eqnarray}
One can actually use Theorem 18 to study the spectrum of the elements of an optimal USDM. In fact, we can state that, for a standard USD problem, if $\rho_0-\sqrt{\frac{\eta_1}{\eta_0}}F_0$ and $\rho_1-\sqrt{\frac{\eta_0}{\eta_1}}F_1$ are not positive semi-definite then the optimal measurement is such that $E_?$ possesses one eigenvalue equal to $1$ and $E_0$ or $E_1$ too. Let us make this result precise in the following theorem.\\

\begin{corollary}A fourth, incomplete, reduction Theorem\\
\graybox{Consider a standard USD problem defined by two density matrices $\rho_0$ and $\rho_1$ and their respective {\it a priori} probabilities $\eta_0$ and $\eta_1$ (any USD problem of two density matrices can be reduced to such a form according to Chapter 3). Consider also an optimal measurement $\{E_0^{opt},E_1^{opt}, E_?^{opt}\}$ to that problem. Let $F_0$ and $F_1$ be the two operators $\sqrt{\sqrt{\rho_0}\rho_1\sqrt{\rho_0}}$ and $\sqrt{\sqrt{\rho_1}\rho_0\sqrt{\rho_1}}$. The fidelity $F$ of the two states $\rho_0$ and $\rho_1$ is then given by $F=\Tr(F_0)=\Tr(F_1)$.\\
\begin{eqnarray}
\textrm{If}\,\,\, \left\{
\begin{array}{c}
\rho_0-\sqrt{\frac{\eta_1}{\eta_0}} F_0 \geq 0 \\
\rho_1-\sqrt{\frac{\eta_0}{\eta_1}} F_1 \geq 0
\end{array}
\right. \textrm{is violated then there exists}
\end{eqnarray}
\begin{eqnarray*}
|e\rangle \in {\cal S}_{\rho_0}\,\,\textrm{and}\,\, |e'\rangle \in {\cal K}_{\rho_0} \,\,\textrm{such that}\,\,
\left\{
\begin{array}{c}
E_?^{opt} |e\rangle =|e\rangle\\
E_1^{opt} |e'\rangle=|e'\rangle\\
E_0^{opt} |e\rangle =E_0^{opt} |e'\rangle=E_1^{opt}|e\rangle=E_?^{opt} |e'\rangle=0,
\end{array}
\right.
\end{eqnarray*}
\begin{eqnarray*}
\textrm{or} \nonumber
\end{eqnarray*}
\begin{eqnarray*}
|e\rangle \in {\cal S}_{\rho_1}\,\,\textrm{and}\,\, |e'\rangle \in {\cal K}_{\rho_1} \,\,\textrm{such that}\,\,
\left\{
\begin{array}{c}
E_?^{opt} |e\rangle =|e\rangle\\
E_0^{opt} |e'\rangle=|e'\rangle\\
E_1^{opt} |e\rangle =E_1^{opt} |e'\rangle=E_0^{opt}|e\rangle=E_?^{opt} |e'\rangle=0.
\end{array}
\right.
\end{eqnarray*}}
\end{corollary}

First let us note that this theorem makes this assumption of a {\it standard} USD problem. It is in principle not necessary to make such an assumption to derive the existence of some eigenvector of $E_?$, $E_0$ or $E_1$ with eigenvalue $1$ since Theorem 18 is valid for any pair of density matrices without overlapping supports. Nevertheless, this theorem aims to be a 'fourth' reduction theorem. It means in particular that, for any given USD problem of two density matrices, we would like to apply our 'four' reduction theorems and always end up with the optimal USD measurement.

\begin{figure}[h!]
  \centering
  \includegraphics[width=10cm]{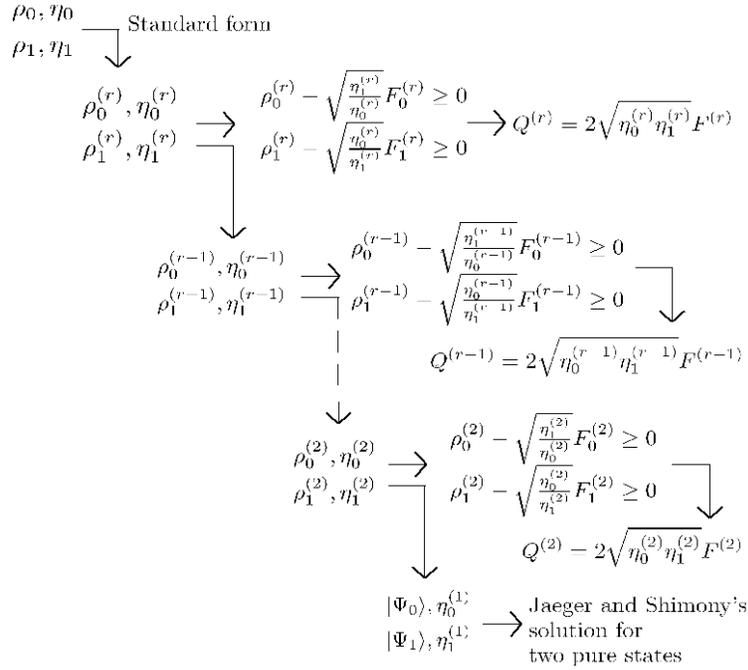}
  \caption{A constructive way to solve any USD problem (the exponent $^{(r)}$ denotes the rank of the density matrices after reduction)}
  \label{4red}
\end{figure}

The above theorem is a kind of incomplete {\it reduction theorem}. A reduction theorem is a theorem that allows us to decrease the size of the USD problem by splitting off some subspace onto which no optimization is needed. To have a complete reduction theorem here, we would need to characterize $|e\rangle$ and $|e'\rangle$ without solving the whole optimization problem. But only the existence of $|e\rangle$ and $|e'\rangle$ is so far ensured. If such a reduction theorem were found then we would have a recipe to solve any USD problem. Let us assume that $|e\rangle$ and $|e'\rangle$ are fully characterized and let us start from a general USD of two mixed states. We use the three first reduction theorems to make it standard. We then check whether the two operators $\rho_0-\sqrt{\frac{\eta_1}{\eta_0}} F_0$ and $\rho_1-\sqrt{\frac{\eta_0}{\eta_1}} F_1$ are positive semi-definite. If yes then we know the optimal failure probability as well as the optimal measurement to perform since this case falls into the first class of exact solutions (middle regime). If the two operators $\rho_0-\sqrt{\frac{\eta_1}{\eta_0}} F_0$ and $\rho_1-\sqrt{\frac{\eta_0}{\eta_1}} F_1$ are not positive semi-definite, we can use our last reduction theorem to get rid of two dimensions. At that point, we check again the positivity of the two operators $\rho_0'-\sqrt{\frac{\eta_1'}{\eta_0'}} F_0'$ and $\rho_1'-\sqrt{\frac{\eta_0'}{\eta_1'}} F_1'$ of the reduced problem. We see here a constructive way to solve any USD problem. If the two operators $\rho_0'-\sqrt{\frac{\eta_1'}{\eta_0'}} F_0'$ and $\rho_1'-\sqrt{\frac{\eta_0'}{\eta_1'}} F_1'$ never happen to be positive, we end up with only two pure states and can finally find the optimal measurement (see Fig.~\ref{4red}). The only problem in that nice picture is that we only know that $|e\rangle$ and $|e'\rangle$ exist but we cannot until now characterize them.\\

Here comes another important remark. There are only two ways to find a complete characterization of the two eigenvectors $|e\rangle$ and $|e'\rangle$. The first is to consider a low dimensional USD problem. The second is to consider a highly symmetric problem. The former case simply is the two pure states case. Indeed, either the operators $\rho_0-\sqrt{\frac{\eta_1}{\eta_0}} F_0$ and $\rho_1-\sqrt{\frac{\eta_0}{\eta_1}} F_1$ are positive semi-definite or we have $|e\rangle \in {{\cal S}_{\rho_{0/1}}}$ and $|e'\rangle \in {{\cal K}_{\rho_{0/1}}}$, eigenvectors of $E_?$ and $E_{1/0}$. In only two dimensions, there is no freedom and $|e\rangle$ and $|e'\rangle$ must be $|\Psi_{0/1}\rangle$ and $|\Psi_{0/1}^\perp\rangle$. If we are interested in higher dimensions, we use some symmetry to give us enough constraint to fully characterize $|e\rangle$ and $|e'\rangle$, we can go up to four dimensions. This is the object of our last section. Before that let us prove Corollary 4.

\paragraph*{\bf Proof of Corollary 4}
To prove this corollary, we begin with the statement given in Theorem 18 for two density matrices $\rho_0$ and $\rho_1$ with same rank $n$ in a $2n$-dimensional Hilbert space. The maximum rank of $E_0$ and $E_1$ then equal $n$. Let us for example consider that $rank(E_0)<n$. The other option corresponding to $rank(E_1)<n$ follows the same argumentation. Because of the completeness relation $E_? +E_1 +E_0= {\mathbb 1}$ fulfilled by the POVM elements, we have, onto the subspace ${\cal S}_{P_0}$, the following equality $P_0E_?P_0 +P_0E_1P_0 +P_0E_0P_0= P_0$. However, ${\cal S}_{E_1} \in {\cal S}_{P_0}^\perp$ so that we are left with
\begin{eqnarray}\label{E?}
P_0E_?P_0 +P_0E_0P_0= P_0.
\end{eqnarray}
Furthermore, in $P_0 E_0 P_0$'s eigenbasis, we have $P_0 E_0 P_0=\sum_{i=1}^{n-1}\lambda_i |\lambda_i \rangle \langle \lambda_i |$ since $E_0$ is of rank $n-1$ and $P_0=\sum_{i=1}^{n-1}|\lambda_i \rangle \langle \lambda_i | + |e \rangle \langle e |$ where $|e \rangle$ completes the $n$ dimensional orthogonal basis of ${\cal S}_{P_0}$.
As a result, $E_? |e \rangle=({\mathbb 1} - E_0 -E_1) |e \rangle=|e \rangle - 0 -0$ and $|e \rangle$ is an eigenvector of $E_?$ with eigenvalue $1$.\\

We can actually go one step further. Since the completeness relation is already fulfilled onto the subspace spanned by $|e \rangle$ and $|e' \rangle$, no optimization is required onto it and we can split it off from the original USD problem. The remaining USD problem to optimize concerns $\rho_0'$ and $\rho_1'$ originated respectively from the density matrix $\rho_0$ and $\rho_1$. Moreover, $\rho_0'$ has rank $n-1$ while $\rho_1'$ has rank $n$. We can indeed denote by ${\cal S}_{|e\rangle}$ the subspace spanned by $|e\rangle$. The reduced Hilbert space is ${\cal H} / {\cal S}_{|e\rangle}$  and ${\cal S}_{\rho_0}$, the support of $\rho_0$, looses one dimension. Thanks to the second reduction theorem, we can reduce this problem to the one of two density matrices of rank $n-1$ in a Hilbert space of dimension $2n-2$. Indeed, the subspace ${\cal K}_{\rho_0'} \cap {\cal S}_{\rho_1'}$ is one dimensional and leads to the detection of $\rho_1'$ with unit probability. We call $| e' \rangle$ the unit vector spanning this $1$-dimensional subspace. We are left with a reduce USD problem in a $2n-2$ dimensional Hilbert space. Importantly, $| e' \rangle$ is in ${\cal K}_{\rho_0'} \cap {\cal S}_{\rho_1'} \subset {\cal S}_{\rho_0'}^\perp = {\cal S}_{\rho_0}^\perp$. Indeed, ${\cal H}={\cal S}_{\rho_0} \oplus {\cal S}_{\rho_0}^\perp = {\cal S}_{\rho_0'} \oplus {\cal S}_{| e \rangle} \oplus {\cal S}_{\rho_0}^\perp$ so that, in ${\cal H'}={\cal H} / {\cal S}_{|e\rangle}$, ${\cal S}_{\rho_0'}^\perp={\cal S}_{\rho_0}^\perp$.\\
In other words, if $\rho_0-\sqrt{\frac{\eta_1}{\eta_0}} F_0$ and $\rho_1-\sqrt{\frac{\eta_0}{\eta_1}} F_1$ are not positive then it exists $|e \rangle$ in ${\cal S}_{P_0}$, eigenvector of $E_?$ with eigenvalue $1$ and $| e' \rangle$ in ${\cal K}_{\rho_0}$, eigenvector of $E_1$ with eigenvalue $1$. Without the assumption that $rk(E_0)<rk(\rho_i)$, we have in general that if $\rho_0-\sqrt{\frac{\eta_1}{\eta_0}} F_0$ and $\rho_1-\sqrt{\frac{\eta_0}{\eta_1}} F_1$ are not positive then there exists $|e \rangle$ in either ${\cal S}_{P_0}$ or ${\cal S}_{P_1}$, eigenvector of $E_?$ with eigenvalue $1$ and $| e' \rangle$ in either ${\cal K}_{\rho_0}$ eigenvector of $E_1$ with eigenvalue $1$ or ${\cal K}_{\rho_1}$, eigenvector of $E_0$ with eigenvalue $1$. The completes the proof. \hfill $\blacksquare$\\

The third consequence of Theorem 18 is the derivation of the optimal USD measurement for any pair of two geometrically uniform states in four dimensions.

\section{Second class of exact solutions}

{\it Geometrically uniform} states, or GU states, are a generalization of symmetric states \cite{eldar01a,eldar02a,eldar03a,eldar03b,eldar03c,eldar04a}. While symmetric state are generated from one generator state and a single unitary transformation, GU states are generated from one generator and a group of unitaries. They are interesting for both practical and theoretical considerations. On the practical side, real applications often exhibit strong symmetries like GU symmetry\footnote{In a cryptographic context, the {\it bit value} states and {\it basis} states in the BB84-type protocol using weak coherent pulses and a phase reference exhibit such a GU symmetry.}. On the theoretical side, this symmetry allows us to seek for simpler conditions and then new results. Actually Eldar proved that the optimal measurement to unambiguously discriminate {\it geometrically uniform} states can be chosen {\it geometrically uniform}, too. This result allows us to derive now the general solution for unambiguously discriminating any pair of GU states in four dimension. Next we give the mathematical definition of the {\it geometrically uniform} states before presenting the optimal failure probability for unambiguously discriminating two {\it geometrically uniform} states in four dimensions and the corresponding optimal measurement.

\subsection{Geometrically uniform states}
A set of GU state is a set of mixed states $\{ \rho_i\}$, $i=1,...,n$ such that $\rho_i=U_i \rho U_i^\dagger$ where $\rho$ is an arbitrary density matrix called the {\it generator} and the set $\{U_i\}$, $i=1,...,n$ is a set of unitary matrices that form an abelian group. In order not to break the symmetry of the states, we assume that all their {\it a priori} probabilities are equal to $\frac{1}{n}$.\\
A consequence of the group structure of the set $\{U_i\}$ is that we can always consider $U_1$ as the identity, and $\rho_1$ as the generator for a given set of GU states. We can therefore always write two GU states as $\rho_0$ and $\rho_1 = U \rho_0 U$ where $U$ is an involution (i.e.\ a unitary transformation $U$ such that $U^2={\mathbb 1}$) with $\eta_0=\eta_1=\frac{1}{2}$. Let us note that two GU states are two symmetric states since only a single unitary is needed.\\
In the next section, we give a second class of exact solutions for USD of two generic density matrices. We provide the optimal failure probability as well as the optimal USD measurement for any two GU states in four dimensions.

\subsection{Optimal unambiguous discrimination of two geometrically uniform states in four dimensions}

\begin{theorem}Optimal unambiguous discrimination of two geometrically uniform states in four dimension\\
\graybox{Consider a USD problem defined by two {\it geometrically uniform} states $\rho_0$ and $\rho_1$ of rank two with equal {\it a priori} probabilities and spanning a four-dimensional Hilbert space. Let $F_0$ and $F_1$ be the two operators $\sqrt{\sqrt{\rho_0}\rho_1\sqrt{\rho_0}}$ and $\sqrt{\sqrt{\rho_1}\rho_0\sqrt{\rho_1}}$. The fidelity $F$ of the two states $\rho_0$ and $\rho_1$ is then given by $F=\Tr(F_0)=\Tr(F_1)$. We denote by $P_0$ and $P_1$, the projectors onto the support of $\rho_0$ and $\rho_1$. The optimal failure probability $Q^{\textrm{opt}}$ for USD then satisfies 
\begin{eqnarray}
1. \,\,\,\, Q^{\mathrm{opt}} &=& F \,\,\,\, \textrm{if} \,\,\,\, \rho_0-F_0 \ge 0\\ \nonumber
\\ \nonumber
2. \,\,\,\, Q^{\mathrm{opt}} &=& 1-\langle x|\rho_0|x\rangle \,\,\,\, \textrm{if} \,\,\,\,
\left\{\begin{array}{l}
\rho_0-F_0 \ngeq 0\\
Spec(P^{\perp}_{1}\,U\,P^{\perp}_{1})=\{a,-b\},\,\,\,\, a,b \in {\mathbb R}^+
\end{array}\right. \\ \nonumber
\\ \nonumber
3.\,\,\,\, Q^{\mathrm{opt}} &=& 1  \,\,\,\, \textrm{otherwise}.
\end{eqnarray}
with $P^{\perp}_{1}\,U\,P^{\perp}_{1}=a|0\rangle\langle0|-b|1\rangle\langle1|$ and $|x \rangle=\frac{1}{\sqrt{a+b}}(e^{-iArg(\langle 1 | \rho_0| 0 \rangle)}\sqrt{b}|0\rangle + \sqrt{a} |1\rangle)$.\\
The POVM elements that realize these optimal failure probabilities are given in the different cases by\\
\begin{eqnarray}
1.\,\,\,\, E_0&=&\Sigma^{-1} \sqrt{\rho_0} \left(\rho_0- F_0 \right) \sqrt{\rho_0}\Sigma^{-1} \\ \nonumber
E_1&=&U E_0 U \\ \nonumber
E_?&=&{\mathbb 1}- E_0 - U E_0 U 
\\ \nonumber
\\ \nonumber
2.\,\,\,\, E_0&=&|x \rangle \langle x |\\ \nonumber
E_1&=&U E_0 U \\ \nonumber
E_?&=&{\mathbb 1}- E_0 - U E_0 U 
\\ \nonumber
\\ \nonumber
3.\,\,\,\, E_0&=&0\\ \nonumber
E_1&=&0\\ \nonumber
E_?&=&{\mathbb 1}.
\end{eqnarray}}
\end{theorem}

\paragraph*{\bf Proof}
We consider a USD problem defined by two {\it geometrically uniform} states $\rho_0$ and $\rho_1=U \rho_0 U$, $U^2={\mathbb 1}$, of rank two, spanning a four-dimensional Hilbert space. This means in particular that ${\mathcal S}_{\rho_0} \cap {\mathcal S}_{\rho_1}=\{0\}$ and $r_{E_0}^{max}=r_{E_1}^{max}=r_{E_?}^{max}=2$.\\

Due to the symmetry of the states, we also notice that $\rho_0-F_0=\rho_1-F_1$. Note that the {\it a priori} probabilities are equal in order not to break the symmetry. Moreover, thanks to Eldar \cite{eldar04a}, we can choose the optimal USD measurement to be GU, too. Thus the POVM elements are such that
\begin{eqnarray}
&E_0&,\\ \nonumber
&E_1&=U E_0 U,\\ \nonumber
&E_?&=U E_? U.
\end{eqnarray}

The statement in Theorem 16 for equal {\it a priori} probability
\begin{eqnarray}
Q^{\mathrm{opt}} = F  & \Leftrightarrow & \, 
\begin{array}{cc}
\rho_0-F_0 \ge 0 \\ 
\rho_1-F_1 \ge 0 \\ 
\end{array}
\end{eqnarray}
then reduces to
\begin{eqnarray}
Q^{\mathrm{opt}} = F  & \Leftrightarrow & \, 
\begin{array}{cc}
\rho_0-F_0 \ge 0. \\ 
\end{array}
\end{eqnarray}
Note that we are not interested in the equivalence. The implication from the right to the left is the only important direction for our purpose here. In that case we need the assumption ${\mathcal S}_{\rho_0} \cap {\mathcal S}_{\rho_1}=\{0\}$ to prove that: If $\rho_0-F_0 \ge 0$ then $Q^{\mathrm{opt}} = F$. Without this assumption, only the other direction is true.\\

If $\rho_0-F_0 \ngeq 0$, Theorem 18 tells us that the ranks of the POVM elements $E_0$ and $E_1$ are not maximum ($E_0$ and $E_1$ have the same rank because of the symmetry). As a consequence, if $\rho_0-F_0 \ngeq 0$ then $rank(E_0)=rank(E_1) < 2$. It follows that if $\rho_0-F_0 \ngeq 0$ then the two POVM elements $E_0$ and $E_1$ have either rank $1$ or rank $0$. If $rank(E_0)=rank(E_1)=0$ then $E_?= {\mathbb 1}$ and $Q=1$. Let us now focus on the remaining case $rank(E_0)=rank(E_1)=1$.\\

Let us now prove that a measurement with $rank(E_0)=rank(E_1)=1$ and $rank(E_?) \le 2$ is necessary a projective measurement with $rank(E_?)=2$. We can introduce the unit vectors and real numbers $| x \rangle \in {\cal K}_{\rho_1}$, $| y \rangle \in {\cal K}_{\rho_0}$, $x$ and $y$ such that
\begin{eqnarray}
E_0= x | x \rangle \langle x|,
E_1 = y | y \rangle \langle y |.
\end{eqnarray}
We call ${\cal S}_{xy}$ the two dimensional subspace spanned by $| x \rangle$ and $| y \rangle$, $P_{xy}$ the projection onto it and $P_{xy}^\perp$ the projector onto its orthogonal complement. By definition of the subspace ${\cal S}_{xy}$,
\begin{eqnarray}
P_{xy}^\perp E_? P_{xy}^\perp= P_{xy}^\perp.
\end{eqnarray}
Therefore $rank (P_{xy}^\perp E_? P_{xy}^\perp) = rank (P_{xy}^\perp)=2$ and $E_?$ must be at least of rank $2$. However $rank(E_?) \le 2$. Therefore $rank (E_?)=2$ and
\begin{eqnarray}
E_?=P_{xy}^\perp.
\end{eqnarray}
We can now consider the subspace ${\cal S}_{xy}$ only. On that subspace, we have
\begin{eqnarray}
E_0 +E_1 =P_{{\cal S}_{xy}}
\end{eqnarray}
that is to say $P_{xy}=x | x \rangle \langle x |+ y | y \rangle \langle y |$. Since $P_{xy}$ is a projector, $P_{xy}=P_{xy}^2$ and it follows that $x | x \rangle \langle x |+ y | y \rangle \langle y | + xy \langle y |x \rangle |y \rangle \langle x| +  xy \langle y |x \rangle |y \rangle \langle x|= x | x \rangle \langle x |+ y | y \rangle \langle y |$. The off-diagonal terms are equal if and only if $\langle y |x \rangle=0$ while the diagonal terms are equal if and only if $x=y=1$. The POVM then is a projective measurement with $rank(E_?)=2$.\\

We now give the optimal USD measurement for a GU projective measurement. Since the measurement is made of projectors, we have $\Tr(E_0 E_1)=0$ which is nothing but $\langle x |U| x \rangle=0$. Because $| x \rangle$ lies in ${\mathcal K}_{\rho_1}$, this relation is equivalent to
\begin{eqnarray}
\langle x |P_1^\perp U P_1^\perp| x \rangle=0.
\end{eqnarray}
$P_1^\perp U P_1^\perp$ is a Hermitian operator and therefore owns real eigenvalues. Note that if $P_1^\perp U P_1^\perp$ must be of rank $2$ since $U$ is full rank. Thus we denote $a$ and $c$ the two eigenvalues of $P_1^\perp U P_1^\perp$ and $| 0 \rangle$ and $| 1 \rangle$ its two eigenvectors. In this eigenbasis,  $| x \rangle \in {\mathcal K}_{\rho_1}$ can be expressed as
\begin{eqnarray}
| x \rangle=
\left(\begin{array}{c}
\alpha\\
\beta
\end{array}
\right)
\end{eqnarray}
This leads to $\langle x |P_1^\perp U P_1^\perp| x \rangle=|\alpha|^2 a + |\beta|^2 c$. Importantly this scalar product can only vanish if $a > 0$ and $c < 0$. We call $-c=b>0$ such that, in $\{| 0 \rangle,| 1 \rangle\},$
\begin{eqnarray}
P_1^\perp U P_1^\perp=
\left(\begin{array}{cc}
a&0\\
0&-b
\end{array}
\right).
\end{eqnarray}
If we include the normalization of $| x \rangle$, we end up with a system of two equations. This system simply is
\begin{eqnarray}
\left\{\begin{array}{c}
|\alpha|^2 a + |\beta|^2 c=0\\
|\alpha|^2 + |\beta|^2 =1
\end{array}
\right.
\end{eqnarray}
and admits a family of solutions parametrized by a phase $\Phi$:
\begin{eqnarray}
\{\alpha=\frac{e^{i\Phi}}{\sqrt{1+a/b}},\beta=\frac{1}{\sqrt{1+b/a}}\}.
\end{eqnarray}
In the basis $\{| 0 \rangle,| 1 \rangle\}$ we can therefore write
\begin{eqnarray}
| x \rangle=
\left(\begin{array}{c}
\frac{e^{i\Phi}}{\sqrt{1+a/b}}\\
\frac{1}{\sqrt{1+b/a}}
\end{array}
\right).
\end{eqnarray}
We can use again the fact that we are interested in  the optimal measurement. Note that we already considered optimality to state than if $\rho_0-F_0 \ngeq 0$ then the POVM is either $\{E_0=E_1=0,E_?={\mathbb 1}\}$ or a projective measurement. Indeed Theorem 18 is only concerned with optimal USD POVM. So far, $|x \rangle$ is valid for any USD measurement such that $E_0=| x \rangle \langle x|$, $E_1=UE_0U$ and $E_?={\mathbb 1}- E_0 - U E_0 U$. Let us now find the optimal one. To do so, we evaluate the success probability $P_{success}^{\mathrm{opt}}$. Because of the symmetry of the two GU states, $\Tr(E_0 \rho_0)=\Tr(E_1 \rho_1)$ and  the success probability $P_{success}^{\mathrm{opt}}=\frac{1}{2}\Tr(E_0 \rho_0)+\frac{1}{2}\Tr(E_1 \rho_1)$ for unambiguously discriminating the two GU state $\rho_0$ and $\rho_1$ takes the form
\begin{eqnarray}
P_{success}^{\mathrm{opt}}=\Tr(E_0 \rho_0)=\langle x |\rho_0| x \rangle.
\end{eqnarray}
After calculation, we obtain
\begin{eqnarray}
P_{success}^{\mathrm{opt}}= \frac{1}{a+b}\left(b\langle0|\rho_0|0\rangle+a\langle1|\rho_0|1\rangle+2\sqrt{ab}Re(\langle0|\rho_0|1\rangle e^{i\Phi})\right).
\end{eqnarray}
We choose the phase $\Phi$ to maximize this success probability $P_{success}^{\mathrm{opt}}$. That is why we choose $\Phi$ such that $Re(\langle0|\rho_0|1\rangle e^{i\Phi})=|\langle0|\rho_0|1\rangle|$. Therefore, $\Phi$ must be $-Arg(\langle0|\rho_0|1\rangle)$ and 
\begin{eqnarray}
| x \rangle=
\left(\begin{array}{c}
\frac{e^{-iArg(\langle0|\rho_0|1\rangle)}}{\sqrt{1+a/b}}\\
\frac{1}{\sqrt{1+b/a}}
\end{array}
\right).
\end{eqnarray} This completes the proof. \hfill $\blacksquare$\\

This theorem leads to a fundamental question: 'Is it possible to find a unified expression for the failure probability $Q$?' In the first class of exact solutions, we can write the three failure probabilities of the three regimes as
\begin{eqnarray}
Q=\alpha \eta_0 F+\frac {1}{\alpha} \eta_1 F \nonumber
\end{eqnarray}
with the above-mentioned $\alpha$. But we do not really expect the bounds in the outer regimes to be often optimal (see discussion in section 5.2) so that this expression does not seem so fundamental. More significatively, for the second class of exact solutions, no unified expression of the failure probability exists. In higher dimension ($dim({\cal H}) > 4$), the number of cases for the optimal failure probability $Q$ might become very large. If this is the case, a unified expression for $Q$ would be a pre-condition to find the general solution to USD of two density matrices.

In the next chapter we analyze an application of both theoretical and practical interest. In fact, we consider the {\it Bennett and Brassard 1984} protocol (BB84 protocol) implemented through weak coherent pulses with strong phase reference. This represents the first solved example of a non reducible USD problem.

%% file: BB84.tex
\chapter{Application of the second class of exact solutions to the BB84 protocol}

In 1984, Bennett and Brassard proposed a protocol to distribute a unconditional secure private key between two parties over a public channel in order to allow a secure communication. This proposed Quantum Key Distribution protocol, the so-called Bennett-Brassard 1984 (or shortly BB84) is here unconditional secure because of the laws of nature (quantum mechanics) and not anymore because of the assumption of a limited computational power of some hypothetical eavesdropper. In the standard BB84 protocol, Alice sends one of the four states $\{0,1,+,-\}$ to Bob. Here $\{0,1\}$ and $\{+,-\}$ are orthogonal pairs and $0$ and $+$ correspond to the {\it bit value} $0$ while $1$ and $-$ correspond to the {\it bit value} $1$. Bob then detects the signal sent in one of the two bases $\{0,1\}$ or $\{+,-\}$.\\
In this thesis, we consider the implementation of a BB84-type protocol that uses weak coherent pulses with a phase reference. In that scenario, Alice sends one of the four states $\{|\frac{\alpha}{\sqrt{2}} \rangle | \frac{\pm \alpha}{\sqrt{2}} \rangle,|\frac{\alpha}{\sqrt{2}} \rangle |\frac{\pm i \alpha}{\sqrt{2}} \rangle \}$. The {\it bit value} is encoded in the sign of the coherent states that is to say $|\frac{ \alpha}{\sqrt{2}} \rangle$ and $|\frac{ i \alpha}{\sqrt{2}} \rangle$ correspond to the {\it bit value} $0$, $|\frac{- \alpha}{\sqrt{2}} \rangle$ and $|\frac{- i \alpha}{\sqrt{2}} \rangle$ correspond to the {\it bit value} $1$. Moreover the phase $i$ plays the role of the basis in the standard BB84 protocol. Firstly let us note that the factor $\frac{1}{\sqrt{2}}$ in the amplitude comes from the technique used to implement the polarized coherent states. Secondly the first mode $|\frac{\alpha}{\sqrt{2}} \rangle$ is common to the four signal states.  This mode is therefore irrelevant for the following analyze. Furthermore it is worth noticing that the states corresponding to the {\it bit value} $0$ and $1$ are not orthogonal since
\begin{eqnarray}
\langle \frac{ \alpha}{\sqrt{2}} | \frac{- \alpha}{\sqrt{2}} \rangle \ne 0,\\
\langle \frac{ i \alpha}{\sqrt{2}}| \frac{-i \alpha}{\sqrt{2}} \rangle \ne 0.
\end{eqnarray}
This QKD protocol is therefore not the standard BB84 protocol. It remains that two important question can be addressed.\\

With what probability can an eavesdropper unambiguously distinguish the {\it basis} of the signal?\\

With what probability can an eavesdropper unambiguously determine which {\it bit value} is sent without being interested in the knowledge of the basis?\\

In fact the first question refers to the unambiguous discrimination of the two {\it basis} $\{| \frac{\pm \alpha}{\sqrt{2}} \rangle\}$ and $\{| \frac{\pm i \alpha}{\sqrt{2}} \rangle\}$. Therefore we can build a mixed state $\rho_0$ that corresponds to the basis $\{| \frac{\pm \alpha}{\sqrt{2}} \rangle\}$ and a mixed state $\rho_1$ for the basis $\{| \frac{\pm i  \alpha}{\sqrt{2}} \rangle\}$. We end up with 
\begin{eqnarray}
\rho_0&=&\frac{1}{2}\left(|\frac{\alpha}{\sqrt{2}} \rangle \langle \frac{\alpha}{\sqrt{2}}|+|\frac{-\alpha}{\sqrt{2}} \rangle \langle \frac{-\alpha}{\sqrt{2}}|\right),\\
\rho_1&=&\frac{1}{2}\left(|\frac{i\alpha}{\sqrt{2}} \rangle \langle \frac{i\alpha}{\sqrt{2}}|+|\frac{-i\alpha}{\sqrt{2}} \rangle \langle \frac{-i\alpha}{\sqrt{2}}|\right).
\end{eqnarray}
where we ignore the irrelevant first mode.\\

The second question refers to the unambiguous discrimination of the two {\it bit value} mixed states. We can for that case build the two density matrices
\begin{eqnarray}
\rho_0&=&\frac{1}{2}\left(|\frac{\alpha}{\sqrt{2}} \rangle \langle \frac{\alpha}{\sqrt{2}}|+|\frac{i\alpha}{\sqrt{2}} \rangle \langle \frac{i\alpha}{\sqrt{2}}|\right),\\
\rho_1&=&\frac{1}{2}\left(|\frac{-\alpha}{\sqrt{2}} \rangle \langle \frac{-\alpha}{\sqrt{2}}|+|\frac{-i\alpha}{\sqrt{2}} \rangle \langle \frac{-i\alpha}{\sqrt{2}}|\right)
\end{eqnarray}
where we again ignore the irrelevant first mode.\\

The states $\{| \frac{\pm \alpha}{\sqrt{2}} \rangle\}, \{| \frac{\pm i \alpha}{\sqrt{2}} \rangle\}$ are four linearly independent pure states. Therefore they span a four dimension Hilbert space. In the next section we will express the four density matrices above in that four dimensional Hilbert space and prove that they are GU states. After that, we will solve the two USD problems arising from the two questions mentioned. It turns out that the first case is reducible to some pure state case while the second one requires our last theorem to be solved. Let us now start with the explicit expression of these four mixed states.

\section{Two geometrically uniform states in a four-dimensional Hilbert space}
A coherent state of amplitude $\alpha$ can be written as a poisson distribution of photon number in the polarization mode $a^\dagger$ as
\begin{eqnarray}
|\alpha \rangle = e^{-\frac{|\alpha|^2}{2}} \sum_{n=0}^{\infty} \frac{(\alpha a^\dagger)}{n!} |0\rangle,
\end{eqnarray}
where $|0\rangle$ denotes the vacuum state. Moreover, the four signal states $|\pm \alpha \rangle,|i\pm \alpha \rangle$ are coherent states in four different polarizations: $\pm 45$° and circular left or right. These polarizations are expressed in terms of two orthogonal polarizations $b_1^\dagger$ and $b_2^\dagger$ as
\begin{eqnarray}
a_0^\dagger&=&\frac{1}{\sqrt{2}}(b_1^\dagger + b_2^\dagger),\\
a_1^\dagger&=&\frac{1}{\sqrt{2}}(b_1^\dagger + i b_2^\dagger),\\
a_2^\dagger&=&\frac{1}{\sqrt{2}}(b_1^\dagger - b_2^\dagger),\\
a_3^\dagger&=&\frac{1}{\sqrt{2}}(b_1^\dagger - i b_2^\dagger).
\end{eqnarray}
Consequently, we can write the four states as
\begin{eqnarray}
|\Psi_0 \rangle&=&|\frac{\alpha}{\sqrt{2}} \rangle |\frac{\alpha}{\sqrt{2}} \rangle, \\
|\Psi_1 \rangle&=&|\frac{\alpha}{\sqrt{2}} \rangle |\frac{i \alpha}{\sqrt{2}} \rangle, \\
|\Psi_0 \rangle&=&|\frac{\alpha}{\sqrt{2}} \rangle |\frac{- \alpha}{\sqrt{2}} \rangle, \\
|\Psi_0 \rangle&=&|\frac{\alpha}{\sqrt{2}} \rangle |\frac{-i \alpha}{\sqrt{2}} \rangle.
\end{eqnarray}

\begin{figure}[h!]
  \centering
  \includegraphics[width=10cm]{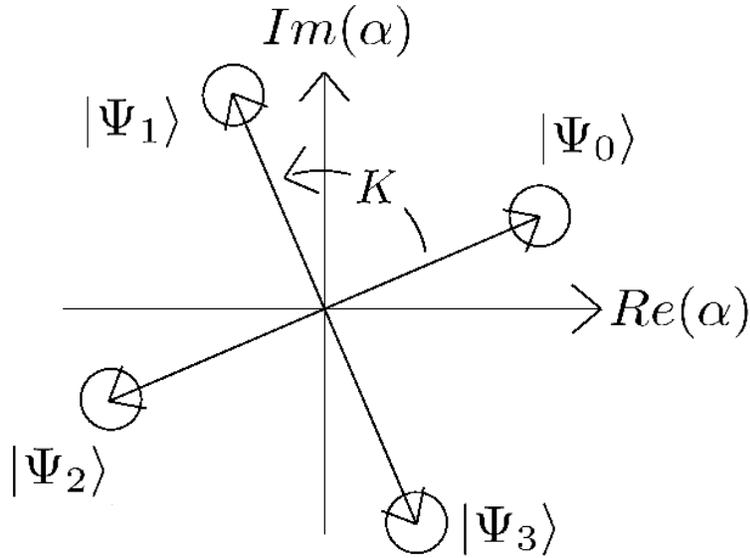}
  \caption{Schematic view of the four symmetric states in the phase space}
  \label{bb84cases}
\end{figure}

The first mode is common to the four states and therefore will be left out. In the phase space, these four states are generated from $|\Psi_0 \rangle$ and a rotation of angle $\frac{\pi}{2}$. This means they are symmetric states and we can write them in a suitable basis following Chefles {\it et al.} \cite{chefles98a}. The idea is that $n$ symmetric states can always be written in an orthonormal basis $\{ |\Phi_j \rangle \}$ as
\begin{eqnarray}
|\Psi_k \rangle= \sum_{j=0}^{n-1}c_j e^{2 i \pi \frac{kj}{n}} |\Phi_j \rangle.
\end{eqnarray}
Note that the phase of the complex numbers $c_j$ is not relevant since we can absorb it in the definition of the basis elements $|\Phi_j \rangle$. Actually the modulus of the coefficients $c_j$s can be expressed \cite{chefles98a} as
\begin{eqnarray}
|c_j|^2=\frac{1}{n^2} \sum_{k,k'} e^{-2 i \pi \frac{j(k-k')}{n}} \langle \Psi_k'|\Psi_{k} \rangle.
\end{eqnarray}
This leads in our case to
\begin{eqnarray}
|c_0|=\frac{1}{\sqrt{2}}e^{-\frac{\mu}{4}}\sqrt{cosh(\frac{\mu}{2})+cos(\frac{\mu}{2})},\\
|c_1|=\frac{1}{\sqrt{2}}e^{-\frac{\mu}{4}}\sqrt{sinh(\frac{\mu}{2})+sin(\frac{\mu}{2})},\\
|c_2|=\frac{1}{\sqrt{2}}e^{-\frac{\mu}{4}}\sqrt{cosh(\frac{\mu}{2})-cos(\frac{\mu}{2})},\\
|c_3|=\frac{1}{\sqrt{2}}e^{-\frac{\mu}{4}}\sqrt{sinh(\frac{\mu}{2})-sin(\frac{\mu}{2})}.
\end{eqnarray}
where $\mu=|\alpha|^2$ stands for the mean photon number. Moreover, in the basis $\{|\Phi_j \rangle\}$, the unitary transformation acting on $|\Psi_0 \rangle$ that generates the other three states is
\begin{eqnarray}
K=\left(
\begin{array}{cccc}
1 & 0 & 0 & 0\\
0 & i & 0 & 0\\
0 & 0 & -1 & 0\\
0 & 0 & 0 & -i \end{array}
\right)\end{eqnarray}
such that $K^4={\mathbb 1}$. The four symmetric states (see Fig.~\ref{bb84cases}) are then expressed as
\begin{eqnarray}
|\Psi_0 \rangle&=&\left(\begin{array}{c}
c_0\\
c_1\\
c_2\\
c_3\end{array}\right),\\
|\Psi_1 \rangle&=&K |\Psi_0 \rangle=\left(\begin{array}{c}
c_0\\
i c_1\\
-c_2\\
-i c_3\end{array}\right),\\
|\Psi_2 \rangle&=&K^2 |\Psi_0 \rangle= \left(\begin{array}{c}
c_0\\
-c_1\\
c_2\\
-c_3\end{array}\right),\\
|\Psi_3 \rangle&=&K^3 |\Psi_0 \rangle= \left(\begin{array}{c}
c_0\\
-i c_1\\
-c_2\\
i c_3\end{array}\right).
\end{eqnarray}


At that point, we are ready to write the four density matrices corresponding to the {\it basis} mixed states and the {\it bit value} mixed states.

\begin{figure}[h!]
  \centering
  \includegraphics[width=10cm]{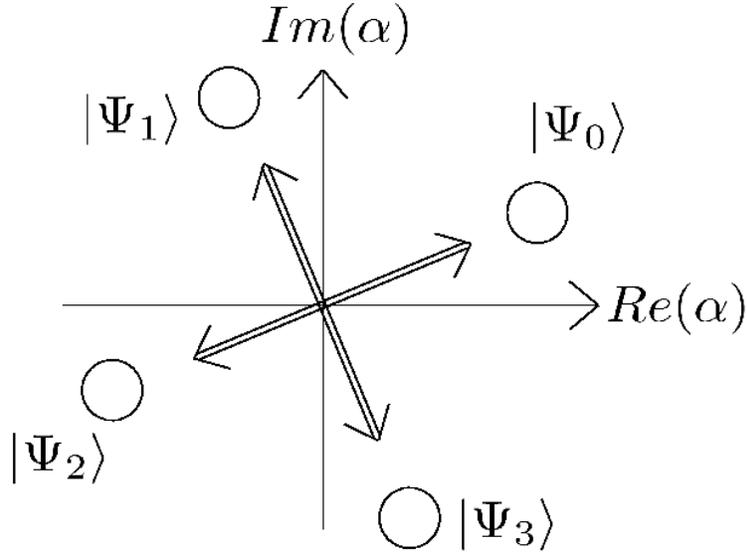}
  \caption{Pairing of the four symmetric states for the {\it basis} mixed states}
  \label{bb84cases2}
\end{figure}

The {\it basis} mixed states (see Fig.~\ref{bb84cases2})
\begin{eqnarray}
\rho_0&=&\frac{1}{2} \left( | \Psi_0 \rangle \langle \Psi_0 | + | \Psi_2 \rangle \langle \Psi_2 | \right),\\
\rho_1&=&\frac{1}{2} \left( | \Psi_1 \rangle \langle \Psi_1 | + | \Psi_3 \rangle \langle \Psi_3 | \right)
\end{eqnarray}
are by construction of rank $2$.

They can be written in a four dimensional Hilbert space spanned by the four linearly independent states $|\Psi_i \rangle$, $i=0,1,2,3$ as
\begin{eqnarray}
\rho_0=\left(
\begin{array}{cccc}
c_0^2 & 0 & c_0c_2 & 0\\
0 & c_1^2 & 0 & c_1c_3\\
c_0c_2 & 0 & c_2^2 & 0\\
0 & c_1c_3 & 0 & c_3^2 \end{array}
\right)\end{eqnarray}
and 
\begin{eqnarray}
\rho_1=\left(
\begin{array}{cccc}
c_0^2 & 0 & -c_0c_2 & 0\\
0 & c_1^2 & 0 & -c_1c_3\\
-c_0c_2 & 0 & c_2^2 & 0\\
0 & -c_1c_3 & 0 & c_3^2 \end{array}\right)\end{eqnarray}
where we choose all the coefficients $c_i$ to be real.

\begin{figure}[h!]
  \centering
  \includegraphics[width=10cm]{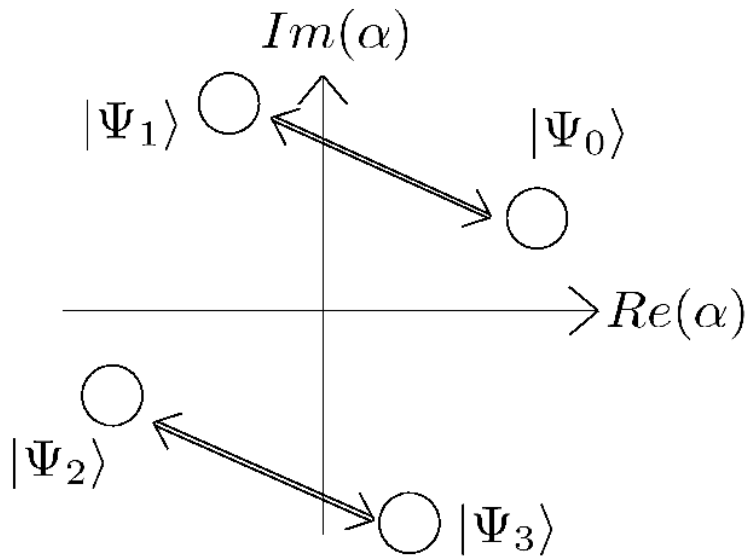}
  \caption{Pairing of the four symmetric states for the {\it bit value} mixed states}
  \label{bb84cases3}
\end{figure}

Thanks to Eqn.~(6.27) and Eqn.~(6.28), we clearly see that
\begin{eqnarray}
\rho_1=K \rho_0 K^\dagger= K^\dagger \rho_0 K.
\end{eqnarray} 
Moreover, we can calculate that $K \rho_0 K = K^\dagger \rho_0 K^\dagger$ in the following calculation.
\begin{eqnarray}
K \rho_0 K &=& \frac{1}{2} \left( K| \Psi_0 \rangle \langle \Psi_0 |K + K| \Psi_2 \rangle \langle \Psi_2 |K \right)\\
&=& \frac{1}{2} \left( | \Psi_1 \rangle \langle \Psi_3 | + | \Psi_3 \rangle \langle \Psi_1 | \right)\\
&=& \frac{1}{2} \left( K^\dagger | \Psi_2 \rangle \langle \Psi_2 |K^\dagger + K^\dagger| \Psi_0 \rangle \langle \Psi_0 |K^\dagger \right)\\
&=&K^\dagger \rho_0 K^\dagger.
\end{eqnarray} 
The consequence is that we can construct two new unitary matrices which are involution\footnote{A unitary transformation $U$ is called an involution if and only if $U^2={\mathbb 1}$.} such that $\rho_1=U_\pm\rho_0U_\pm$. This two involutions are given by
\begin{eqnarray}
U_\pm&=&\frac{K+K^\dagger}{2}\pm i \frac{K-K^\dagger}{2}=U_{\pm}^\dagger.\\
\end{eqnarray}
We can choose to use in the following calculation
\begin{eqnarray}
U=U_-=\left(
\begin{array}{cccc}
1 & 0 & 0 & 0\\
0 & 1 & 0 & 0\\
0 & 0 & -1 & 0\\
0 & 0 & 0 & -1 \end{array}\right).\end{eqnarray}
We have finally written the {\it basis} mixed states as $\rho_0$ and $\rho_1=U\rho_0U$ where $U^2={\mathbb 1}$. This means that the question 'With what probability can an eavesdropper unambiguously distinguish the {\it basis} of the signal?' is related to the unambiguous discrimination of two geometrically uniform mixed states in dimension four. The choice of such a involution matrix will simplify the next calculations.
Finally, in the four dimensional Hilbert space, we see that
\begin{eqnarray}
\rho_0+\rho_1=\left(
\begin{array}{cccc}
c_0^2& 0 & 0 & 0\\
0 & c_1^2 & 0 & 0\\
0 & 0 & c_2^2 & 0\\
0 & 0 & 0 & c_3^2 \end{array}\right)\end{eqnarray}
such that $rank(\rho_0+\rho_1)=4=rank(\rho_0)+rank(\rho_1)$. The two GU states $\rho_0$ and $\rho_1$ do not have overlapping supports and we can apply Theorem 19 about USD of such a pair of states.
The {\it bit value} mixed states (see Fig.~\ref{bb84cases3}) are also rank two matrices by construction. They can be written as
\begin{eqnarray}
\rho_0&=&\frac{1}{2} \left( | \Psi_0 \rangle \langle \Psi_0 | + | \Psi_1 \rangle \langle \Psi_1 | \right),\\
\rho_1&=&\frac{1}{2} \left( | \Psi_2 \rangle \langle \Psi_2 | + | \Psi_3 \rangle \langle \Psi_3| \right).
\end{eqnarray}

In terms of the coefficients $c_i$'s, we obtain the following form in the four dimensional Hilbert space spanned by the states $|\Psi_i \rangle$, $i=0,1,2,3$:
\begin{eqnarray}
\rho_0=\left(
\begin{array}{cccc}
c_0^2 & \frac{1-i}{2} c_0c_1 & 0 & \frac{1+i}{2} c_0c_3\\
\frac{1+i}{2} c_1c_0 & c_1^2 & \frac{1-i}{2} c_1c_2 & 0\\
0 & \frac{1+i}{2} c_2c_1 & c_2^2 & \frac{1-i}{2} c_2c_3\\
\frac{1-i}{2} c_3c_0 & 0 & \frac{1+i}{2} c_3c_2 & c_3^2 \end{array}\right)
\end{eqnarray}
and
\begin{eqnarray}
\rho_1=\left(
\begin{array}{cccc}
c_0^2 & -\frac{1-i}{2} c_0c_1 & 0 & -\frac{1+i}{2} c_0c_3\\
-\frac{1+i}{2} c_1c_0 & c_1^2 & -\frac{1-i}{2} c_1c_2 & 0\\
0 & -\frac{1+i}{2} c_2c_1 & c_2^2 & -\frac{1-i}{2} c_2c_3\\
-\frac{1-i}{2} c_3c_0 & 0 & -\frac{1+i}{2} c_3c_2 & c_3^2 \end{array}\right).\end{eqnarray}
It is unfortunately impossible to choose the phase of the coefficient $c_i$ so that $\rho_0$ and $\rho_1$ are real matrices. Therefore we simply choose all the coefficient $c_i$ to be real and we end up with

\begin{eqnarray}
\rho_0=\left(
\begin{array}{cccc}
c_0^2 & \frac{1-i}{2} c_0c_1 & 0 & \frac{1+i}{2} c_0c_3\\
\frac{1+i}{2} c_1c_0 & c_1^2 & \frac{1-i}{2} c_1c_2 & 0\\
0 & \frac{1+i}{2} c_2c_1 & c_2^2 & \frac{1-i}{2} c_2c_3\\
\frac{1-i}{2} c_3c_0 & 0 & \frac{1+i}{2} c_3c_2 & c_3^2 \end{array}\right)
\end{eqnarray}
and
\begin{eqnarray}
\rho_1=\left(
\begin{array}{cccc}
c_0^2 & -\frac{1-i}{2} c_0c_1 & 0 & -\frac{1+i}{2} c_0c_3\\
-\frac{1+i}{2} c_1c_0 & c_1^2 & -\frac{1-i}{2} c_1c_2 & 0\\
0 & -\frac{1+i}{2} c_2c_1 & c_2^2 & -\frac{1-i}{2} c_2c_3\\
-\frac{1-i}{2} c_3c_0 & 0 & -\frac{1+i}{2} c_3c_2 & c_3^2 \end{array}\right).\end{eqnarray}

The involution connected $\rho_0$ and $\rho_1$ simply is
\begin{eqnarray}
K^2=\left(
\begin{array}{cccc}
1 & 0 & 0 & 0\\
0 & -1 & 0 & 0\\
0 & 0 & 1 & 0\\
0 & 0 & 0 & -1 \end{array}\right).
\end{eqnarray}
Of course, the question 'With what probability can an eavesdropper unambiguously determine which {\it bit value} is sent without being interested in the knowledge of the basis?' is also related to the unambiguous discrimination of two geometrically uniform mixed states in dimension four. Here again, the sum $\rho_0+\rho_1$ in the four dimensional Hilbert space is given by
\begin{eqnarray}
\rho_0+\rho_1=\left(
\begin{array}{cccc}
c_0^2& 0 & 0 & 0\\
0 & c_1^2 & 0 & 0\\
0 & 0 & c_2^2 & 0\\
0 & 0 & 0 & c_3^2 \end{array}\right)\end{eqnarray}
implying that the two GU states $\rho_0$ and $\rho_1$ do not have overlapping supports. Consequently Theorem 19 can be used.

\begin{figure}[h!]
  \centering
  \includegraphics[width=10cm]{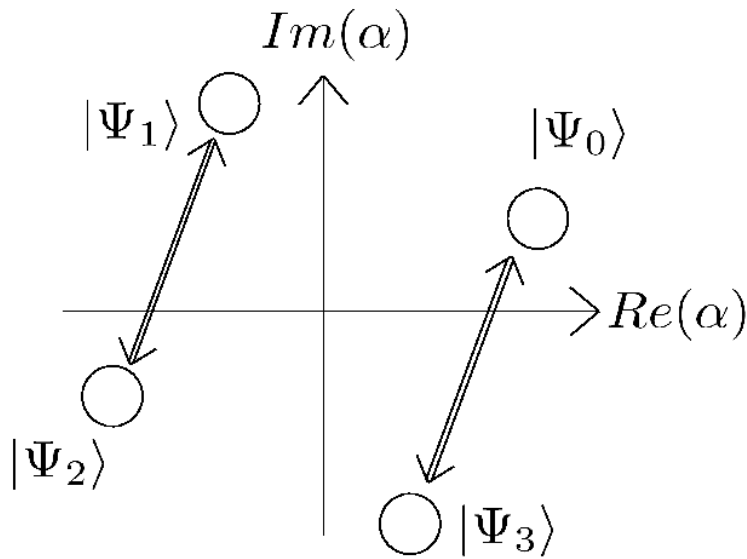}
  \caption{Third possible pairing of the four symmetric states}
  \label{bb84cases4}
\end{figure}

Actually one could consider a third USD problem coming from the pairing of the four states $\Psi_i$ (see Fig.~\ref{bb84cases4}). This last case is concerned with the unambiguous discrimination of the two mixed states $\rho_0=\frac{1}{2} \left( | \Psi_0 \rangle \langle \Psi_0 | + | \Psi_3 \rangle \langle \Psi_3 | \right)$  and $\rho_1=\frac{1}{2} \left( | \Psi_1 \rangle \langle \Psi_1 | + | \Psi_2 \rangle \langle \Psi_2| \right)$ but this case is similar\footnote{unitary equivalent} to the previous case. Indeed one can go from the former to the later case by using the unitary $K^2$. This is not the case between the two problems of unambiguously discriminating the {\it basis} states and the {\it bit value} states.

\section{USD of the {\it basis} mixed states}
Let us repeat that the two density matrices to unambiguously discriminate are
\begin{eqnarray}
\rho_0=\left(
\begin{array}{cccc}
c_0^2 & 0 & c_0c_2 & 0\\
0 & c_1^2 & 0 & c_1c_3\\
c_0c_2 & 0 & c_2^2 & 0\\
0 & c_1c_3 & 0 & c_3^2 \end{array}
\right)\end{eqnarray}
and 
\begin{eqnarray}
\rho_1=U\rho_0U=\left(
\begin{array}{cccc}
c_0^2 & 0 & -c_0c_2 & 0\\
0 & c_1^2 & 0 & -c_1c_3\\
-c_0c_2 & 0 & c_2^2 & 0\\
0 & -c_1c_3 & 0 & c_3^2 \end{array}\right).\end{eqnarray}
with 
\begin{eqnarray}
U=\left(
\begin{array}{cccc}
1 & 0 & 0 & 0\\
0 & 1 & 0 & 0\\
0 & 0 & -1 & 0\\
0 & 0 & 0 & -1 \end{array}\right).
\end{eqnarray}

With a bit of concentration, one can realize that these two density matrices are block diagonal. Indeed, we can use the permutation matrix
\begin{eqnarray}
P=\left(
\begin{array}{cccc}
1 & 0 & 0 & 0\\
0 & 0 & 1 & 0\\
0 & 1 & 0 & 0\\
0 & 0 & 0 & 1 \end{array}\right)\end{eqnarray}
and obtain
\begin{eqnarray}
P\rho_{0,1}P=\left(
\begin{array}{cccc}
c_0^2 & \pm c_0c_2 & 0  & 0\\
\pm c_0c_2 & c_2^2 & 0 &  0\\
0 & 0 & c_1^2 & \pm c_1c_3\\
0 & 0  & \pm c_1c_3 & c_3^2 \end{array}\right).
\end{eqnarray}
This already tells us that we can analytically solve this problem which is reducible to some pure states case. Indeed $\rho_0$ and $\rho_1$ are block diagonal where each block is two dimensional. We will nevertheless use the non reduced density matrices to find the optimal USD measurement. The reason is that, as we will in the next paragraph, we can compute the operator $\rho_0 - F_0$ and check its positivity for any value of the amplitude $\alpha$. Note here that the spectra of $\rho_0 - F_0$ and $\rho_1 -F_1$ are identical since $\rho_1-F_1=\rho_0-F_0$ for two GU states. With that, we have the optimal failure probability as soon as the optimal measurement. Again, we could use the second and third reduction theorems but the present example gives us the opportunity to use other tools.

We now focus our attention onto $\rho_0$ only since $\rho_1$ is similar to it. The density matrix
\begin{eqnarray}
P\rho_0P=\left(
\begin{array}{cccc}
c_0^2 & c_0c_2 & 0  & 0\\
 c_0c_2 & c_2^2 & 0 &  0\\
0 & 0 & c_1^2 &  c_1c_3\\
0 & 0  &  c_1c_3 & c_3^2 \end{array}\right)
\end{eqnarray}
can be easily diagonalized using the block diagonal unitary matrices
\begin{eqnarray}
PU_0P=\left(
\begin{array}{cccc}
\frac{c_0}{\sqrt{c_0^2 + |c_0^2}} & \frac{c_2}{\sqrt{c_0^2 + c_0^2}} & 0  & 0\\ \frac{c_2}{\sqrt{c_0|^2 + c_0^2}} & \frac{-c_0}{\sqrt{c_0^2 + c_0^2}} & 0 &  0\\
0 & 0 & \frac{c_1}{\sqrt{c_1^2 + c_3^2}} &  \frac{c_3}{\sqrt{c_1^2 + c_3^2}}\\
0 & 0  &  \frac{c_3}{\sqrt{c_1^2 + c_3^2}} & \frac{-c_1}{\sqrt{c_1^2 + c_3^2}} \end{array}\right).
\end{eqnarray}
If is not too difficult to find that the eigenvalues of $P\rho_0P$ are therefore given by
\begin{eqnarray}
\lambda_0=c_0^2+c_2^2\\
\lambda_1=c_1^2+c_3^2
\end{eqnarray}
which gives, in terms of the mean photon number $\mu$
\begin{eqnarray}
\lambda_{0,1}=\frac{1\pm e^{-\mu}}{2}.
\end{eqnarray}
If we undo everywhere the permutation matrix $P$, the density matrices $\rho_{0,1}$ can obviously be diagonalized with the help of the unitary transformation
\begin{eqnarray}
U_0=\left(
\begin{array}{cccc}
\frac{c_0}{\sqrt{c_0^2 + c_0^2}} & 0 &  \frac{c_2}{\sqrt{c_0^2 + c_0^2}} & 0\\
0 & \frac{c_1}{\sqrt{c_1^2 + c_3^2}} & 0 &   \frac{c_3}{\sqrt{c_1^2 + c_3^2}}\\
 \frac{c_2}{\sqrt{c_0^2 + c_0^2}} & 0 & \frac{-c_0}{\sqrt{c_0^2 + c_0^2}} & 0\\
0 &  \frac{c_3}{\sqrt{c_1^2 + c_3^2}} & 0 & \frac{-c_1}{\sqrt{c_1^2 + c_3^2}} \end{array}\right).
\end{eqnarray}
and its square root takes the form
\begin{eqnarray}
\sqrt{\rho_0}=\left(
\begin{array}{cccc}
\frac{c_0^2}{\sqrt{c_0^2 + c_0^2}} & 0 & \frac{c_0c_2 }{\sqrt{c_0^2 + c_0^2}}& 0\\
0 &  \frac{c_1^2}{\sqrt{c_1^2 + c_3^2}}& 0 &  \frac{c_1c_3}{\sqrt{c_1^2 + c_3^2}}\\
 \frac{c_0c_2}{\sqrt{c_0^2 + c_0^2}} & 0 &  \frac{c_2^2}{\sqrt{c_0^2 + c_0^2}}& 0\\
0 & \frac{c_1c_3 }{\sqrt{c_1^2 + c_3^2}}& 0 &  \frac{c_3^2}{\sqrt{c_1^2 + c_3^2}}\end{array}.
\right)\end{eqnarray}
The next step is to calculate the operator $F_0=\sqrt{\sqrt{\rho_0}\rho_1\sqrt{\rho_0}}$. Our two GU states are related through the relation $\sqrt{\rho_1}=U\sqrt{\rho_0}U$. As a result, the equality $\sqrt{\rho_0}\sqrt{\rho_1}=F_0V$ leads to
\begin{eqnarray}
\sqrt{\rho_0}U \sqrt{\rho_0}=F_0VU.
\end{eqnarray}
In the $\rho_0$'s eigenbasis, we obtain
\begin{eqnarray}
U_0\sqrt{\rho_0}U\sqrt{\rho_0}U_0=U_0F_0VUU_0=U_0F_0U_0T
\end{eqnarray}
where $T=U_0VUU_0$ is a unitary transformation.
One can calculate the operator $U_0\sqrt{\rho_0}U\sqrt{\rho_0}U_0$ and find
\begin{eqnarray}
\left(
\begin{array}{cccc}
c_0^2 - c_2^2 & 0 & 0  & 0\\
0 & c_1^2 - c_3^2 & 0 &  0\\
0 & 0 & 0 & 0\\
0 & 0 & 0& 0 \end{array}\right)
\end{eqnarray}
which is always positive if multiplied by some signature matrix
\begin{eqnarray}
T=\left(
\begin{array}{cccc}
\pm 1 & 0 & 0  & 0\\
0 & \pm 1 & 0 &  0\\
0 & 0 & 0 & 0\\
0 & 0 & 0& 0 \end{array}\right).
\end{eqnarray}
Note here that, in terms of the mean photon number $\mu$, the quantities $c_0^2-c_2^2=e^{\frac{-\mu}{2}}cos\frac{\mu}{2}$ and $c_1^2-c_3^2=e^{\frac{-\mu}{2}}sin\frac{\mu}{2}$ are not always positive.
In the end, the positive operator $F_0$ is of the form
\begin{eqnarray}
U_0FU_0=\left(
\begin{array}{cccc}
|c_0^2 - c_2^2| & 0 & 0  & 0\\
0 & |c_1^2 - c_3^2| & 0 &  0\\
0 & 0 & 0 & 0\\
0 & 0 & 0& 0 \end{array}\right).
\end{eqnarray}
The explicit form of the unitary $V$ is only relevant to calculate the elements of the optUSDM. But our first goal is to find the spectrum of the operator $\rho_0-F_0$. For that, four cases are to take into account depending on the sign of $c_0^2-c_2^2$ and $c_1^2-c_3^2$.\\

\begin{figure}[h!]
  \centering
  \includegraphics[width=10cm]{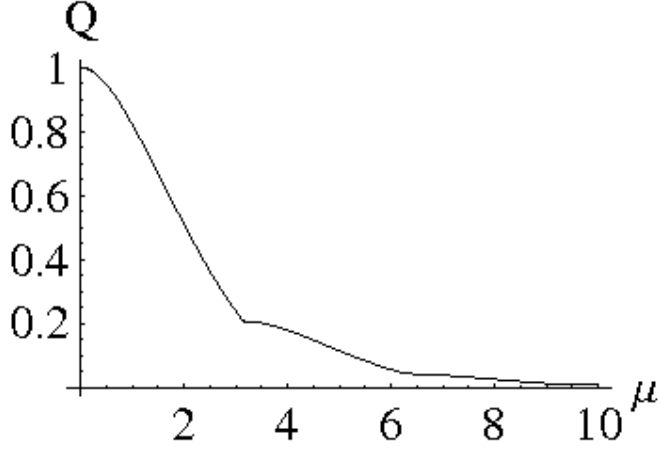}
  \caption{Optimal failure probability for USD of the {\it basis} mixed states}
  \label{bb84q1}
\end{figure}

Everything is gathered to obtain the explicit form the operator $\rho_0-F_0$ in the eigenbasis of $\rho_0$. Indeed, we have
\begin{eqnarray}
U_0(\rho_0-F_0)U_0&=&\left(
\begin{array}{cccc}
c_0^2 + c_2^2 & 0 & 0  & 0\\
0 & c_1^2 + c_3^2 & 0 &  0\\
0 & 0 & 0 & 0\\
0 & 0 & 0& 0 \end{array}\right)
+
\left(
\begin{array}{cccc}
|c_0^2 - c_2^2| & 0 & 0  & 0\\
0 & |c_1^2 - c_3^2| & 0 &  0\\
0 & 0 & 0 & 0\\
0 & 0 & 0& 0 \end{array}\right)\\
&=&2\left(
\begin{array}{cccc}
max\{c_0^2, c_2^2\} & 0 & 0  & 0\\
0 & max\{c_1^2, c_3^2\} & 0 &  0\\
0 & 0 & 0 & 0\\
0 & 0 & 0& 0 \end{array}\right) \ge 0.
\end{eqnarray}
The spectrum of the operator $\rho_0-F0$ is positive for any value of the mean photon number $\mu$. As a consequence, the optimal failure probability $Q$ reaches the lower bounds $F=\Tr(F_0)=|c_0^2 - c_2^2| +|c_1^2 - c_3^2|$. In terms of the mean photon number $\mu$ (see Fig.~\ref{bb84q1}), the optimal failure probability is given by
\begin{eqnarray}
Q=e^{\frac{-\mu}{2}}\left(|cos\frac{\mu}{2}|+|sin\frac{\mu}{2}|\right).
\end{eqnarray}

Let us note here that if we were interested in the unambiguous discrimination of
\begin{eqnarray}
\rho_0&=&\frac{1}{2}\left(|\alpha \rangle \langle \alpha|+|-\alpha \rangle \langle -\alpha|\right)\\
{\textrm and} \,\, \rho_1&=&\frac{1}{2}\left(|i\alpha \rangle \langle i\alpha|+|-i\alpha \rangle \langle -i\alpha|\right).
\end{eqnarray} then we would find
\begin{eqnarray}
Q=e^{-\mu}\left(|cos\mu|+|sin\mu|\right).
\end{eqnarray}

Let us conclude this section and this example by adding that we can give the optimal measurement to achieve $Q=F$. Indeed, the useful matrix $\Sigma$ is diagonal and therefore its inverse simply is
\begin{eqnarray}
\Sigma^{-1}=\left(
\begin{array}{cccc}
c_0^{-2} & 0 & 0 & 0\\
0 & c_1^{-2} & 0 & 0\\
0 & 0 & c_2^{-2} &0\\
0 & 0 & 0 & c_3^{-2} \end{array}\right).
\end{eqnarray}
In the four different cases parametrized by the signature $T$, the elements of the optimal POVM are finally given by
\begin{eqnarray}
E_0&=&\Sigma^{-1}\sqrt{\rho_0}(\rho_0-F_0)\sqrt{\rho_0}\Sigma^{-1}\\
E_1&=&UE_0U\\
E_?&=&\Sigma^{-1}(\sqrt{\rho_0}+\sqrt{\rho_1}V^\dagger)F_0(\sqrt{\rho_0}+V\sqrt{\rho_1})\Sigma^{-1}
\end{eqnarray}
where all the different matrices involved in these equations are perfectly known. This concludes this section and the first example.

\section{USD of the {\it bit value} mixed states}

The second case corresponds to the unambiguous discrimination of the two density matrices
\begin{eqnarray}
\rho_0=\left(
\begin{array}{cccc}
c_0^2 & \frac{1-i}{2} c_0c_1 & 0 & \frac{1+i}{2} c_0c_3\\
\frac{1+i}{2} c_1c_0 & c_1^2 & \frac{1-i}{2} c_1c_2 & 0\\
0 & \frac{1+i}{2} c_2c_1 & c_2^2 & \frac{1-i}{2} c_2c_3\\
\frac{1-i}{2} c_3c_0 & 0 & \frac{1+i}{2} c_3c_2 & c_3^2 \end{array}\right)
\end{eqnarray}
and
\begin{eqnarray}
\rho_1=U\rho_0U=\left(
\begin{array}{cccc}
c_0^2 & -\frac{1-i}{2} c_0c_1 & 0 & -\frac{1+i}{2} c_0c_3\\
-\frac{1+i}{2} c_1c_0 & c_1^2 & -\frac{1-i}{2} c_1c_2 & 0\\
0 & -\frac{1+i}{2} c_2c_1 & c_2^2 & -\frac{1-i}{2} c_2c_3\\
-\frac{1-i}{2} c_3c_0 & 0 & -\frac{1+i}{2} c_3c_2 & c_3^2 \end{array}\right)\end{eqnarray}
with 
\begin{eqnarray}
U=K^2=\left(
\begin{array}{cccc}
1 & 0 & 0 & 0\\
0 & -1 & 0 & 0\\
0 & 0 & 1 & 0\\
0 & 0 & 0 & -1 \end{array}\right).
\end{eqnarray}
This USD task is far more complicated than the first one. It is difficult to find the unitary transformations to diagonalize $\rho_0$ and $\rho_1$ and therefore the square root of those states as well as $F_0$ and $F_1$ cannot be easily expressed. We have to resort to a particular decomposition of the two states $\rho_0$ and $\rho_1$. This decomposition allows us to diagonalize the operator $\rho_0-F_0$ in an unknown basis and find its spectrum. First we review some relevant properties of the density matrices $\rho_0$ and $\rho_1$. Next, we solve the unambiguous discrimination of these two GU states.\\

Actually one can write
\begin{eqnarray}
\rho_0=APA
\end{eqnarray}
where $A$ is a real diagonal matrix and $P=\frac{P^2}{2}$ a pseudo projector. They are defined as
\begin{eqnarray}
A=\left(
\begin{array}{cccc}
c_0 & 0 & 0 & 0\\
0 & c_1 & 0 & 0\\
0 & 0 & c_2 &0\\
0 & 0 & 0 & c_3 \end{array}\right)
\end{eqnarray}
and
\begin{eqnarray}
P=\left(
\begin{array}{cccc}
1 & \frac{1-i}{2} & 0 & \frac{1+i}{2}\\
\frac{1+i}{2}  & 1 & \frac{1-i}{2} & 0\\
0 & \frac{1+i}{2} & 1 & \frac{1-i}{2} \\
\frac{1-i}{2} & 0 & \frac{1+i}{2} & 1\end{array}\right).
\end{eqnarray}
Here come three remarks arising from this decomposition. First of all, let us note that they commute since they are both diagonal. Due to the symmetry between $\rho_0$ and $\rho_1$ and to the commutation between $A$ and $U$, we have $\rho_1=UAPAU=AUPUA$.
Second of all, we can consider the sum of the two density matrices $\rho_0$ and $\rho_1$. We have $\rho_0+\rho_1=APA+AUPUA=A(P+UPU)A$ and $P+UPU=2{\mathbb 1}$. Thus
\begin{eqnarray}
\rho_0+\rho_1=2A^2
\end{eqnarray}
and we could denote $A=\sqrt{\frac{\Sigma}{2}}$. The last remark is the more important. Actually $\Tr(P)=4$. This is not a lot but it implies that $P$ is equal to twice a two-dimensional projector. As a matter of fact, there exists a unitary transformation $W$ so that
\begin{eqnarray}
WPW^\dagger=\left(
\begin{array}{cccc}
2 &  & 0 & 0\\
0 & 0 & 0 & 0\\
0 & 0& 0 & 0 \\
0 & 0 & 0 & 2\end{array}\right).
\end{eqnarray}
Such a unitary matrix can be given by the Discrete Fourier Transform
\begin{eqnarray}
W=\frac{1}{2}\left(
\begin{array}{cccc}
1 & 1& 1 & 1\\
1 & i & -1 & -i\\
1 & -1& 1 & -1 \\
1 & -i & -1 & i\end{array}\right).
\end{eqnarray}

The interest of the decomposition provide in Eqn.(6.78) is that it allows us to write $\rho_0=APA$ in an unknown basis better suited to investigate the spectrum of the operator $\rho_0-F_0$. Indeed we can write
\begin{eqnarray}
\rho_0&=&APA\\
&=&\frac{AP}{\sqrt{2}}\frac{PA}{\sqrt{2}}\\
&=&\sqrt{\rho_0}R_0^\dagger R_0\sqrt{\rho_0}
\end{eqnarray}
where we introduce the unitary transformation $R_0$ such that $\frac{PA}{\sqrt{2}}=R_0\sqrt{\rho_0}$. Consequently, we obtain
\begin{eqnarray}
R_0\rho_0R_0^\dagger &=&\frac{PA}{\sqrt{2}}\frac{AP}{\sqrt{2}}\\
&=&\frac{PA^2P}{2}
\end{eqnarray}
For the unambiguous discrimination of the two {\it basis} mixed states, we knew the unitary transformation $U_0$ that diagonalizes $\rho_0$. It was possible to write $F_0$ and finally express the operator $\rho_0-F_0$. Here we can not directly work with the eigenbasis of $\rho_0$. Instead, we try to use the matrix $R_0\rho_0R_0^\dagger$, knowing only the existence of this unitary transformation $R_0$. We are only interested in the spectrum of $\rho_0-F_0$ and the precise form of $R_0$ is finally irrelevant as long as it permits us to find the spectrum of $\rho_0-F_0$. Nevertheless, we must say, that the explicit expression of the POVM elements will not be provided since, it that case, we do need to know $R_0$. Moreover, as we will soon see, we will not be able to calculate the complete expression of $Q$ for all the regime of the mean photon number $\mu$.

\begin{figure}[h!]
  \centering
  \includegraphics[width=10cm]{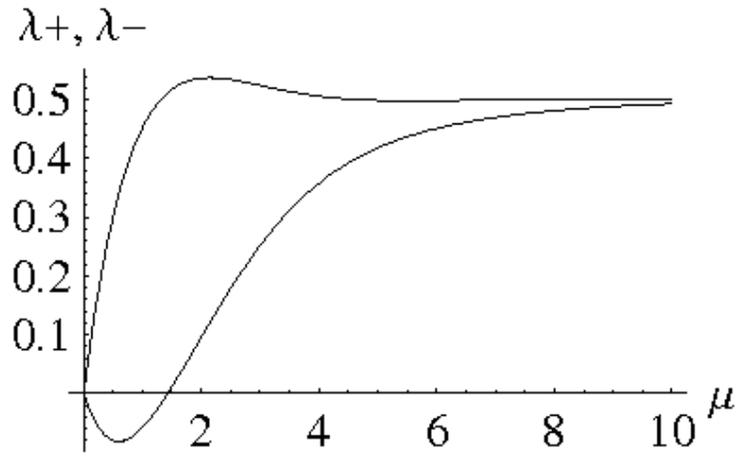}
  \caption{Spectrum of the operator $\rho_0-F_0$ for USD of the {\it bit value} mixed states}
  \label{spectrum1}
\end{figure}

Let us now calculate the spectrum of $\rho_0-F_0$. We first apply the Fourier Transform $W$ onto $R_0\rho_0R_0^\dagger$ to end up with
\begin{eqnarray}
WR_0\rho_0R_0^\dagger W^\dagger&=&\frac{1}{2}WPA^2PW^\dagger\\
&=&\frac{1}{2}\left(
\begin{array}{cccc}
c_0^2+c_1^2+c_2^2+c_3^2 & 0& 0 & c_0^2+i c_1^2-c_2^2-ic_3^2\\
0 & 0 & 0 & 0\\
0 & 0& 0 & 0 \\
c_0^2-ic_1^2-c_2^2+i c_3^2& 0 & 0 & c_0^2+c_1^2+c_2^2+c_3^2\end{array}\right).
\end{eqnarray}
Actually, since the states $\rho_0$ is normalized, we have $c_0^2+c_1^2+c_2^2+c_3^2=1$ and therefore
\begin{eqnarray}
WR_0\rho_0R_0^\dagger W^\dagger=\frac{1}{2}\left(
\begin{array}{cccc}
1 & 0& 0 & \Lambda \\
0 & 0 & 0 & 0\\
0 & 0& 0 & 0 \\
\Lambda^* & 0 & 0 & 1\end{array}\right).
\end{eqnarray}
where
\begin{eqnarray}
\Lambda=(c_0^2-c_2^2)+i (c_1^2-c_3^2).
\end{eqnarray}
In fact, a Hermitian matrix of the form
\begin{eqnarray}
\left(\begin{array}{cc}
a &  b e^{i \phi}\\
b e^{-i \phi}& a\end{array}\right)
\end{eqnarray}
with $a$, $b$ and $\phi$ real and positive, has for eigenvalues
\begin{eqnarray}
\lambda_{\pm}=a \pm b
\end{eqnarray} and for eigenvectors
\begin{eqnarray}
|v_\pm \rangle=\frac{1}{\sqrt{2}}\left(\begin{array}{c}
\pm e^{i \phi}\\
1 \end{array}\right).
\end{eqnarray}
Here we are only interested in the spectrum of $\rho_0$. The formula above gives us its eigenvalues as
\begin{eqnarray}
\lambda_\pm&=&\frac{1\pm|\Lambda|}{2}\\
&=&\frac{1\pm e^{-\frac{\mu}{2}}}{2}.
\end{eqnarray}
As for the {\it basis} mixed states case where we calculate the operator $U_0\sqrt{\rho_0}U\sqrt{\rho_0}U_0$, we now consider the operator $WR_0\sqrt{\rho_0}U\sqrt{\rho_0}R_0^\dagger W^\dagger$. Actually this operator is of a similar form than $WR_0\rho_0R_0^\dagger W^\dagger$. Indeed we obtain
\begin{eqnarray}
WR_0\sqrt{\rho_0}U\sqrt{\rho_0}R_0^\dagger W^\dagger=\frac{1}{2}\left(
\begin{array}{cccc}
(c_1^2+c_3^2)-(c_0^2+c_2^2) & 0& 0 & - \Lambda^*\\
0 & 0 & 0 & 0\\
0 & 0& 0 & 0 \\
- \Lambda & 0 & 0 & (c_1^2+c_3^2)-(c_0^2+c_2^2)\end{array}\right).
\end{eqnarray}
Thanks to Eqn.(6.94), we find that its eigenvalues are
\begin{eqnarray}
\gamma_\pm=(c_1^2+c_3^2)-(c_0^2+c_2^2)\pm|\Lambda|.
\end{eqnarray}
Moreover, with the help of Eqn.(6.95), we obtain the unitary that diagonalizes the operator $WR_0\sqrt{\rho_0}U\sqrt{\rho_0}R_0^\dagger W^\dagger$. This unitary is of form
\begin{eqnarray}
K^\dagger=\frac{1}{\sqrt{2}}\left(
\begin{array}{cccc}
-\frac{\Lambda^*}{|\Lambda|} & 0& 0 & \frac{\Lambda^*}{|\Lambda|}\\
0 & \sqrt{2} & 0 & 0\\
0 & 0& \sqrt{2} & 0 \\
1 & 0 & 0 & 1\end{array}\right).
\end{eqnarray}
If we replace the coefficients $c_i$ by their expressions in term of the mean photon number $\mu$, we end up with
\begin{eqnarray}
KWR_0\sqrt{\rho_0}U\sqrt{\rho_0}R_0^\dagger W^\dagger K^\dagger=\frac{1}{2}\left(
\begin{array}{cccc}
-e^{-\mu} + e^{\frac{-\mu}{2}}& 0& 0 & 0\\
0 & 0 & 0 & 0\\
0 & 0& 0 & 0 \\
0 & 0 & 0 & -e^{-\mu} - e^{\frac{-\mu}{2}}\end{array}\right).
\end{eqnarray}
The eigenvalues in the top left corner is always positive while the eigenvalue in the bottom right corner is always negative. Therefore the operator $F_0$ in its eigenbasis is of the form
\begin{eqnarray}
K W R_0 F_0 R_0 W^\dagger K^\dagger&=&KWR_0\sqrt{\rho_0}U\sqrt{\rho_0}R_0^\dagger W^\dagger K^\dagger T\\
&=&\frac{1}{\sqrt{2}}\left(
\begin{array}{cccc}
e^{\frac{-\mu}{2}} -e^{-\mu}& 0& 0 & 0\\
0 & 0 & 0 & 0\\
0 & 0& 0 & 0 \\
0 & 0 & 0 & e^{\frac{-\mu}{2}} + e^{-\mu} \end{array}\right)
\end{eqnarray}
and the unitary matrix $V$ equals $R_0^\dagger W^\dagger K^\dagger TKWR_0U$, where $T$ is the signature
\begin{eqnarray}
T=\left(
\begin{array}{cccc}
1 & 0 & 0  & 0\\
0 & 1 & 0 &  0\\
0 & 0 & 1 & 0\\
0 & 0 & 0& -1 \end{array}\right).
\end{eqnarray}
We have now all the necessary matrices to calculate the operator $\rho_0-F_0$ in the $F_0$'s eigenbasis. We obtain
\begin{eqnarray}
K W R_0 (\rho_0-F_0) R_0 W^\dagger K^\dagger&=&KWR_0\rho_0R_0^\dagger W^\dagger K^\dagger -KWR_0\sqrt{\rho_0}U\sqrt{\rho_0})R_0^\dagger W^\dagger K^\dagger T\\ \nonumber
&=&KWPA^2P W^\dagger K^\dagger -KWPAUAP W^\dagger K^\dagger T\\ \nonumber
&=&e^{\frac{-\mu}{2}}\left(
\begin{array}{cccc}
Cosh(\frac{\mu}{2})-Cos^2(\frac{\mu}{2})& 0& 0 & -iSin^2(\mu)\\
0 & 0 & 0 & 0\\
0 & 0& 0 & 0 \\
iSin^2(\mu) & 0 & 0 & Sinh(\frac{\mu}{2})-Sin^2(\frac{\mu}{2})\end{array}\right).
\end{eqnarray}
We are very closed to find the spectrum of $\rho_0-F_0$. We can denote by $M$ the previous matrix. The eigenvalues of this matrix $M$ are given by the roots of the polynomial $P(x)=x^2-\Tr(M)x+Det(M)$ which simply are
\begin{eqnarray}
x{\pm}=\frac{1}{2}\left(\Tr(M)\pm\sqrt{\Tr(M)^2-4 Det(M)}\right).
\end{eqnarray}
All this complicated construction was necessary to obtain the spectrum of the operator $\rho_0-F_0$. We victoriously end up with
\begin{eqnarray}
Spect(\rho_0-F_0)=\frac{1}{2}\left(1-e^{\frac{-\mu}{2}} \pm e^{-\mu}\sqrt{1+e^{\mu}-2e^{\frac{\mu}{2}}Cos(\mu)}\right).
\end{eqnarray}
This spectrum is not always positive (see Fig.~\ref{spectrum1}).

Only in the regime of relatively large $\mu$, the quantity $\frac{1}{2}(1-e^{\frac{-\mu}{2}} - e^{-\mu}\sqrt{1+e^{-\mu}-2e^{\frac{-\mu}{2}}Cos(\mu)}$ is greater than $0$. More precisely,
\begin{eqnarray}
Spect(\rho_0-F_0) \ge 0 \Leftrightarrow \mu \ge \mu_0\thickapprox1.4386
\end{eqnarray}
where $\mu_0$ is the solution of the equation $\frac{1}{2}\left(1-e^{\frac{-\mu}{2}} - e^{-\mu}\sqrt{1+e^{-\mu}-2e^{\frac{-\mu}{2}}Cos(\mu)}\right)=0$.\\

\begin{figure}[h!]
  \centering
  \includegraphics[width=10cm]{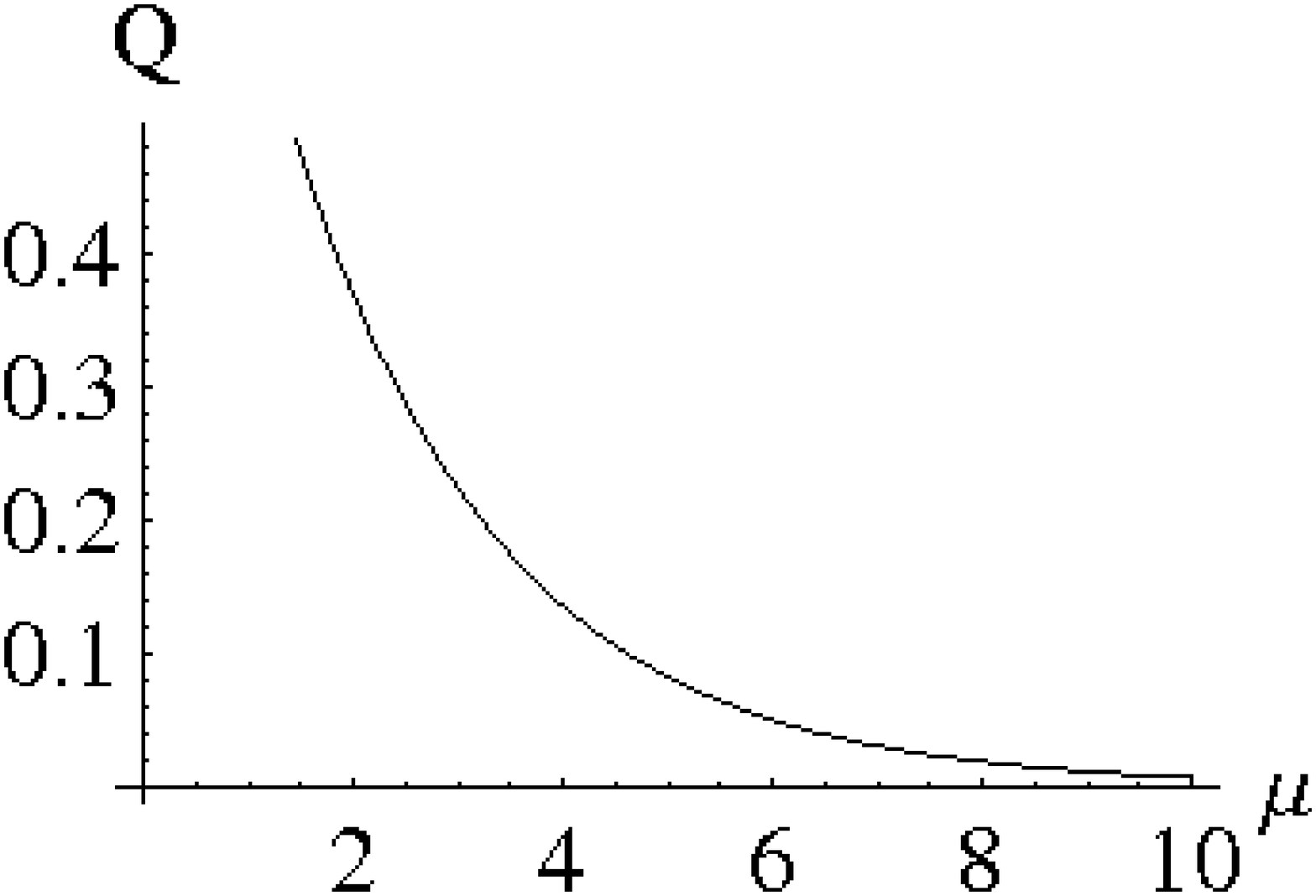}
  \caption{Optimal failure probability for USD of the {\it bit value} mixed states for $\mu \ge \mu_0$}
  \label{bb84q2}
\end{figure}

In the regime $\mu \ge \mu_0$ (see Fig.~\ref{bb84q2}), the optimal failure probability reaches the overall lower bound and we therefore get
\begin{eqnarray}
Q=F=\Tr(F_0)=e^{\frac{-\mu}{2}}.
\end{eqnarray}

The corresponding optimal measurement is moreover given by
\begin{eqnarray}
E_0&=&\Sigma^{-1} \sqrt{\rho_0} \left(\rho_0- F_0 \right) \sqrt{\rho_0}\Sigma^{-1} \\ \nonumber
E_1&=&U E_0 U \\ \nonumber
E_?&=&{\mathbb 1}- E_0 - U E_0 U. 
\end{eqnarray}
Note that for $\mu=\mu_0$, the POVM elements $E_0$ and $E_1$ have rank $1$ since one eigenvalue of $\rho_0-F_0$ vanishes.\\

We can remark here again that if we wanted to unambiguously discriminate
\begin{eqnarray}
\rho_0&=&\frac{1}{2}\left(|\alpha \rangle \langle \alpha|+|i\alpha \rangle \langle i\alpha|\right)\\
{\textrm and} \,\, \rho_1&=&\frac{1}{2}\left(|-\alpha \rangle \langle -\alpha|+|-i\alpha \rangle \langle -i\alpha|\right).
\end{eqnarray} then we would find for $\mu \ge 0.7193$
\begin{eqnarray}
Q=e^{-\mu}.
\end{eqnarray}\\

\begin{figure}[h!]
  \centering
  \includegraphics[width=10cm]{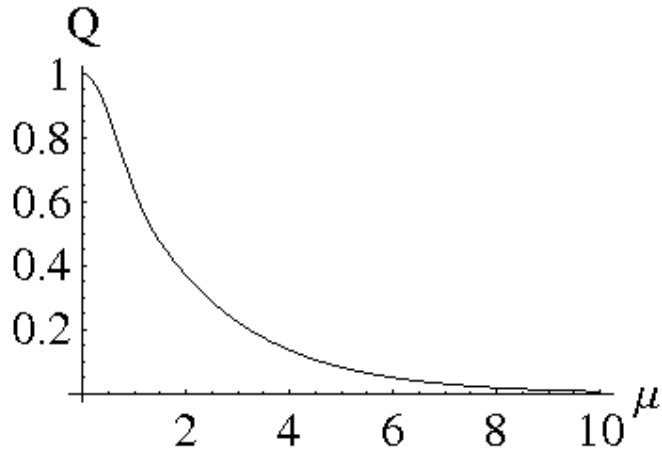}
  \caption{Optimal failure probability for USD of the {\it bit value} mixed states}
  \label{bb84q4}
\end{figure}

In the regime $\mu \le \mu_0$ where the operator $\rho_0-F_0$ is not positive, we have to check the spectrum of the operator $P_1^\perp U P_1^\perp$.
It is actually, as far as we know, not possible to calculate analytically its spectrum. Even if it is not really satisfying, we compute numerically the spectrum of $P_1^\perp U P_1^\perp$. It turns out that it always has two eigenvalues of opposite sign in the regime $\mu \le \mu_0$. Consequently, we can write the operator $P^{\perp}_1 U P^{\perp}_1$ in its eigenbasis $\{|0\rangle , |1\rangle \}$ as
\begin{eqnarray}
P^{\perp}_1 U P^{\perp}_1=a|0\rangle\langle0|-b|1\rangle\langle1|, \,\,\, a,b \in {\mathbb R}^+.
\end{eqnarray}
And in virtue of Theorem 19, the optimal failure probability (see Fig.~\ref{bb84q4}) for unambiguously discriminating the {\it bit value} mixed states is
\begin{eqnarray}
Q^{\mathrm{opt}} = 1-\frac{1}{a+b}(b\langle0|\rho_0|0\rangle+a\langle1|\rho_0|1\rangle+2\sqrt{ab}|\langle0|\rho_0|1\rangle|).
\end{eqnarray}

\begin{figure}[h!]
  \centering
  \includegraphics[width=10cm]{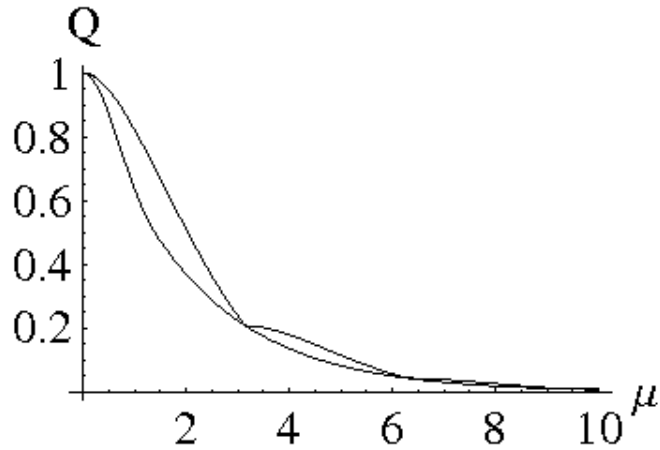}
  \caption{Comparison between the optimal failure probabilities for USD of the {\it basis} and the {\it bit value} mixed states}
  \label{bb84qccl}
\end{figure}

So far, no neat expression in terms of $\mu$ is known for this optimal failure probability $Q^{\mathrm{opt}}$ for $\mu \le \mu_0$ even if we do know its structure. This comes from the rather complicated form of the states $\rho_0$ and $\rho_1$. As a final word, let us add that the optimal USD measurement is of form
\begin{eqnarray}
\begin{array}{l}
E_0=|x \rangle \langle x |\\
E_1=U E_0 U \\
E_?={\mathbb 1}- E_0 - U E_0 U 
\end{array}
\,\,{\textrm with}\,\,
|x \rangle=\left(
\begin{array}{c}
\frac{e^{-iArg(\langle 1 | \rho_0| 0 \rangle)}}{\sqrt{1+a/b}}\\
\frac{1}{\sqrt{1+b/a}}\\
0\\
0
\end{array}
\right),
\end{eqnarray}
even here also, we can note write them in term of the mean photon number $\mu$. On the last graph \ref{bb84qccl}, we can show and compare the two optimal failure probabilities derived in this chapter.

This conclude the last chapter of this thesis.\\

This last example might appear a bit unsatisfactory to the reader since no analytical expression for $P^{\perp}_1 U P^{\perp}_1$ is known. However this is exactly the contrary. During my work on Unambiguous State Discrimination, I was guided by the four density matrices presented in this chapter. They were my inspiration as well as my life ring. They are actually at the core of the derivation of the two classes of exact solutions and the numerous theorems derived in this thesis would not have been found without them. 

%% file: conclusion.tex
\chapter{Epilogue} \label{epilogue}
The main results of this thesis are, first, the two classes of exact solutions, second the reduction theorems, and finally the solution to unambiguous comparison of $n$ pure states having some simple symmetry and the application of our results on USD to a BB84-type protocol.\\

There are actually two directions for research in USD. The first path is of course the derivation of new solutions. The second is to find new applications of the already known solutions. In this thesis, we have tried to follow both paths. On one hand, we have derived new tools and new classes of exact solutions. On the other hand, we have given two examples of application for our tools.\\
 
With respect to the newly developed tools, we have presented the notion of parallel addition $\rho_0 \Sigma^{-1} \rho_1$ in the context of unambiguous state discrimination. We have also shown the relevance of the two operators $\sqrt{\sqrt{\rho_0}\rho_1\sqrt{\rho_0}}$ and $\sqrt{\sqrt{\rho_1}\rho_0\sqrt{\rho_1}}$. We have finally provided two new classes of exact solutions as well as the three reduction theorems as we now discuss.\\

The two classes of exact solutions derived in this thesis are the only two analytical solutions for unambiguous discrimination of two generic density matrices known so far. There now exist six analytical solutions for optimal unambiguous discrimination of quantum states. They correspond to the unambiguous discrimination of:\\

1. Any set of linearly independent symmetric pure states \cite{chefles98a}.\\

2. Any pair of nonoverlapping mixed states\footnote{Any USD problem of two density matrices can be reduced to such a form according to Theorem 9.} such that the two operators $\rho_0-\alpha \sqrt{\sqrt{\rho_0}\rho_1\sqrt{\rho_0}}$ and $\rho_1-\frac{1}{\alpha} \sqrt{\sqrt{\rho_1}\rho_0\sqrt{\rho_1}}$ are positive semi-definite, and where $\alpha$ depends on the regime of the ratio $\sqrt{\frac{\eta_1}{\eta_0}}$ [chapter 4]. Note that the case of 'Any pair of two pure states' solved by Jaeger and Shimony \cite{jaeger95a} is included in this class of solutions.\\

3. Any pair of geometrically uniform mixed states of rank two in a four-dimensional Hilbert space [chapter 5]. We find that only three options for the expression of the failure probability exist. First, if the operator $\rho_0-\sqrt{\frac{\eta_1}{\eta_0}} F_0$ is positive semi-definite, then the pair of density matrices falls in the first class of exact solutions. If this is not the case, either the operator $P_1^\perp U P_1^\perp$ has one positive and one negative eigenvalue 
or it has two eigenvalues of the same sign. In the former case, we can give the optimal failure probability in terms of the eigenvalues and eigenvectors of $P_1^\perp U P_1^\perp$. In the later case, no unambiguous discrimination is possible and the failure probability simply equals unity.\\

4. A pure state and a density matrix with arbitrary {\it a priori} probabilities \cite{herzog05b}.\\

5. Any pair of mixed states with one-dimensional kernel {\cite{rudolph03a}.\\

6. Any pair of subspaces \cite{bergou06}.\\

Note that for the classes 2 and 3, we provide the optimal failure probability as well as the optimal measurement. Moreover, the solutions 4, 5 and 6 are reducible to some pure-state solutions. As we showed in this thesis, the reduction theorems and the solution for USD of two pure states are sufficient to derive those three solutions.\\

The three reduction theorems allow us to reduce USD problems to simpler cases for which the solution might be known. This is the case, as we showed in chapter 3, for the {\em unambiguous comparison of two pure states} {\cite{barnett03a,kleinmann05,herzog05a}}, the {\em unambiguous comparison of $n$ pure states having some simple symmetry\footnote{$n$ linearly independent pure states with equal {\it a priori} probabilities and equal and real overlaps.}}, {\em state filtering} \cite{bergou03a,herzog05b} and the {\em unambiguous discrimination of two subspaces} \cite{bergou06}. The reduction theorems also permit us to define a so-called standard USD problem. This problem is concerned with two density matrices of the same rank $r$ in a $2r$-dimensional Hilbert space. This is proposed as a starting point for further investigations in unambiguous state discrimination in order to avoid trivial cases or unnecessary complexity. The reductions come from simple geometrical considerations and can be summarized as follows. With the first reduction theorem, we split off any common subspace between the supports of the two density matrices $\rho_0$ and $\rho_1$. Thanks to the second reduction theorem, we eliminate, if present, the part of the support of $\rho_1$ which is orthogonal to the support of $\rho_0$ and {\it vice versa}. With the third reduction theorem, if two density matrices are block diagonal, we decompose the global USD problem into decoupled unambiguous discrimination tasks on each block. These three reduction theorems are also used to derive general theorems on unambiguous state discrimination. For example, the first reduction theorem is required to derive the two classes of exact solutions since the assumption of two density matrices without overlapping supports is made.\\
 
With respect to the applications, we have used our new tools for the unambiguous comparison of $n$ pure states with a simple symmetry\footnote{$n$ linearly independent pure states with equal {\it a priori} probabilities and equal and real overlaps.} and to answer two crucial questions\footnote{First 'With what probability can an eavesdropper unambiguously distinguish the {\it basis} of the signal?' and second 'With what probability can an eavesdropper unambiguously determine which {\it bit value} is sent without being interested in the knowledge of the basis?'.} related to the implementation of the Bennett-Brassard 1984 Quantum Key Distribution protocol. In fact we prove that the comparison of $n$ linearly independent pure states with equal {\it a priori} probabilities and equal and real overlaps, a task related to the USD of two density matrices, can be reduced to $n$ unambiguous discriminations of two pure states and can then be solved. The question to know whether any unambiguous comparison of pure states is always reducible to some pure state cases remains open\footnote{while the unambiguous comparison of mixed states is generally not reducible to some pure states case \cite{kleinmann05}}. With respect to the BB84-type protocol implemented with weak coherent pulses and a phase reference, we give the probability with which an eavesdropper can unambiguously distinguish the {\it basis} of the signal as well as the probability with which an eavesdropper can unambiguously determine which {\it bit value} is sent without being interested in the knowledge of the basis.\\

Finally, as we discussed in chapter 5, a unified expression for the failure probability for the second class of exact solutions might be a pre-condition to find new solutions in unambiguous discrimination of two density matrices. Moreover new consequences of Theorem 18 should be investigated.

%% file: Appendix.tex
\chapter{Appendix}

\section{Appendix A}

\begin{theorem} Theorem
For any operator $A$,
\begin{eqnarray}
A^\dagger A|x \rangle=0 \Leftrightarrow A^\dagger |x \rangle=0.
\end{eqnarray}
\end{theorem}

\paragraph{\bf Proof}
We show this equivalence by proving separately the two implications.\\

$\Leftarrow ] $ This direction is trivial. If $A |x \rangle=0$ then $A^\dagger A|x \rangle=0$.\\

$\Rightarrow ]$ Here we make use of a fundamental theorem of linear algebra for any linear map $A$, the kernel of $A\dagger$ equals the orthogonal complement of the image of $A$ that is to say $Ker(A^\dagger)=Im(A)^\perp$.
Let us start with a vector $|x \rangle$ such that $A^\dagger A|x \rangle=0$. $A |x \rangle$ is in the kernel of $A^\dagger$ so that $A |x \rangle$ is in $Im(A)^\perp$. Moreover, by definition, $A |x \rangle$ is in $Im(A)$. It implies that $A |x \rangle=0$.
This completes the proof. \hfill $\blacksquare$

\section{Appendix B}

\paragraph*{\bf Proof of Lemma 2}
For any operator $A$, we can introduce a polar decomposition $A=|A| \, V$ with $|A|=\sqrt{A A^\dagger}=V \, \sqrt{A^\dagger A} \, V^\dagger$. Note that $V$ is unitary and not necessarily unique, while $\sqrt{A A^\dagger}$ and $\sqrt{A^\dagger A}$ are unique and positive semi-definite. Moreover, since $|A|$ might not have full rank, let us introduce the unitary transformation $V'=ZV$ where $Z$ is a unitary matrix of the form
\begin{eqnarray}\label{form}
Z = \left( \begin{array}{cc} {\mathbb 1}_{\cal{S_{|A|}}} & 0 \\ 0 & T \end{array} \right)
\end{eqnarray}
and $T$, a unitary matrix having support on $\cal{S^\perp_{|A|}}$. From this remark, it follows that if $A=|A| \, V$ is a valid polar decomposition then $A=|A| \, V'$ is as well a valid polar decomposition. Indeed, $A=|A| \,V'=|A| \, V$ and $|A|=V'\sqrt{A^\dagger A}V'^\dagger=V\sqrt{A^\dagger A}V^\dagger$. 

We can now introduce a polar decomposition of $A$ in the quantity $\Tr(AW)$ and find
\begin{eqnarray}
|\Tr(AW)|=|\Tr(|A|VW)|=|\Tr(|A|^{1/2}|A|^{1/2}VW)|.
\end{eqnarray}
We denote $X=|A|^{1/2}=X^{\dagger}$ and $Y=|A|^{1/2}VW$ ($W$ and $V$ are both unitary matrices) and apply the Cauchy-Schwarz inequality (Theorem 2) to obtain
\begin{eqnarray}
|\Tr(AW)|=|\Tr(X^{\dagger}Y)| \le \sqrt{\Tr(|A|)} \,\, \sqrt{\Tr(W^{\dagger}V^{\dagger}|A|VW))}=\Tr(|A|) \; .
\end{eqnarray}
Equality holds if and only if $|A|^{1/2}=\beta |A|^{1/2}VW$, for some $\beta \in {\mathbb C}$. This is possible if and only if $\beta V W= R$, where $R$ is of the same form than the unitary $Z$ in Eqn. (\ref{form}). We can multiply each side with its adjoint and then find $|\beta|^2=1$. This implies that $\beta=e^{-\imath \phi}$ for some angle $\phi$ so that we find the connection $W =V^{\dagger}Re^{\imath \phi}$. Since $V$ comes from a polar decomposition of $|A|$ and $R$ is of the form of $T$, $W^\dagger$ is a valid unitary for a polar decomposition of $|A|$. This completes the proof. \hfill $\blacksquare$

\section{Appendix C}

\paragraph*{\bf Proof of Lemma 3}
To complete the proof, we see two basic properties of the supports of two positive semi-definite matrices $M$ and $N$
\begin{eqnarray}\label{basics1}
{\mathcal S}_{MN} \subset {\mathcal S}_M,
\end{eqnarray}
\begin{eqnarray}\label{basics2}
{\mathcal S}_M \subset {\mathcal S}_{M+N}.
\end{eqnarray}

The first ingredient is to see that $A:B$ is Hermitian. Indeed, we can write
\begin{eqnarray}
A(A+B)^{-1}B&=&A(A+B)^{-1}(B+A-A)\\
&=&A(A+B)^{-1}(A+B)-A(A+B)^{-1}A.
\end{eqnarray}
Let us underline that $A(A+B)^{-1}(A+B)=A \Pi_{{\cal S}_{A+B}}=A$ since ${\cal S}_{A} \subset {\cal S}_{A+B}$. Similarly $(A+B)(A+B)^{-1}A=\Pi_{{\cal S}_{A+B}} A=A$. As a result,
\begin{eqnarray}
A(A+B)^{-1}B&=&A-A(A+B)^{-1}A\\
&=&(A+B)(A+B)^{-1}A-A(A+B)^{-1}A\\
&=&A(A+B)^{-1}B.
\end{eqnarray}

Now we can prove that ${\mathcal S}_{A:B} \subset {\mathcal S}_A \cap {\mathcal S}_B$.
Indeed ${\mathcal S}_{A(A+B)^{-1}B} \subset {\mathcal S}_{A}$ and ${\mathcal S}_{B(A+B)^{-1}A} \subset {\mathcal S}_{B}$. Since $A(A+B)^{-1}B=B(A+B)^{-1}A$, it follows that ${\mathcal S}_{A:B} \subset {\mathcal S}_A \cap {\mathcal S}_B$.

The last step is to prove that ${\mathcal S}_A \cap {\mathcal S}_B \subset {\mathcal S}_{A:B}$. To do so, let $x$ be in ${\mathcal S}_A \cap {\mathcal S}_B$ and find a vector $y \in {\mathcal S}_A \cup {\mathcal S}_B$ such that $(A:B)y=x$. Actually, such a $y$ is given by $(A^{-1}+B^{-1})x$. Indeed
\begin{eqnarray}
(A:B)y&=&A(A+B)^{-1}B (A^{-1}+B^{-1})x\\
&=&B(A+B)^{-1}A A^{-1} + A(A+B)^{-1}BB^{-1}x\\
&=&B(A+B)^{-1}x + A(A+B)^{-1} x
\end{eqnarray}
since, $\forall x \in {\mathcal S}_{A} \cap {\mathcal S}_{B}$, $AA^{-1}x=x$ and $BB^{-1}x=x$. Finally we can write $(A:B)y= (B+A) (A+B)^{-1} x=x$ since $x \in {\mathcal S}_A \cap {\mathcal S}_B \subset {\mathcal S}_{A+B}$. These completes the proof. \hfill $\blacksquare$

%% file: Lebenslauf.tex
\chapter*{Curriculum Vitae}

\subsubsection{Persönliche Daten}
Geburtsdatum: 	\hfill 05.02.1978\\
Geburtsort:		\hfill Lyon, Frankreich\\

\subsubsection{Studium und Praktika}
\textbf{Abitur}	\hfill 1995\\
\noindent \textbf{Diplom-Ingenieur} an der Ecole Centrale de Marseille, Marseille, Frankreich \hfill 1998-2001\\
ex Ecole Nationale Sup\'erieure de Physique de Marseille\\
\noindent \textbf{Praktikum} im Laboratoire de Physique Nucl\'eaire et des Hautes \'Energies, \'Ecole Polytechnique, Paris, Frankreich \hfill Jul.-Sept. 2000\\
Betreuer des Praktikums: Dr. Arnd Specka\\
Thema des Praktikums: Elaboration of a calorimeter for HERA (DESY, Hamburg, Deutschland)\\
\noindent \textbf{Praktikum} im Institut Fresnel, Marseille, Frankreich \hfill Sept. 1999-Jul. 2000\\
Betreuer des Praktikums: Prof. Dr. Michel Lequime\\
Thema des Praktikums: Experimental Study of the laser ablation of a PVC surface\\
\noindent \textbf{Praktikum} im Laboratoire d'Astronomie Spatiale, Marseille, Frankreich \hfill Jul.-Sept. 2000\\
Betreuer des Praktikums: Dipl.-Ing. Philippe Lamy\\
Thema des Praktikums: Energy Cartography of the Sun\\
\noindent \textbf{Diplomarbeit} am Centre de Physique Th\'eorique, Marseille, Frankreich \hfill 2000-2001\\
Diplomvater: Dr. C. Rovelli\\
Thema der Diplomarbeit: Introduction to 2d manifold-independent spinfoam theory\\
\noindent \textbf{Promotionsstudium} am Institut für Theoretische Physik I und am Insitut für Optik, Max-Planck-Forschungsgruppe, Universität Erlangen-Nürnberg, Erlangen, Deutschland \hfill seit 2002\\